\documentclass[twocolumn,epjc3]{svjour3}  
\journalname{Eur. Phys. J. A}

\usepackage{amsfonts,amsmath,graphicx,epsfig,amssymb}
\usepackage{epsfig}
\usepackage{mathrsfs}
\usepackage{latexsym}
\usepackage{amsxtra}
\usepackage{amsbsy}
\usepackage{amscd}
\usepackage{color}
\usepackage{bbm}
\usepackage{bm}
\usepackage{hyperref}
\usepackage{multirow}
\usepackage{makecell}
\usepackage{graphics}
\usepackage{amsfonts}
\usepackage{subfigure}
\usepackage{longtable}
\usepackage{color}
\usepackage{cite}
\allowdisplaybreaks

\renewcommand{\theenumi}{\roman{enumi}}

\newcommand{\ee}{e^+e^-}
\newcommand{\NNb}{\bar NN}
\newcommand{\BBb}{\bar BB}
\newcommand{\YYb}{\bar YY}
\newcommand{\ppb}{\bar pp}
\newcommand{\LLb}{\bar \Lambda\Lambda}
\newcommand{\LcLcb}{\bar \Lambda_c\Lambda_c}
\newcommand{\SSb}{\bar \Sigma\Sigma}
\newcommand{\LSb}{\bar \Lambda\Sigma}
\newcommand{\LSbz}{\bar \Lambda\Sigma^0}
\newcommand{\SLb}{\bar \Sigma\Lambda}
\newcommand{\NXb}{\bar N\Xi}
\newcommand{\XNb}{\bar \Xi N}
\newcommand{\NLb}{\bar N\Lambda}
\newcommand{\LNb}{\bar \Lambda N}
\newcommand{\NSb}{\bar N\Sigma}
\newcommand{\SNb}{\bar \Sigma N}
\newcommand{\LXb}{\bar \Lambda\Xi}
\newcommand{\XLb}{\bar \Xi\Lambda}
\newcommand{\SXb}{\bar \Sigma\Xi}
\newcommand{\XSb}{\bar \Xi\Sigma}
\newcommand{\XXb}{\bar \Xi\Xi}

\newcommand{\OOb}{\overline \Omega\Omega}
\newcommand{\SSbp}{\bar\Sigma^-\Sigma^+}
\newcommand{\SSbz}{\bar\Sigma^0\Sigma^0}
\newcommand{\SSbm}{\bar\Sigma^+\Sigma^-}

\def\Tr{\mathop{\rm Tr}\nolimits}
\def\M{{\cal{M}}}
\def\Re{\mathop{\rm Re}\nolimits}
\def\Im{\mathop{\rm Im}\nolimits}

\begin{document}

\title{The antibaryon-baryon interaction based on chiral EFT\\ with a focus
on $\bar pp$ and $\bar\Lambda\Lambda$ reactions}

\author{Qin-He Yang\thanksref{addr1,e1}
\and Johann Haidenbauer\thanksref{addr1,e2}
\and Ulf-G. Mei{\ss}ner\thanksref{addr2,addr1,addr3,e3}
}
\thankstext{e1}{e-mail: qi.yang@fz-juelich.de}
\thankstext{e2}{e-mail: j.haidenbauer@fz-juelich.de}
\thankstext{e3}{e-mail: meissner@hiskp.uni-bonn.de}

\institute{{Institute for Advanced Simulation (IAS-4), 
Forschungszentrum J\"ulich, D-52425 J\"ulich, Germany} \label{addr1}
           \and
           {Helmholtz-Institut~f\"{u}r~Strahlen,~Kernphysik~and~Bethe~Center~for~Theoretical~Physics~and\\~Cluster~of~Excellence~``Color~meets~Flavor'',
~Universit\"{a}t~Bonn,~D-53115~Bonn,~Germany} \label{addr2}
           \and 
           {Peng Huanwu Collaborative Center for Research and Education,
International Institute for Interdisciplinary and Frontiers, Beihang University, Beijing 100191, China} \label{addr3}
%           \and 
%          {CASA, Forschungszentrum J\"ulich, D-52425 J\"ulich, Germany} \label{addr4}
}

\date{\today}

\maketitle

\begin{abstract}
The antibaryon-baryon ($\bar BB$) interaction is reviewed in the context
of the hyperon(Y)-production reactions $\bar NN\to\bar YY$ with special emphasis 
on the $\bar \Lambda\Lambda$ system. 
The experimental situation for $\bar BB\to\bar YY$ from threshold up to 
antiproton lab momenta of $2$~GeV/c is discussed and likewise 
past theoretical approaches and results. 
The formalism for a description of the $\bar BB$ 
interaction in terms of potentials derived within
SU(3) chiral effective field theory (EFT) is outlined. 
Constraints from SU(3) flavor symmetry on the
$\BBb$ interaction in the various reaction channels are discussed.  
As concrete application a new analysis of the reaction 
$\bar pp\to\bar \Lambda\Lambda$ is performed
utilizing potentials for the $\bar NN$ and $\bar \Lambda\Lambda$ channels,
and for the $\bar NN\to\bar\Lambda\Lambda$ transition, 
up to next-to-leading order in the chiral 
expansion, supplemented by a phenomenological part to account for 
annihilation processes. An excellent reproduction of data from the 
PS185 experiment at CERN up to momenta of $1.55$~GeV/c is
achieved. When including higher-order contact terms in the analysis 
the range of agreement can be extended up to momenta around $1.65$~GeV/c, 
where a complete set of scattering observables is available.
Predictions for $\LLb$ scattering are provided and compared 
to results from the literature. 
\end{abstract}

\section{Introduction}
The antinucleon-nucleon ($\NNb$) interaction has been
fair\-ly extensively investigated over the years, experimentally, see
\cite{Amsler:1991wbb,Dover:1992vj,Klempt:2002ap,Zhou:2012ui} and
references therein, as well as theoretically 
\cite{Dover:1980pd,Dover:1981pp,Cote:1982gr,Timmers:1984xv,Hippchen:1991rr,Mull:1991rs,Mull:1994gz,Entem:2006dt,El-Bennich:2008ytt,Kang:2013uia,Dai:2017ont,Ortega:2024zjx}.
For other anti\-baryon-baryon ($\BBb$) systems, like those 
involving the hyperons $\Lambda$, $\Sigma$, and $\Xi$ (collectively called $Y$) , the situation is
much less satisfying. Indeed, due to their short lifetime
their interaction can be only studied in production 
reactions where these baryons appear in the final state. 
With regard to that, pioneering measurements have been performed
in the course of the PS185 experiment at the LEAR facility at CERN in
the 1990ies where not only integrated cross sections for 
$\ppb\to\LLb$ have been established but also
a wealth of differential and spin-dependent observables \cite{Klempt:2005pp}. 
The reaction $\ppb\to\LSbz$+c.c. has been explored in detail as well. 

In recent times the $\YYb$ interaction received again increased attention 
due to experiments by the BaBar \cite{BaBar:2007fsu,BaBar:2014cco}, 
Belle \cite{Belle:2008kpj}, 
and BESIII \cite{BESIII:2012jve,BESIII:2017hyw,BESIII:2021fqx,BESIII:2022cxi,BESIII:2022tvj,BESIII:2022fhe,BESIII:2026btb}
Collaborations and by the GlueX experiment at JLab \cite{GlueX:2025yql}, 
where $\YYb$ pairs are produced in the decay of heavy mesons 
or in $\ee$- and photon-induced reactions. 
Additionally, there have been measurements of $\LLb$
correlation functions in heavy-ion collisions
and in high energy $pp$ collisions by the ALICE Collaboration
\cite{ALICE:2019igo,ALICE:2021cyj}. 
In $e^+e^-$-induced experiments also the $\BBb$ interaction 
involving baryons with strangeness $|S|>1$ have been studied, i.e. 
for $\XXb$ \cite{BESIII:2020ktn,BESIII:2021aer}
and $\OOb$~\cite{BESIII:2022kzc}, and even 
$\LcLcb$~\cite{Belle:2008xmh,BESIII:2023rwv} has been explored. 
Moreover, not only $\BBb$ systems with total strangeness $S=0$ have
been investigated but also some with $S=1$ and $-1$
like $\bar\Lambda p$ etc., see Ref.~\cite{Haidenbauer:2024smo} 
for pertinent references. 
Among the various experiments the strongest interest among theorists 
was evoked by the reactions $\ee\to\LLb$ and $\ee\to\LSbz$+c.c. since 
they allow one to determine the electromagnetic form factors 
of the $\Lambda$ in the time-like region and of the $\Lambda$-$\Sigma^0$ transition form factors, respectively, 
see the reviews \cite{Denig:2012by,Dai:2024lau}. 

In the present manuscript we revisit the reaction $\NNb\to\YYb$ in
light of those recent developments. Given the already mentioned impressive
effort by the PS185 experiment the quality and diversity of their data on
strangeness production in $\ppb$ collisions constitutes 
a substantial and, in our opinion, a decisive constraint on the $\YYb$ 
interaction, in particular for the $\LLb$ system.  
First, we provide a short review of
the PS185 experiments and of the theoretical approaches that were 
applied to study the data on the $\LLb$ and $\LSb$+c.c. production.
For the main part we focus on a re-analysis of those data 
utilizing state-of-the-art techniques. Specifically, instead of the 
phenomenological $\BBb$ potentials used in the past now the interactions 
employed in the analysis are constructed along the lines of chiral
effective field theory (EFT), see e.g. \cite{Epelbaum:2008ga},
in order to meet modern standards. Thereby, 
we can build upon our successful application of this approach to 
the $\Lambda N$, $\Sigma N$ \cite{Haidenbauer:2013oca,Haidenbauer:2019boi,Haidenbauer:2023qhf}, 
and $\NNb$ systems \cite{Kang:2013uia,Dai:2017ont}.
For $\ppb\to\LLb$, where a wealth of spin-dependent observables is available, a detailed analysis is
performed with the aim to pin down the partial-wave contributions to the reaction
amplitude and to establish the interaction in the $\LLb$ system. 

The paper is structured in the following way: 
In the subsequent section an historical overview of past studies of the 
reactions $\NNb\to\YYb$ is given. 
Sect.~\ref{sec:PS185} provides a summary of the $\NNb\to\YYb$ experiments
performed in the PS185 experiment at LEAR. 
The treatment of the $\BBb$ interaction within SU(3) chiral EFT is described
in detail in Sect.~\ref{sec:BBbar}. Further details such as the SU(3) 
factors and the isospin coefficients are compiled in appendices. 
The formalism used for calculating $\YYb$ observables is outlined in Sect.~\ref{sec:Obs}. The achieved results are presented in two sections.
Those for $\NNb$ scattering in the momentum range relevant for $\YYb$
production are covered in Sect.~\ref{sec:NNb}. 
Results for $\NNb\to\LLb$ and for $\LLb$
scattering are presented and discussed in 
Sect.~\ref{sec:LLb}.  
The paper ends with a brief summary and an outlook. 
%%%%%%%%%%%%%%%%%%%%%%%%%%%%%%%%%%%%%%%%%%%%%%%%%
\section{Historical reminiscences}
\label{sec:history}

The Low Energy Antiproton Ring (LEAR) at the CERN facility
in Switzerland was built in 1982 and operated until 1996. 
A comprehensive summary of the physical achievements at
LEAR can be found in the reviews by Klempt et al.~\cite{Klempt:2002ap,Klempt:2005pp}.
The production of strange baryons ($\Lambda$, $\Sigma$) was
investigated in the PS185 experiment. The results of PS185 are summarized in the review from 2002~\cite{Klempt:2002ap}, which provides also an overview of the theoretical investigations. 
A much shorter review was published in 2022 by Zhou et al.~\cite{Zhou:2022jwr}.

A proposal in the 1990ties to extend the momentum range of
LEAR to momenta beyond $2$~GeV/c so that $\Xi$'s 
but also charmed baryons like $\Lambda_c$
can be produced (SuperLEAR)~\cite{Dover:1991gc} was not approved. Unfortunately, also
more recent plans for $\ppb\to\YYb$ experiments at the FAIR facility
by the PANDA Collaboration
\cite{Papenbrock:2016bum,PANDA:2020zwv,PANDA:2021ozp}
could not be realized so far. An overview of the kinematics is provided
in Table~\ref{tab:thr}. 

\begin{table}[htbp!]
    \centering
        \caption{$\NNb$ kinematics at the thresholds of different $\YYb$ channels. 
    For $\SSb$ and $\XXb$ averaged masses are used.
    All quantities are in MeV or MeV/c, respectively.}
    \begin{tabular}{c| c c c c c|c}
    \hline\hline
         & $\LLb$ & $\LSb$& $\SSb$ & $\XXb$ & $\OOb$ & $\LcLcb$ \\
        \hline
        $\varepsilon$          & 355  & 432  & 510 & 760 & 1468 & 2697\\
        $T_{\rm lab}$          & 777  & 963  & 1155 & 1827 & 4086 & 9269\\
        $p_{\rm cm}$           & 604  & 672  & 737 & 926 & 1384 & 2085\\
        $p_{\rm lab}$          & 1435  & 1655  & 1874 & 2601 & 4936 & 10164 \\
        \hline
    \end{tabular}

    \label{tab:thr}
\end{table}

The reactions $\ppb\to \LLb$ and $\ppb\to\LSb$+c.c. were investigated already 
in the early 1960s, 
experimentally \cite{Button:1961zzb} as well as theoretically 
\cite{Sopkovich:1962zz}. However, the prospect of the LEAR facility and, 
specifically, the PS185 experiment
boosted the theoretical interest in those reactions significantly. 
The experiments at LEAR not only offered
unprecedented accuracy of the data on integrated and differential cross
sections but also on spin-dependent observables,
due to the self-analyzing property of the weak decay
of the hyperons. Moreover, it enabled access to momenta extremely
close to the thresholds of the reactions, i.e. for excess
energies $\varepsilon = \sqrt{s}-m_{\bar Y}-m_Y$ less than
$1$~MeV. The latter was
seen as a doorway to the understanding of the production
mechanism of hyperons. Indeed, several works published just
before the emergence of the first PS185 data \cite{Barnes:1987aw} considered 
the reaction mechanism in the quark picture where, in the simplest case, a 
$\bar uu$ (or $\bar dd$) pair is annihilated and a $\bar ss$ pair
is created \cite{Genz:1984ig,Rubinstein:1985bb,Kohno:1986mg,Furui:1987cc}. 
In the works \cite{Genz:1984ig,Rubinstein:1985bb,Kohno:1986mg}
it was assumed that the annihilation/creation of the $\bar qq$ pair is
mediated by a gluon exchange and the process was referred to as $^3S_1$
model, reflecting the quantum numbers of the gluon. In Ref.~\cite{Furui:1987cc}
it was assumed that the quark pair is annihilated/created with the quantum
numbers of the vacuum, the so-called $^3P_0$ model. Later works 
considered actually both mechanisms~\cite{Burkardt:1988pk}, while 
Alberg and collaborators \cite{Alberg:1988zr,Alberg:1993tw} argued that 
one should even employ a combination of the two production mechanisms. 
 
In addition more sophisticated production mechanisms were considered, 
like by Roberts~\cite{Roberts:1990yh}, which involve four-quark states 
as intermediates, called the ``diquonium intermediate'' model. 
In Ref.~\cite{Kroll:1987pj} a
model was employed where, besides quarks, diquarks are introduced as elementary
constituents of the baryons. Strangeness production proceeds then via quark
and diquark annihilation/creation through one gluon. Finally, an entirely
different ansatz is followed in the constituent quark model by the Salamanca group
\cite{Ortega:2011zza,GarciaOrtega:2012fmb}. In this case Goldstone bosons are
exchanged between quarks and prompt the strangeness production. 

In the conventional meson-exchange picture the transition from $\NNb$ to 
$\YYb$ is realized by the exchange of strange mesons. Here, first of 
all the $K$(495) is of relevance, a Goldstone boson and the lightest strange meson,
but others like the $K^*$ (890, $J^P=1^-$), $K_2^*$ (1430, $2^+$), and the
$K_0^*$ (700, $0^+$) (formerly $\kappa$) have been considered too. 
Calculations within the meson-exchange framework 
have been presented by various authors, see Refs.~\cite{Tabakin:1985yv,Niskanen:1985gr,Kohno:1986mg}, 
as well as by the J\"ulich group~\cite{Hippchen:1996zv}, 
in anticipation of the PS185
experiments. Note, however, that calculations in the 
meson-exchange framework had been already performed at a much earlier stage,
namely at the time when the
first data on the reaction $\ppb\to\LLb$ became available around 1960
\cite{Sopkovich:1962zz,Bessis:1963}. Results for meson-exchange potentials 
that include a comparison with the PS185 data can be found in 
\cite{Kohno:1987be,Kohno:1987uj,LaFrance:1990kb,Timmermans:1990tt,Haidenbauer:1991kt,Timmermans:1992fu,Wang:2025cex}. 

An essential aspect of the strangeness production is the role played by
the interaction in the initial ($\NNb$) and final ($\YYb$) states. Indeed,
it was already realized at a very early stage that cross sections calculated
in the (plane-wave) Born approximation are way too large and completely
unrealistic
\cite{Sopkovich:1962zz}, see also \cite{Kohno:1986mg,LaFrance:1988if}. 
Specifically, the $\NNb$ interaction characterized by sizable annihilation processes leads to a drastic reduction of the transition cross section,
usually by more than one order of magnitude \cite{Kohno:1986mg}. 
Note also that spin-dependent observables like the polarization are simply
zero in Born approximation.  
In some of the works done in the Born approximation effects of the
initial-state and final-state interactions (ISI and FSI, respectively) are accounted 
for in an approximate way by means of the so-called Sopkovich factor
\cite{Sopkovich:1962zz}. These factors simulate the S-matrices in the
initial- and final state and are multiplied to the Born results. They  
are equipped with free parameters that are adjusted to the empirical 
production cross section in the application. 
For a meaningful comparison with data it is preferable to
evaluate the strangeness-production process at least 
in distorted wave Born approximation (DWBA) or, better,
within a coupled-channels treatment. 
Of course, this requires also a realistic
description/potential of the $\NNb$ interaction (in the relevant energy range)
and of the $\YYb$ systems. 

Amplitude- and/or partial-wave analyses for $\ppb\to \LLb$
were performed by several authors. The
earliest one is from
1989 \cite{Kudryavtsev:1988gk} and considered the energy
range up to $\varepsilon\approx 25$~MeV ($1507.6$~MeV/c), where
only differential cross sections and polarizations are 
available. $S$- and $P$-wave amplitudes were included in a 
scattering-length approximation. A comprehensive analysis was
carried out in Ref.~\cite{Tabakin:1991ct}
in 1991 who considered data up to $1695$~MeV/c. An amplitude-
as well as a partial-wave analysis was performed (the latter
in a scattering length approach), and it was found that
partial waves up to $L=3$ are needed to describe the PS185
data up to $1546$~MeV/c and up to $L=6$ at $1695$~MeV. 
The analysis in Ref.~\cite{Schneider-Neureither:1992iwo}
was again performed in the scattering-length approximation
(up to $1546$~MeV/c) and included partial waves up to $L=2$.
An attempt was made to determine not only $\ppb\to \LLb$
but also the $\LLb\to\LLb$ amplitudes. Results of a ``model-independent''
partial-wave analysis based on data up to $1546$~MeV/c were
also reported in Ref.~\cite{Timmermans:1992fu} utilizing the 
so-called Nijmegen method \cite{Bergervoet:1990zy}. Partial 
waves up to $L=2$ in the $\LLb$ system were included.  
Finally, an extensive partial-wave analysis was presented in Ref.~\cite{Bugg:2004rj} 
in 2004, considering data up to
$1990.6$~MeV/c, where for the first time results for the spin-transfer coefficients
$D_{nn}$ and $K_{nn}$ as well triple-spin observables were included.
In this work partial waves up to $L=4$ were found to be 
of relevance. 

Among the issues most prominently discussed in the literature regarding the
reaction $\ppb \to \LLb$ is the dominance of $P$-waves
at very low energies. Indeed, already at the lowest
measured energy, less than $1$~MeV from the threshold, the differential cross
section is anisotropic and the polarization is 
non-zero \cite{Barnes:1989je,Barnes:1994bh,Barnes:2000be}, signaling 
the presence of $P$-wave contributions. Additionally, in the empirical
analysis in Ref.~\cite{Barnes:1989je} but also in some of the model
calculations \cite{Haidenbauer:1991kt} $P$-waves even dominate the
reaction cross section from around $\varepsilon \approx 15$~MeV onward. In 
fact, in some of the calculations this is the case already for 
$\varepsilon \gtrsim 7$~MeV~\cite{LaFrance:1990kb}. An initial interpretation 
considered this as
a $P$-wave enhancement and as indication for $P$-wave resonances
close to or at the threshold \cite{Dalkarov:1987jt,Dalkarov:1990ra}. 
Later investigations
attributed the observation to a more conventional and unspectacular reason,
namely an $S$-wave suppression caused by the presence of a strong
annihilation in the $\NNb$ and $\LLb$ systems. Indeed, practically all
calculations, where the effects of the ISI and FSI are taken into account
in a realistic way, can describe the energy dependence of the PS185 data 
close to the threshold without any exotic ingredients. 

Another aspect of the production mechanism that was widely discussed in
the literature is the dominance of the spin-triplet configurations. 
Experimental evidence for the dominance of the triplet contributions comes
from the so-called spin-singlet fraction, a linear combination of the 
spin-correlation
parameters $C_{xx}$, $C_{yy}$, and $C_{zz}$ (see below for their definition), 
\begin{equation}
    F_S = \frac{1}{4} \left( 1 + C_{xx}-C_{yy}+C_{zz}\right)\, ,
    \label{eq:FS}
\end{equation}
which can take up values between $0$ and $1$ but was found to be close to
$0$ in the PS185 measurements~\cite{Barnes:1990bs}. Interestingly, 
meson exchange as well as the production mechanisms in the simple quark model
lead to a dominance of spin-triplet states, though for quite different
reasons. In the former the tensor force
from $K$ exchange together with the centrifugal barrier in the final-state
favors transitions where the orbital angular momentum changes by two units,
i.e. $L_{\LLb} = L_{\ppb}-2$. Such transitions can only occur for triplet
states. This feature is further enhanced by inclusion of
$K^*$ exchange where then the corresponding tensor components add up whereas
the corresponding central transition potentials cancel each other to a 
certain extent~\cite{Haidenbauer:1991kt}. 
In the simple quark picture the singlet fraction 
is even identically zero, in the $^3S_1$ and in the $^3P_0$ model.  
Because of the Pauli principle, the two spectator
quarks in the $\Lambda$ have to be in a spin-singlet state. Therefore, 
these models act as a triplet filter also for the $\LLb$ system itself. 

In this context let us mention that in Refs.~\cite{Haidenbauer:1992gv,Haidenbauer:1992wp} it was suggested to explore
additional observables that are sensitive to the differences in the underlying
production mechanisms of the meson-exchange picture and of the simple
quark model pair-creation process. The observables allegedly appropriate
for that are the spin-transfer coefficients $D_{nn}$ and $K_{nn}$. However, 
already in 1996 some doubts on the conclusiveness of the suggestion were
raised in Ref.~\cite{Richard:1996bb}, based on the interrelation between
those quantities and the spin-correlation parameters $C_{ij}$ and the fact 
that the latter are not well reproduced by the interactions considered in
Refs.~\cite{Haidenbauer:1992gv,Haidenbauer:1992wp}. The experimental 
results, finally published in 2004 \cite{Bassalleck:2004dz} and 
2006 \cite{PS185:2006yyx}, respectively, confirmed the scepticism since the
actual measurement was neither in line with the prediction from meson exchange
nor with those from simple quark models. 

One issue that comes up periodically is the possibility 
of $S$-wave bound states and/or resonances near the $\LLb$ threshold. 
For example, the  
$\xi(2230)$ (now called $f_J(2220)$ \cite{ParticleDataGroup:2024cfk}),  
a narrow state observed in $J/\psi$ radiative decays to $\bar KK$, whose mass
practically coincides with the $\LLb$ threshold, was suggested as a possible
candidate for that \cite{Ono:1986rj}. 
The state has been observed in experiments by the Mark III 
\cite{MARK-III:1985qfw} and by the BES \cite{BES:1996upd} Collaborations
but was not confirmed
in more recent measurements by BaBar \cite{BaBar:2010bwc}. 
In Ref.~\cite{Carbonell:1993dt} a possible $\LLb$ subthreshold state was
deduced directly from the PS185 data \cite{Barnes:1989je,Barnes:1994bh}. 
Indeed, those early
measurements at very low energies exhibited a shoulder in the reaction
cross section around $\epsilon=0.5-1.0$~MeV which one could describe by 
introducing a bound state. However, a more precise re-measurement of the
energy region in question did not confirm the presence of a structure \cite{Barnes:2000be}. 
A state likewise deduced from the PS185 data has been suggested
by Bugg~\cite{Bugg:2004rj}, as a result of a partial-wave analysis.
The parameters of the resonance are $m=2290\pm 20$~MeV, 
$\Gamma = 270\pm 35$~MeV. Given the rather large width, it
was speculated in \cite{Haidenbauer:2023zcu} that this resonance could be
simply an effective parameterization of the FSI effects in the 
$^3S_1$-$^3D_1$ partial wave rather than a genuine physical state.
More recently, a large non-zero cross section was observed
in the reaction $\ee\to\LLb$ in a BESIII experiment \cite{BESIII:2017hyw},  
barely $1$ MeV away from the $\LLb$ threshold. Its description  
requires the inclusion of a so far unobserved narrow resonance, as
shown in Refs.~\cite{Cao:2018kos,Xiao:2019qhl}. 
Last but not least, the $X(2239)$ and $\eta(2225)$ have been interpreted as
hidden-strange molecular states from the $\LLb$ interaction
\cite{Zhu:2019ibc}. 

Finally, some works addressed the question of 
how well is the $\LLb$ interaction constrained by the production data.
Results for $\LLb$ scattering (mainly integrated cross sections) are shown 
in Refs.~\cite{LaFrance:1990kb,Timmermans:1992fu,Haidenbauer:1991kt,GarciaOrtega:2012fmb}. 
For $\LLb$ results deduced from a
partial-wave analysis see Ref.~\cite{Schneider-Neureither:1992iwo}.

There are far less theoretical studies of other hyperon-production reactions
like $\ppb\to\LSbz$+c.c., $\ppb\to\SSb$, or $\ppb\to\XXb$
\cite{Haidenbauer:1992hv,Haidenbauer:1993ws,Ortega:2012HI,Haidenbauer:1993wca}.
This simply reflects the fact that there have been and, for the latter two, 
still are much less data available for those channels. 
Regarding the production of charmed baryons, the reaction $\ppb\to\LcLcb$ 
has been extensively investigated at some stage, see Refs.~\cite{Haidenbauer:2009ad,Haidenbauer:2016pva}
and references therein, 
namely at a time where pertinent experiments by the PANDA Collaboration \cite{PANDA:2009yku} seemed imminent.

%%%%%%%%%%%%%%%%%%%%%%%%%%%%%%%%%%%%%%%%%%%%%%%
\section{The PS185 data}
\label{sec:PS185}

Data from LEAR by the PS185 experiment have been published
in 
\cite{Barnes:1987aw,Barnes:1989je,Barnes:1990bs,Barnes:1994bh,Barnes:1996si,Barnes:2000be,PS185:2002tao,Bassalleck:2004dz,PS185:2006yyx,Johansson:1999emx}.
An over\-view of the reactions, the momenta/excess energies
and the observables is provided in Table~\ref{tab:La}.

\begin{table}[tbp]
\renewcommand{\arraystretch}{1.2}
\centering
\caption{
Summary of PS185 data on $\ppb\to\YYb$. The excess energy in
the $\YYb$ channels is denoted by $\varepsilon$. 
}
\begin{tabular}{cccc}
\hline \hline
$p_{lab}$ (MeV/c) & $\varepsilon$ (MeV) & Observables & Ref. \\
\hline \hline
\multicolumn{4}{c}{$\ppb\to\LLb$} \\
\hline
$1435.9$-$1445.3$ & $0.85-4.05$ & $\sigma$, $d\sigma/d\Omega$, $P_y$ &
\cite{Barnes:1989je} \\
$1439.5$-$1567.4$ & $0-45.5$ & $\sigma$, $d\sigma/d\Omega$ &
\cite{Barnes:1994bh} \\
$1436.9$-$1448.9$ & $0-5.5$ & $\sigma$, $d\sigma/d\Omega$,
$P_y$ & \cite{Barnes:2000be} \\
$1476.5$, $1507.5$ & $14.6$, $25.5$ & $\sigma$, $d\sigma/d\Omega$, $P_y$  &
\cite{Barnes:1987aw} \\
$1546$, $1695$ & $39.1$, $91.7$ & $\sigma$, $d\sigma/d\Omega$, $P_y$,
$C_{ij}$&
\cite{Barnes:1990bs} \\
$1642$, $1918$ & $73$, $170$ & $\sigma$, $d\sigma/d\Omega$, $P_y$, $C_{ij}$ &
\cite{Barnes:1996si} \\
$1726$, $1771$ & $103$, $119$ & $\sigma$, $d\sigma/d\Omega$, $P_y$, $C_{ij}$ &
\cite{Barnes:1996cf} \\
$1637$ & $71$ & $D_{nn}$, $K_{nn}$ &
\cite{PS185:2002tao} \\
$1525$ & $31.5$ & $P_y$, $D_{nn}$, $K_{nn}$ &
\cite{Bassalleck:2004dz} \\
$1637$ & $71$ & $\sigma$, $d\sigma/d\Omega$, etc. &
\cite{PS185:2006yyx} \\
\hline \hline
\multicolumn{4}{c}{$\ppb\to\LSbz$+c.c.} \\
\hline 
$$ & $0-2.5$ & $\sigma$ &
\cite{Barnes:2000be} \\
$1695$ & $14.8$ & $\sigma$, $d\sigma/d\Omega$ &
\cite{Barnes:1990wh} \\
$1726$, $1771$ & $26$, $40$ & $\sigma$, $d\sigma/d\Omega$, $P_y$, $C_{ij}$ &
\cite{Barnes:1996cf} \\
\hline \hline 
\multicolumn{4}{c}{$\ppb\to\SSbp\,(\SSbm)$} \\
\hline
$1922$ & $23\, (7)$ & $\sigma$, $d\sigma/d\Omega$ &
\cite{PS185:1997oge,Johansson:1999emx} \\
\hline \hline
\end{tabular}
\label{tab:La}
\renewcommand{\arraystretch}{1.0}
\end{table}

First data, on $\ppb\to\LLb$, were 
published in 1987 and consisted of $\sigma$, $d\sigma/d\Omega$, $P_n$
at $1.4765$ and $1.5075$~MeV/c \cite{Barnes:1987aw}. The first data
on $\ppb\to\LSb$+c.c. appeared in 1990 \cite{Barnes:1990wh} and
on the $\SSb$ channels in 1997 \cite{PS185:1997oge}. 
The last PS185 publication appeared in 2006 (for $1637$~MeV/c) and included 
the first data on triple-spin observables~\cite{PS185:2006yyx}. 
For the excess energies of $14.6$, $25$, and $40$~MeV data on $\LLb$ as well
as $\LSb$+c.c. production are available which allows to compare 
the observables of the two reactions at the same $\varepsilon$. 
At $25$~MeV even data on $\SSbp$ are available. 
The upper limit of the LEAR facility was $2$~GeV/c 
($\varepsilon_{\LLb} = 199$~MeV).

Besides the officially published results there are data 
at further energies which can only be found in doctoral theses. 
Regarding $\LLb$, 
%Fritz Stinzing~\cite{Stinzing:1991} ($1.4522$-$1.5064$~GeV/c),
%(only data at lower momenta have been published!)\\
in the thesis of Ralf Todenhagen~\cite{Todenhagen:1995} results for 
$1653$~MeV/c and $1658$~MeV/c ($\varepsilon=77$ and $79$~MeV) are reported.
Energies around the $\SSbm$ threshold ($1889.7-1906.8$~MeV/c) are
covered by the thesis of Thomas Sefzick~\cite{Sefzick:1992}.  
Note that the $\SSbp$, $\SSbz$, and $\SSbm$
thresholds are at $1853.2$, $1871.1$, and $1898.9$~MeV/c,
respectively. The thesis of 
Ralph-Andre~Kraft~\cite{Kraft:1994} presents data at $1835$~MeV/c 
($\varepsilon=141$~MeV) while in the one by Ralf W.~Br\"oders~\cite{Broders:1995} results for $1949$~MeV/c and $1992$~MeV/c 
($\varepsilon=181$ and $196$~MeV) can be found. 

Additional data on $\LSbz$+c.c. can be found in the thesis of
Kirsten~Sachs~\cite{Sachs:1996}, namely for the momenta  
$1949$~MeV/c and $1992$~MeV/c ($\varepsilon=104$ and $119$~MeV).

Regarding the reaction $\ppb\to\XXb$ there are no data from PS185 because, 
as already mentioned, the upper limit of the LEAR facility was $2$~GeV/c.
In any case, as far as we know for this reaction only upper limits have been
ever published in the past. 

Pre-LEAR $\LLb$ cross-section data in the momentum region studied by us 
\cite{Button:1961zzb,Oh:1973ny,Jayet:1978yq} will be included in the 
figures shown below, but are not taken into account in our fitting procedure.
 
\subsection{Remarks on the $\LLb$ data}
\label{sec:LLbdata}
\begin{itemize}
\item[1)] 
The data on the $\LLb$ cross sections in Ref.~\cite{Barnes:2000be} and in 
the last PS185 publication \cite{PS185:2006yyx} differ in the overall
magnitude from the earlier results. %, see Fig.~\ref{fig:LLb}. 
The discrepancy was first discussed in Ref.~\cite{Barnes:2000be}, see 
page~5, and is in part due an incomplete (pre-GEANT) Monte Carlo used
in older analyses. Furthermore, in that work it is argued that the observed
difference in the cross sections is largely accounted for by the approximately $15$~\% overall normalization uncertainty estimated for the earlier experiments. 
Specifically, an upward adjustment/correction of approximately $8$-$12$~\% is 
suggested. Thus, in the present work we have re-scaled all pre-2000 
cross-section values uniformly by a factor of $1.1$.
In addition, following 
the work of the Nijmegen group \cite{Timmermans:1992fu}, we allow for an
individual re-normalization of each data set in the
fitting procedure, i.e. the normalization is ``floated''. This is required
for a consistent analysis because, for example as pointed out in
Ref.~\cite{PS185:2006yyx}, the data at $1637$~MeV/c from 2006 require 
a factor $1.26$ for reconciliation with those at the near-by momentum of
$1642$~MeV/c from 1996~\cite{Barnes:1996si}.
 
Note also that the near-threshold data from Ref.~\cite{Barnes:2000be} 
supersede the old results from Refs.~\cite{Barnes:1989je,Barnes:1994bh}. 
The latter two data sets are obsolete and are left out in our fitting procedure. 
This concerns also the differential cross sections and the analyzing
powers in those publications.

\item[2)]
There has been a new determination of the weak decay constant of the 
$\Lambda$~\cite{BESIII:2018cnd,Ireland:2019uja}, $\alpha_\Lambda$, after 
the LEAR experiments. The new value is around $0.75$
\cite{ParticleDataGroup:2024cfk}, while the old one was 
around $0.64$. Accordingly the polarizations should be scaled by 
a factor $0.64/0.75\approx 0.85$, and the spin-correlation parameters 
by a factor
$(0.64/0.75)^2\approx 0.73$. The spin-singlet fraction $F_S$ has to be determined anew, based on Eq.~(\ref{eq:FS}). 
In any case, the latter quantity is not included in the fitting procedure
because it is not an independent observable. 
In this context let us mention that there were tensions between the measured
values and the physically allowed range for several observables published
from the PS185 experiment. With the new value for $\alpha_\Lambda$ the
situation has significantly improved as we will see below. 
Also, for curiosity reasons we want to
mention that an attempt to determine the decay constant directly from the
$\ppb\to\LLb$ data has been made in the thesis of 
Paschke \cite{Paschke:2001}, by imposing all physical constraints among
others, see Appendix~A.4 of that work. May be not surprisingly, the result achieved for
$\alpha_\Lambda$ there agrees pretty well with the new value. 
\item[3)]
There has been a change in the mass of the $\Lambda$ between 1992 and 1994, 
namely from $m_\Lambda =1115.63\pm 0.05$~MeV to now 
$m_\Lambda=1115.683\pm 0.006$~MeV \cite{ParticleDataGroup:2024cfk}.
which in principle could have an impact on the data very close to the threshold \cite{Barnes:1989je,Barnes:1994bh}.
However, the resulting shift in the $\LLb$ excess energy of about
$0.1$~MeV is smaller than the accuracy of the beam momentum
at LEAR, which produces an uncertainty of $0.35$~MeV  \cite{Barnes:1994bh}.
In fact, the uncertainty in the mass from 1992 itself induces a spread
of $0.1$~MeV. 
In any case, as already mentioned in 1), with the new near-threshold data 
from 2000 \cite{Barnes:2000be} the old ones have become obsolete anyway.
\end{itemize}

\subsection{Data on $\ppb$ scattering}

Data on $\ppb$ scattering have been obtained at 
LEAR \cite{Klempt:2002ap,Klempt:2005pp}
but also in many other experiments. We focus here only on those that
are in the momentum region relevant for our study of $\YYb$ production,
which means between $1.4$ and $2$ GeV/c. 
Specifically, we consider 
integrated (total, elastic, charge-exchange) cross sections from Refs.
\cite{Abrams:1969jm,Coupland:1977zr,Parker:1971pea,Cutts:1974vi}
and 
differential $\ppb$ cross sections from Refs.
\cite{Eisenhandler:1976kf,Kunne:1988kg,Daum:1968xea}. 
Experimental results for analyzing powers are taken from 
Refs.~\cite{Kunne:1988kg,Daum:1968xea}. 

%%%%%%%%%%%%%%%%%%%%%%%%%%%%%%%%%%%%
\section{The $\bar{B}B$ interaction potential}
\label{sec:BBbar}

The employed $\bar{B}B$ potentials consist of an elastic part and
a part which takes into account effectively the presence of annihilation
processes. 
The elastic part of the $\bar BB$ interaction is derived within SU(3) 
chiral effective field theory ($\chi$EFT) up to next-to-leading order (NLO),
while the annihilation part is described phenomenologically
in terms of an optical potential. 
Below we start with describing the potential from $\chi$EFT and then
we proceed to the phenomenological part. 

\subsection{Elastic part of the $\bar BB$ interaction}

\begin{figure*}[t]
 \centering
 \includegraphics[width=0.11\linewidth]{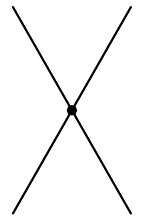}
 \includegraphics[width=0.11\linewidth]{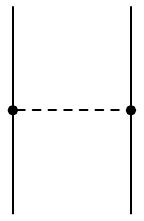}
 
 \includegraphics[width=0.11\linewidth]{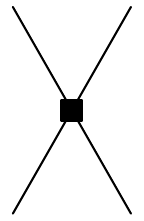}
 \includegraphics[width=0.11\linewidth]{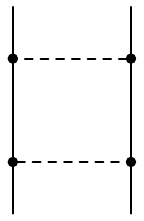}
 \includegraphics[width=0.11\linewidth]{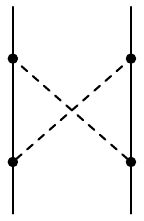}
 \includegraphics[width=0.11\linewidth]{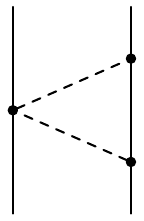}
 \includegraphics[width=0.11\linewidth]{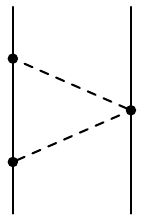}
 \includegraphics[width=0.11\linewidth]{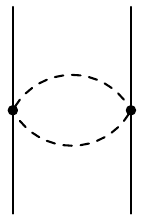}
 \caption{Relevant diagrams up-to-and-including next-to-leading order. Solid and dashed lines denote octet baryons and pseudoscalar mesons, respectively. The square symbolizes a contact vertex with two derivatives. 
First line: LO contact term, one-meson exchange.
Second line: NLO contact term, planar box, crossed box, left triangle, right triangle, football diagram.} \label{fig:feynman}
\end{figure*}

As already mentioned above, the elastic part of the $\bar BB$ interaction 
is derived within SU(3) $\chi$EFT up to NLO, i.e.
up to the chiral order $\nu=2$ ~\cite{Epelbaum:2008ga}. Thus,
there are contributions from one- and two-meson exchanges, involving 
all possible combinations of the Goldstone bosons $\pi$, $\eta$, and $K$.  
In addition $\bar BB$ contact terms arise, without derivatives (at LO)
and with two derivatives (at NLO). An overview of the possible diagrams
is provided in Fig.~\ref{fig:feynman}. 
 
  \subsubsection{Single-meson exchange potential} 

At LO there are contribution from one pseudoscalar-meson exchange (OPE). 
The pertinent interaction potential for 
the reaction $\bar{B}_1B_2\to \bar{B}_3B_4$ can be written as 
 \begin{eqnarray}
     V^{\rm OPE}_{\bar{B}B}
    &=&-f_{B_1 B_3 P} f_{B_2 B_4 P}\frac{\bm{{\sigma}_1\cdot q\,\bm{\sigma}_2\cdot q}}{ {\rm q}^2+M_P^2}\,
     \mathcal{I}_{\overline{B}_1B_2\to \overline{B}_3B_4}\, , \nonumber \\
     &&
\end{eqnarray}
where ${\bf q}={\bf p}'-{\bf p}$ is the transferred momentum defined in terms 
of the 
final (${\bf p}'$) and initial (${\bf p}$) center-of-mass momenta of the 
baryons (or antibaryons). Furthermore, $f_{B_i B_j P}$ are the 
meson-baryon-baryon coupling constants,  
$\mathcal{I}_{\overline{B}_1B_2\to \overline{B}_3B_4}$ are the isospin factors,
and $M_P$ is the mass of the exchanged meson. 
For $\bar BB$ systems with strangeness $S=0$  
($\NNb\to\NNb$, $\NNb\to\LLb$ and $\LLb\to\LLb$, etc.)  
 the isospin factors of one pseudoscalar meson exchange are listed in 
 Table~\ref{Tab:IFOBE} in~\ref{app:Iso}. Here and below we use 
 $\bar BB$ generically for all $\bar{B}_iB_j\to \bar{B}_kB_l$ reactions.

 In line with Ref.~\cite{Kang:2013uia} for one-pion exchange 
 we include the relativistic $1/m^2$ corrections to the static potential
 \cite{Epelbaum:2004fk}. In the application presented below this concerns only 
 the $\NNb$ channel. Accordingly, we use 
\begin{equation}
\label{opep_full}
\frac{\mbox{\boldmath $\sigma$}_1 \cdot{\bf q}\,\mbox{\boldmath$\sigma$}_2\cdot{\bf q}}{{\bf q}^2 + M_\pi^2} \
\to
\left(1-\frac{p^2+p'^2}{2m^2} \right) 
\frac{\mbox{\boldmath $\sigma$}_1 \cdot{\bf q}\,\mbox{\boldmath$\sigma$}_2\cdot{\bf q}}
{{\bf q}^2 + M_\pi^2} \ ,
\end{equation}
with $m$ the nucleon mass. For $\BBb$ channels with unequal masses, like for 
$\LSb$+c.c., one could follow  Ref.~\cite{Epelbaum:2004fk} and take
the reduced mass in the first approximation.  

 Under the assumption of SU(3) flavor symmetry, the various coupling
 constants $f_{B_iB_jP_k}$ are related to each other by \cite{deSwart:1963pdg}
\begin{equation}
\begin{array}{rlrl}
f_{NN\pi}  = & f, 
& f_{NN\eta_8}  = & \frac{1}{\sqrt{3}}(4\alpha -1)f, \\
f_{\Lambda NK} = & -\frac{1}{\sqrt{3}}(1+2\alpha)f, 
& f_{\Xi\Xi\pi}  = & -(1-2\alpha)f, \\
f_{\Xi\Xi\eta_8}  = & -\frac{1}{\sqrt{3}}(1+2\alpha )f, 
& f_{\Xi\Lambda K} = & \frac{1}{\sqrt{3}}(4\alpha-1)f, \\
f_{\Lambda\Sigma\pi}  = & \frac{2}{\sqrt{3}}(1-\alpha)f, 
& f_{\Sigma\Sigma\eta_8}  = & \frac{2}{\sqrt{3}}(1-\alpha )f, \\
f_{\Sigma NK} = & (1-2\alpha)f, 
& f_{\Sigma\Sigma\pi}  = & 2\alpha f, \\ f_{\Lambda\Lambda\eta_8}  = & -\frac{2}{\sqrt{3}}(1-\alpha )f, & f_{\Xi\Sigma K} = & -f.
\end{array}
\label{eq:SU3}
\end{equation}
Accordingly, all coupling constants are given in terms of $f\equiv g_A/(2f_0)$
and the ratio $\alpha=F/(F+D)$. Here, $f_0$ is the Goldstone boson decay
constant, $g_A$ is the axial-vector strength measured in neutron $\beta$-decay, 
and $F+D = g_A$ (strictly speaking in the chiral limit). 
In the present work we use $f_0\equiv f_\pi = 92.4$~{\rm MeV},
i.e. we identify $f_0$ with the empirical decay constant of the pion, and 
ignore a possible SU(3) breaking \cite{Haidenbauer:2023qhf}. Regarding
$g_A$, a slightly larger value than the experimental 
value is used, $g_A = 1.29$, in order to account for the Goldberger-Treiman
discrepancy \cite{Epelbaum:2004fk}. 
As before \cite{Haidenbauer:2013oca},
for the $F/(F+D)$ ratio, we adopt $\alpha = 0.4$ which is the SU(6) value. 
Further, the $\eta$ meson is identified with the octet-state $\eta_8$.

Note that some of the contributions from pseudosca\-lar meson exchange to 
the $\bar BB$ interaction are of opposite sign as those in
the $BB$ case. This sign difference arises from transforming the $BBP$ vertices
to the $\bar B \bar BP$ vertices via charge conjugation and a rotation
in the isospin space and is commonly referred to as $G$-parity transformation. Specifically, for the interaction due to OPE and 
two pseudoscalar-meson exchange (TPE) 
$G$-parity implies simply that
\begin{equation}
V^{\rm OPE}_{\bar BB} = (-1)^I \, V^{\rm OPE}_{BB}, \quad 
V^{\rm TPE}_{\bar BB} = (-1)^{I_1+I_2} \, V^{\rm TPE}_{BB}
\label{eq:G}
\end{equation}
where $I$, $I_1$, and $I_2$ are the isospins of the exchanged pseudoscalar
mesons. Some of
the contributions of two-boson exchange involving pions and/or etas, discussed
in the next subsection, can be obtained through 
the $G$-parity transformation from the corresponding $BB$ potentials. 
However, Eq.~(\ref{eq:G}) is not applicable for contributions involving $K$ and $\bar K$ 
mesons because these do not have a well defined $G$-parity. In that case a
separate calculation is necessary, see Appendix A of Ref.~\cite{Yang:2024iuc}.
In the present work changes in the sign are incorporated into the
isospin factors, while for the coupling constants at the $\bar B\bar BP$
vertices the corresponding values from Eq.~(\ref{eq:SU3}) are used.  

\subsubsection{Two-meson exchange potential} 
A detailed derivation of the two-meson exchange potential in SU(3) $\chi$EFT
can be found in Refs.~\cite{Haidenbauer:2013oca,Yang:2024iuc}. Here we
summarize the final expressions as used in the numerical calculation.
Note that the divergent part from dimensional regularization and the renormalization scale dependent 
part are already left out in the formulae given below. Their effects
are absorbed into the contact terms (more precisely, into the corresponding low-energy constants).

For the irreducible part of the planar box (pb) diagrams, the potential for 
the reaction
$\bar{B}_1B_2\to \bar{B}_3B_4$ can be written as
\begin{eqnarray}
    V_{\bar{B}B}^{\rm pb}&=&N
    \left(V^{\rm pb}_{C}
    +V^{\rm pb}_{S}\bm{\sigma_2\cdot\sigma_1}
    +V^{\rm pb}_{T}\bm{\sigma}_2\cdot \bm{q} \,\bm{\sigma}_1\cdot \bm{q}\right)
    \nonumber\\
    &&\times\,\mathcal{I}^{\mathrm{pb}}_{\bar{B}_1B_2\to\bar{B}_3B_4}\,,
    \label{eq:pb}
\end{eqnarray}
with $N=f_{B_1B_{i_1}P_1}f_{B_3B_{i_1}P_2}f_{B_2B_{i_2}P_1}f_{B_4B_{i_2}P_2}$ 
and an isospin factor $\mathcal{I}^{\mathrm{pb}}_{\bar{B}_1B_2\to \bar{B}_3B_4}$. The latter can
be found again in \ref{app:Iso}, for all two-meson
exchange contributions. 
In Eq.~(\ref{eq:pb}) one has
\begin{eqnarray}
    V^{\rm pb}_{C}
    &=&\frac{1}{192\pi^2}\Bigg[\frac{5}{3}q^2+\frac{(M_{1}^2-M_{2}^2)^2}{q^2}+16(M_{1}^2+M_{2}^2)\nonumber\\
    &&+\frac{M_{1}^2-M_{2}^2}{q^4}\Big(12q^4+(M_{1}^2-M_{2}^2)^2\nonumber\\
    &&-9q^2(M_{1}^2+M_{2}^2)\Big)\ln\frac{M_{1}}{M_{2}}\nonumber\\
    &&+\frac{2}{w^2(q)}\Big(23q^4-\frac{(M_{1}^2-M_{2}^2)^4}{q^4}
      \nonumber\\ 
    &&+56(M_{1}^2+M_{2}^2)q^2+
    8\frac{M_{1}^2+M_{2}^2}{q^2}(M_{1}^2-M_{2}^2)^2
    \nonumber\\
    &&+2(21M_{1}^4 +22M_{1}^2M_{2}^2+21M_{2}^4)\Big)L(q)\Bigg]\,,
    \end{eqnarray}
    \begin{eqnarray}
    V^{\rm pb}_{T}&=&-\frac{1}{8\pi^2}\left[L(q)-\frac{1}{2}-\frac{M_{1}^2-M_{2}^2}{2q^2}\ln\frac{M_{1}}{M_{2}}\right]\nonumber\\
    &=&-\frac{1}{q^2}V^{\rm pb}_{S}\,,
\end{eqnarray}
where ${q}=|{\bf p}'-{\bf p}|$. 
The subscripts \lq $1$' and \lq $2$' denote the relevant pseudoscalar 
mesons appearing in the intermediate states. The functions $L$ and $\omega$ 
are given by 
\begin{eqnarray}
    L(q)&=&\frac{w(q)}{2q}\ln\frac{(qw(q)+q^2)^2-(M_1^2-M_2^2)^2}{4M_1M_2 q^2}\,,
    \nonumber \\
    w(q)&=&\frac{1}{q}\sqrt{(q^2+(M_1+M_2)^2)(q^2+(M_1-M_2)^2)}\,,
\end{eqnarray}
when the masses of the exchanged mesons are different. 
For the equal-mass case, these simplify to 
\begin{eqnarray}
   L(q)&=&\frac{w(q)}{q}\ln\frac{w(q)+q}{2M}\,, \nonumber\\
   w(q)&=&\sqrt{q^2+4M^2}\,.
\end{eqnarray}

For the crossed-box (cb) diagrams, the potential can be written as
\begin{eqnarray}
    V_{\bar{B}B}^{\rm cb}&=&N\,\left(V^{\rm cb}_{C}+V^{\rm cb}_{S}\bm{\sigma_2\cdot\sigma_1}
    +V^{\rm cb}_{T} \bm{\sigma}_2\cdot \bm{q} \,\bm{\sigma}_1\cdot \bm{q}\right)
    \nonumber \\
    &&\times\,\mathcal{I}^{\mathrm{cb}}_{\bar{B}_1B_2\to\bar{B}_3B}\,,
\end{eqnarray}
with $N=f_{B_1B_{i_1}P_1}f_{B_3B_{i_1}P_2}
    f_{B_2B_{i_2}P_2}f_{B_4B_{i_2}P_1}$.
Here, one has 
\begin{eqnarray}
    V^{\rm cb}_{C}&=&-V^{\rm pb}_{C}\,,\nonumber\\
     V^{\rm cb}_{T}&=&-\frac{1}{q^2}V^{\rm cb}_{S}=V^{\rm pb}_{T}\,.
\end{eqnarray}

For the triangle (tr) diagrams, the pertinent potential can be written as 
\begin{eqnarray}
    V^{\rm tr}_{\bar{B}B}= N\, V^{\rm tr}_C \, \mathcal{I}^{\mathrm{tr}}_{\bar{B}_1B_2\to\bar{B}_3B_4}\,,\label{eq:Vtc}
\end{eqnarray}
with $N=f_{B_1B_{i}P_1}f_{B_3B_{i}P_2}$ (right triangle)
or $N=f_{B_2B_{i}P_1}\\f_{B_4B_{i}P_2}$ (left triangle). 
Here, one has, 
\begin{align}
 V^\text{tr}_\mathrm C (q) = &\frac{1}{768 \pi ^2 f_0^2} 
 \Bigg[2 \left(M_1^2+M_2^2\right)-\frac{\left(M_1^2-M_2^2\right)^2}{q^2}
 +\frac{13}{3} q^2 \nonumber \\
 & \!\!\!\!\!\!\! -\Big(8 \left(M_1^2+M_2^2\right)-\frac{2 \left(M_1^2-M_2^2\right)^2}{q^2}+10 q^2\Big) L\left(q\right)\notag\\
 & \!\!\!\!\!\!\!\!\!\!\!\!\!\!\!\!\!
 -\frac{M_1^2-M_2^2}{q^4} \left(\left(M_1^2-M_2^2\right)^2-3 \left(M_1^2+M_2^2\right) q^2\right) \ln \frac{M_1}{M_2}\Bigg]\,. 
\end{align}

Finally, for the football (fb) diagram, the potential can be written as
\begin{eqnarray}
    V^{\mathrm{fb}}_{\bar{B}B}=V^{\rm fb}_C\, \mathcal{I}^{\mathrm{fb}}_{\bar{B}_1B_2\to\bar{B}_3B_4}\,.
\end{eqnarray}
The potential $V^{\rm fb}_C$ has the following form
\begin{eqnarray}
 V^\text{fb}_\mathrm C (q)&=&\frac{1}{3072 
 \pi^2 f_0^4}
 \Bigg[-2 \left(M_1^2+M_2^2\right)-\frac{\left(M_1^2-M_2^2\right)^2}{2 q^2} \nonumber \\
 &-& \frac{5}{6} q^2 
 -\frac{M_1^2-M_2^2}{2 q^4} \Big(\left(M_1^2-M_2^2\right)^2 \nonumber \\
 &+&3 \left(M_1^2+M_2^2\right) q^2\Big) \ln \frac{M_1}{M_2} 
 +w^2\left(q\right) L\left(q\right) \Bigg]\,.
    \label{eq:Vfc}
\end{eqnarray}

In the actual calculations only
the non-polynomial part is taken into account, i.e. the pieces
proportional to $L(q)$ and to $1/q^2$ and $1/q^4$. The polynomial
part only renormalizes the LO and NLO contact terms and, therefore, 
is not considered. 

\subsubsection{Contact terms}
The potential due to the LO contact terms is given by
\cite{Weinberg:1990rz}
\begin{eqnarray}
V^{(0)}_{\bar{B}B}&=& C_{S} + C_{T}\,
\mbox{\boldmath $\sigma$}_1\cdot\mbox{\boldmath $\sigma$}_2\,,
\label{V0}
\end{eqnarray}
and consists of a central term and a spin-spin term. 
The parameters $C_{S}$ and $C_{T}$ are low-energy
constants (LECs) depending on the considered antibaryon-baryon channel
${\bar{B}_1B_2\to\bar{B}_3B_4}$.
These need to be determined by a fit to data.
At NLO the spin- and momentum-dependence of the contact terms reads \hfill
%\begin{widetext}
\begin{eqnarray}
&&V^{(2)}_{BB}= C_1 {\bf q}^{\,2}+ C_2 {\bf k}^{\,2} + (C_3 {\bf q}^{\,2}+ C_4 {\bf k}^{\,2})
\,\mbox{\boldmath $\sigma$}_1\cdot\mbox{\boldmath $\sigma$}_2
 \nonumber \\
&& \quad \quad + \frac{i}{2} C_5 (\mbox{\boldmath $\sigma$}_1+\mbox{\boldmath $\sigma$}_2)\cdot
 ({\bf q} \times {\bf k})
 + C_6 ({\bf q} \cdot \mbox{\boldmath $\sigma$}_1) ({\bf q} \cdot \mbox{\boldmath $\sigma$}_2)
 \nonumber \\ 
&&\quad \quad + C_7 ({\bf k} \cdot \mbox{\boldmath $\sigma$}_1) ({\bf k} \cdot \mbox{\boldmath $\sigma$}_2)
 + \frac{i}{2} C_8 (\mbox{\boldmath $\sigma$}_1-\mbox{\boldmath $\sigma$}_2)\cdot ({\bf q} \times {\bf k}) \ ,
  \nonumber \\
\label{V1}
\end{eqnarray}
%\end{widetext}
where the $C_i$ ($i=1,\dots,8$) are additional LECs and ${\bf k}$  is
the average momentum defined by ${\bf k}=({\bf p}'+{\bf p})/2$.
When performing a partial-wave projection, these terms contribute to the two $S$--wave
($^1S_0$, $^3S_1$) potentials, the four $P$--wave
($^1P_1$, $^3P_0$, $^3P_1$, $^3P_2$) potentials, and the $^3S_1$-$^3D_1$ and $^1P_1$-$^3P_1$
transition potentials in the following way \cite{Epelbaum:2004fk}:
\begin{eqnarray}
\label{VC0}
V(^1S_0) 
&=& \tilde{C}_{^1S_0} + {C}_{^1S_0} ({p}^2+{p}'^2)~, \nonumber\\
V(^3S_1) 
&=& \tilde{C}_{^3S_1} + {C}_{^3S_1} ({p}^2+{p}'^2)~, \nonumber\\
V(^3D_1 -\, ^3S_1) &=& {C}_{\epsilon_1}\, {p'}^2~, \nonumber\\
V(^3S_1 -\, ^3D_1) &=& {C}_{\epsilon_1}\, {p}^2~, \nonumber \\
V(^3P_0) &=& {C}_{^3P_0}\, {p}\, {p}'~,\nonumber\\
V(^1P_1) &=& {C}_{^1P_1}\, {p}\, {p}'~,\nonumber\\
V(^3P_1) &=& {C}_{^3P_1}\, {p}\, {p}'~, \nonumber\\ 
V(^3P_1 - {^1P_1}) &=& {C}_{^3P_1-^1P_1}\, {p}\, {p}'~,\nonumber\\
V(^1P_1 - {^3P_1}) &=& {C}_{^1P_1-^3P_1}\, {p}\, {p}'~,\nonumber\\
V(^3P_2) &=& {C}_{^3P_2}\,  {p}\, {p}'~,
\label{VC}
\end{eqnarray}
with $p = |{\bf p}\,|$ and ${p}' = |{\bf p}\,'|$.
$\tilde C_\alpha$ and $C_\alpha$ are appropriate combinations of the
$C_i$'s appearing in Eqs.~(\ref{V0}) and (\ref{V1}), see Ref.~\cite{Haidenbauer:2013oca}. 
Note that the terms involving $C_{^1P_1-{}^3P_1}$ are absent in the $NN$
interaction due to the Pauli principle, and also in the present case
they would only contribute to $\LSb$/$\SLb$, i.e. to channels where the
baryon masses are different. 
The leading contact terms in higher partial waves, included in an extension 
of the NLO potential, are given by
\begin{eqnarray}
\label{VCP}
V(^{1}D_2) &=& {D}_{^{1}D_2}\, {p^2}\, {p}'^2~, \nonumber\\
V(^3D_{J}) &=& {D}_{^3D_{J}}\, {p^2}\, {p}'^2~; \
J=1,2,3 \ , \nonumber\\
V(^3F_2 -\, ^3P_2) &=& {D}_{\epsilon_2}\, p^3\,{p'}^{1}~ \ , \nonumber \\
V(^3P_2 -\, ^3F_2) &=& {D}_{\epsilon_2}\, p^1\,{p'}^{3}~ \ ,
\end{eqnarray}
corresponding to contributions at fourth order in the chiral 
expansion (N$^3$LO) \cite{Epelbaum:2008ga},
and analogous expressions for the sixth order contributions 
(N$^5$LO) \cite{Reinert:2017usi}.  
%%%%%%%%%%%%%%%%%%%%%%%%%%%%%%%%%%%%%%%%%%%%%%%%%

\subsection{Annihilation part of the $\bar BB$ potential}

In Refs.~\cite{Kang:2013uia,Dai:2017ont} the annihilation part of
the $\NNb$ potential was constructed in a similar way as the 
short-range component of the elastic part, namely via contact
terms. Assuming that annihilation proceeds via a coupling of $\NNb$ to some
effective door-way states $X$ with threshold significantly below
that of $\NNb$, the transition potential was parameterized likewise
in the form of Eqs.~(\ref{VC}) and the contribution of the annihilation
then results from $V_{\NNb\to X} G_X V_{X \to \NNb}$ where
the Green’s function $G_X$ is set equal to the unitarity cut, i.e. 
to $-i$ times a constant, see Ref.~\cite{Kang:2013uia} for details. 
A similar treatment of annihilation has been also followed in Ref.~\cite{Yang:2022qoy} in a study of the electromagnetic form factors
of the nucleons in the timelike region, where only the $^3S_1$-$^3D_1$
partial wave of $\NNb$ enters. 

The treatment of annihilation described above is not practical anymore in the
present case. Here we aim at a description of the $\NNb$ interaction,
i.e. a reproduction of $\NNb$ data, in the energy region where the $\YYb$
channels open. For that, effects of annihilation have to be included also in higher
partial waves, and not only in the $S$- and $P$-waves. Then we would need 
to introduce many more LECs which, however, are difficult to fix given the 
fact that there is no $\NNb$ partial-wave analysis available at such high
energies. 
Therefore, we decided to follow past studies of the
$\NNb$ interaction and assume 
that the bulk of annihilation can be accounted for by a phenomenological
optical potential, where we consider a central,
a spin-orbit, and -- for exploratory reasons -- also a tensor-type term:  
\begin{eqnarray}
    V_{\BBb}^{\rm opt} =&& \Big[ U_C + i W_C +
    (U_{LS}+ i W_{LS}) {\bf L\cdot S} \nonumber \\
    +&& (U_T+i W_T)\, {\bm \sigma}_{\bar B}{\bf \cdot r}
    {\bm \sigma}_{B}{\bf \cdot r}
    \Big] \exp{(-r^2/2r^2_0)} \ .
    \label{eq:Vopt}
\end{eqnarray}
This guarantees annihilation effects in all relevant partial waves. 
In principle, in the low partial waves additional contributions to the annihilation part from contact terms could be included, in line with the power
counting and in close analogy to our $\NNb$ potentials derived 
within $\chi$EFT up to N$^2$LO \cite{Kang:2013uia} and 
N$^3$LO \cite{Dai:2017ont}. However, we found that 
there is practically no improvement of the results 
when such terms are added.  
For consistency reasons in the present work we use phenomenological optical
potentials also for describing the annihilation in the $\YYb$ channels. 
Regarding $\NNb$ the parameters of the phenomenological part and the LECs 
of the contact terms are determined by a fit to available
$\NNb$ data (integrated cross sections, differential cross
sections, analyzing powers) in the relevant momentum region
of $1.4$ to $1.9$~GeV/c, see Figs.~\ref{fig:ppsigma}
to \ref{fig:ppbay}.
The parameters in the phenomenological part of the $\YYb$
interaction and the LECs in $\YYb$ and in the transition
potentials $\NNb\to\YYb$ are determined in a fit to the
PS185 data. 

For completeness and for information, a general over\-view of the SU(3)
relations for the $\BBb$ interaction of octet baryons from strangeness $S=2$
to $S=-2$ is provided in \ref{app:SU3}. Of course, as pointed out there,
since we consider only the channels $\NNb$ and $\LLb$ in the present
calculation, restrictions on the LECs from SU(3) flavor symmetry do not 
have an impact yet.  

\subsection{Scattering equation}

The reaction amplitudes are obtained from the solution of a relativistic 
coupled-channels Lippmann-Schwinger (LS) equation. For that purpose 
a partial-wave projection of the interaction 
potentials is performed, as described in detail in Ref.~\cite{Polinder:2006zh},
so that the LS equation reads 
\begin{eqnarray}
T_{\nu''\nu'}^{L''L'}(p'',p';\sqrt{s})&&=V_{\nu''\nu'}^{L''L'}(p'',p')
 \nonumber\\&&
+\sum_{L,\nu}\int_0^\infty \frac{dpp^2}{(2\pi)^3} \, V_{\nu''\nu}^{L''L}(p'',p)
\nonumber \\ &&
\times\frac{1}{\sqrt{s}-2E^\nu_p+i0^+}T_{\nu\nu'}^{LL'}(p,p';\sqrt{s}).
\nonumber\\
\label{eq:LS} 
\end{eqnarray}
Here, the labels $\nu$, $\nu'$, and $\nu''$ indicate 
the particle channels and $L$, $L'$, and $L''$ the orbital angular momenta.
Like in the standard $NN$ case we have either uncoupled spin-singlet and 
triplet waves (where $L''=L'=L=J$) or coupled partial waves 
(where $L'',L',L=J-1,J+1$). For $\BBb$ systems like 
$\NNb$, $\LLb$, ..., the total energy $\sqrt{s}$ is given by 
$\sqrt{s}= 2E^\nu_q=2\sqrt{m_\nu^2+q_\nu^2}$, where $m_\nu$ and $q_\nu$ 
are the masses and the on-shell momenta. For $\LSb$ it will be
$\sqrt{s}= E^{\nu,1}_q+E^{\nu,2}_q=\sqrt{m_{\nu,1}^2+q_\nu^2}+
\sqrt{m_{\nu,2}^2+q_\nu^2}$ and a similar substitution for $2E^\nu_p$. 
We solve the LS equation in the isospin basis, i.e. for $I=0$ and 
$I=1$ separately. The Coulomb interaction in the $\NNb$ channel is not
taken into account because we are far away from the threshold of that
system where it would be relevant. Of course, for a dedicated study of
the $\SSb$ system near its threshold the inclusion of the Coulomb interaction
would be mandatory, as well as the implementation of the mass differences
between $\Sigma^+$, $\Sigma^0$, and $\Sigma^-$. 

The elastic part of the potential, derived within $\chi$EFT, is cut off 
with a regulator function, 
\begin{equation}
f^\Lambda(p',p) = \exp\left[-\left(p'^4+p^4\right)/\Lambda^4\right], 
\label{cutoff}
\end{equation}
when inserted into the LS equation, 
in order to remove high-energy components \cite{Epelbaum:2004fk}.
The cutoff values used in applications of $\chi$EFT to the $NN$, $\NNb$ 
and $YN$ interactions are typically in the order of $400-600$~MeV
\cite{Epelbaum:2014efa,Kang:2013uia,Dai:2017ont,Haidenbauer:2013oca}.
In the present work the choice of the cutoff range is dictated
by the on-shell momenta in the initial $\NNb$ channel in the region 
where the considered $\YYb$ channels open.
For $\LLb$, $\LSb$, and 
$\SSbm$ the corresponding % laboratory momenta are $p_{lab}=
% 1.4353$, $1.6531$, and $1.8989$~GeV/c, respectively. 
on-shell momenta are $604$, $672$, and $750$ MeV/c,
respectively. Thus, we take the latter value as guideline and as
a lower limit. Low values like in studies of the $NN$ or $YN$ interactions
would simply distort the transition potentials from the $\NNb$ channel
to the $\YYb$ channels too strongly and, therefore, do not allow
for realistic results. In practice we consider values of $800-900$~MeV.
Note that there is basically no risk that we encounter difficulties with
unphysical bound states or resonances, contrary to what happens for $NN$ \cite{Epelbaum:2004fk}, etc. The presence of annihilation more or less
guarantees that such states would be far out in the complex plane, if
they are generated at all.

%%%%%%%%%%%%%%%%%%%%%%%%%%%%%%%%%%%%
\section{Observables} 
\label{sec:Obs}

\begin{figure}[t]
    \centering
    \includegraphics[width=0.99\linewidth]{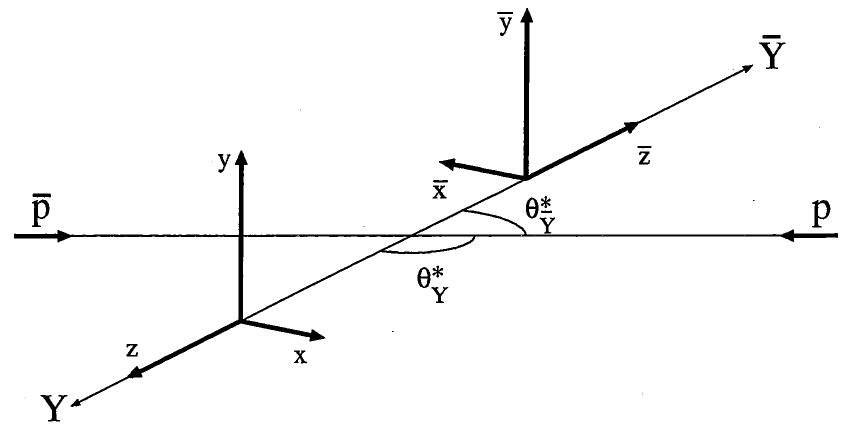}
    \caption{Coordinate system used by PS185 for the
    reaction $\ppb\to\YYb$ \cite{PS185:2006yyx}}
    \label{fig:coord}
\end{figure}

Scattering observables are evaluated via the spin-scatter\-ing
matrix $\M$, for which we take the usual parameterization 
\cite{LaFrance:1980,Bystricky:1984,Holzenkamp:1989tq}, 
\begin{eqnarray} 
\M=&&
\frac{1}{2}\left[(a+b)I +(a-b) 
{\bm\sigma}_1.{\bf\hat{n}}\, {\bm\sigma}_2{\bf\hat{n}}  
 +(c+d) {\bm\sigma}_1. {\bf\hat{k}} \, 
 {\bm\sigma}_2 . {\bf\hat{k}} \right.
 \nonumber\\ 
 &&{}~~+(c-d) {\bm \sigma}_1 .{\bf\hat{p}} \, {\bm\sigma}_2. {\bf\hat{p}}  
+e( {\bm\sigma}_1+{\bm \sigma}_2 ). {\bf\hat{n}} \nonumber \\
&&{}~~+f( {\bm\sigma}_1-{\bm \sigma}_2 ). {\bf\hat{n}} 
+g( {\bm\sigma}_1 .{\bf\hat{k}} \, 
{\bm\sigma}_2 .{\bf\hat{p}}+  
{\bm\sigma}_1 .{\bf\hat{p}} \, {\bm\sigma}_2. {\bf\hat{k}}) \nonumber \\
&&\left.{}+~~h {\bm\sigma}_1\times{\bm\sigma}_2 . {\bf\hat{n}} 
\right],
\label{eq:M}
\end{eqnarray}
with 
\begin{equation}
{\bf\hat{p}}={{\bf{p'}}\over{|{\bf{p'}}|}}, \qquad     
{\bf\hat{ n}}={ {\bf{p}} \times  {\bf{p'}}
            \over |{\bf{p}}\times {\bf{p'}}|    },  
  \qquad {\bf\hat{k}}={\bf\hat{n}} \times {\bf\hat{p}}\,. 
  \label{eq:Mkin}
\end{equation}
Here, {$\bf p$} ({$\bf p'$}) are the centre-of-mass (c.m.s) momenta in the initial (final) state. 
The coefficients $a$, ..., $h$ are obtained via the 
helicity-state matrix elements $\varphi_i$, $i=1,8$, based on 
the $BB$ reaction amplitudes (Eq.~(\ref{eq:LS})), 
see Appendix C of Ref.~\cite{Holzenkamp:1989tq}. 
The calculation of the spin observables in the $\LLb$
coordinate system, i.e. in the convention of PS185 see 
Fig.~\ref{fig:coord}, requires that the 
direction of the $\bar \Lambda$ is defined as $z$ direction.
In this case, the angle in Eq.~(C.3) of Ref.~\cite{Holzenkamp:1989tq} has to be set to $\theta=0$ so that 
$\sin\theta=0$ and $\cos\theta=1$. 
There is also a change in the sign for some 
observables because in the usual convention 
$\overline{\Lambda}$, 
$\{\bf\hat{x},\bf\hat{n},\bf\hat{z}\}$ coincides with 
$\{\bf\hat{k},\bf\hat{n},\bf\hat{p}\}$, while for 
$p$ or $\Lambda$,
the axes $\{\bf\hat{x},\bf\hat{n},\bf\hat{z}\}$ coincides with 
$\{-\bf\hat{k},\bf\hat{n},-\bf\hat{p}\}$. 

Note that $f=h=0$ when the particle masses are identical
like for the $\ppb$ or $\LLb$ states. These amplitudes are, however, 
nonzero for the $\LSb$+c.c. system. Detailed discussions of the
scattering formalism with emphasis on the reaction $\ppb\to\LLb$
can be found in 
Refs.~\cite{Durand:1964,Tabakin:1991ct,Elchikh:1999ir,Paschke:2000mx}. 
Explicit expressions used by us can be found in 
Ref.~\cite{Haidenbauer:1991kt} ($\LLb$) 
and in Ref.~\cite{Haidenbauer:1993ws} (for $\LSb$+c.c.).
For convenience we summarize them in the following: 

With the spin-scattering matrix $\M$ of Eq.~(\ref{eq:M}) 
the differential cross section is given by 
\begin{eqnarray}
\frac{d\sigma}{d\Omega} = \frac{p'}{p}I_{0}&&= 
      \frac{1}{4}\frac{p'}{p}\Tr[\M \M^{\dag}] 
=\frac{1}{2}\frac{p'}{p}\left(|a|^2+|b|^2+|c|^2 \right. \nonumber\\
&&\left.\quad+|d|^2+|e|^2+|f|^2+|g|^2+|h|^2\right)\,.
\end{eqnarray}
The polarizations and analyzing powers are given by 
\begin{eqnarray} 
 P_y^{(i)}&&=\frac{\Tr[{\bm\sigma}_{(i)}.\bf\hat{n} \M \M^{\dag}]}
 {\Tr[\M \M^{\dag}]}\,,
 \nonumber \\
P_y^{(1)}I_0&&=\Re (ae^*+bf^*)+\Im (dg^*+ch^*),\nonumber \\
P_y^{(2)}I_0&&=\Re (ae^*-bf^*)+\Im (dg^*-ch^*), \\
\nonumber \\
A_y^{(i)}&&=\frac{\Tr[\M{\bm\sigma}_{(i)}.\bf\hat{n}\M^{\dag}]}
 {\Tr[\M \M^{\dag}]} \,,\nonumber\\
A_y^{(1)}I_0&&=\Re (ae^*+bf^*)-\Im (dg^*+ch^*),\nonumber \\
A_y^{(2)}I_0&&=\Re (ae^*-bf^*)-\Im (dg^*-ch^*),
 \end{eqnarray}
 while the spin-correlation parameters are given by
\begin{eqnarray}
C_{ij}&&=
     \frac{\Tr[\vec{\sigma}_{(1)}. \hat{\imath}\,\vec{\sigma}_{(2)} .
           \hat{\jmath} \M \M^{\dag}]}
            {\Tr[\M \M^{\dag}]} \nonumber \,, \\
C_{xx}I_0&&=-\Re (ad^*+bc^*)-\Im (ge^*+hf^*),\nonumber \\
C_{yy}I_0&&=\frac{1}{2}(|a|^2-|b|^2-|c|^2+|d|^2+|e|^2-|f|^2\nonumber\\
&&\quad+|g|^2-|h|^2),\nonumber \\
C_{zz}I_0&&=\Re (ad^*-bc^*)+\Im (ge^*-hf^*),\nonumber \\
C_{xz}I_0&&=-\Re (ag^*-bh^*)-\Im (ed^*+cf^*), 
\nonumber \\
C_{zx}I_0&&=-\Re (ag^*+bh^*)-\Im (ed^*-cf^*). 
\end{eqnarray} 
Finally, the expressions for the spin-transfer parameters and the
triple-spin observables are
\begin{eqnarray}
D_{ij}&&= 
       \frac{\Tr[\bm{\sigma}_{(2)}. \hat{\imath} \M  {\bm{\sigma}}_{(2)} .
           \hat{\jmath} \M^{\dag}]}
           {\Tr[\M \M^{\dag}]}, \nonumber \\  
D_{nn}I_0&&=\frac{1}{2}(|a|^2+|b|^2-|c|^2-|d|^2+|e|^2+|f|^2\nonumber \\
          &&\quad -|g|^2 -|h|^2), \nonumber \\ 
D_{xx}\,I_0 &=& \Re (a^*b + c^*d + e^*f + g^*h), \nonumber \\
D_{zz}\,I_0 &=& \Re (a^*b - c^*d + e^*f - g^*h), \nonumber \\
D_{xz}\,I_0 &=& \Re (c^*g - d^*h) + \Im (a^*f - b^*e), \nonumber \\ 
D_{zx}\,I_0 &=& \Re (c^*g - d^*h) - \Im (a^*f - b^*e), \\ 
\nonumber \\
K_{ij}&&=
  \frac{\Tr[\bm{\sigma}_{(1)}. \hat{\imath} \M {\bm{\sigma}}_{(2)} .
         \hat{\jmath} \M^{\dag}]}
         {\Tr[\M \M^{\dag}]}, \nonumber \\
K_{nn}I_0&&=\frac{1}{2}(|a|^2-|b|^2+|c|^2-|d|^2+|e|^2-|f|^2\nonumber \\
          &&\quad -|g|^2 +|h|^2), \nonumber \\
K_{xx}\,I_0 &=& -\Re (a^*c + b^*d) - \Im (e^*h + f^*g) \nonumber \\
K_{zz}\,I_0 &=& -\Re (a^*c - b^*d) - \Im (e^*h - f^*g) \nonumber \\
K_{xz}\,I_0 &=& \Re (a^*h - b^*g) + \Im (c^*e - d^*f) \nonumber \\ 
K_{zx}\,I_0 &=& -\Re (a^*h + b^*g) - \Im (c^*e + d^*f)\,,
\end{eqnarray}
and
\begin{eqnarray}
C_{0\alpha ij}&=&\frac{\Tr[\bm{\sigma}_{(1)}.\hat{\imath} \, 
 \bm{\sigma}_{(2)}.\hat{\jmath} \M \bm{\sigma}_{(2)} .\hat{\alpha} 
                    \M^{\dag}]}
                               {\Tr[\M \M^{\dag}]},\nonumber \\
C_{0nnn}\,I_0 &=& \Re (a^*e + b^*f) + \Im (c^*h + d^*g) \nonumber\\
C_{0nxx}\,I_0 &=& \Re (c^*f - d^*e) - \Im (a^*g - b^*h) \nonumber\\
C_{0nzz}\,I_0 &=& \Re (c^*f + d^*e) + \Im (a^*g + b^*h) \nonumber\\
C_{0nzx}\,I_0 &=& -\Re (e^*g - f^*h) + \Im (a^*d - b^*c) \nonumber\\
C_{0nxz}\,I_0 &=& -\Re (e^*g + f^*h) + \Im (a^*d + b^*c)\,,
\end{eqnarray}
where $C_{0nnn} = A_n$. Note that in our convention $n\equiv y$. We use
one or the other notation simply in order to be in line with the ones 
in the respective experimental papers.

As already indicated above, in case of the reaction $\ppb\to\LLb$ there
are six independent (complex) amplitudes. Thus, for a complete determination of those
one needs at least 11 independent observables. With the publication of the
triple-spin observables by Paschke et al.~\cite{PS185:2006yyx} this condition
has been fulfilled, at least at the momentum $1637$~MeV/c. Indeed, with 
$d\sigma/d\Omega$, $P_y$, and the $C_{ij}$, together with $A_n$, $D_{nn}$,
$K_{nn}$, and the $C_{nij}$, one has already 12 observables. 
In fact, in
Ref.~\cite{PS185:2006yyx} results for more spin-transfer and triple-spin 
observables have been presented. However, these are redundant and, following
Bugg~\cite{Bugg:2004rj}, we include only the aforementioned set in our
fitting procedure. 
In this context let us also point to Ref.~\cite{Elchikh:1999ir} where relations
between the various spin observables and resulting constraints are discussed.
For example, in forward ($\theta=0$) and in backward ($\theta=\pi$)
direction one has $C_{xz}=0$ and $C_{xx}=-C_{yy}$
which follows from rotational invariance.
The authors also note that $D_{nn}=K_{nn}$
when $F_S=0$, which suggests that $D_{nn}$ and $K_{nn}$
cannot differ too much when the singlet fraction is very small. 

For the reaction $\ppb\to\LSbz$+c.c. there are 8 independent amplitudes. 
However, in this case only 8 independent observables have been measured so 
far, see Table~\ref{tab:La}. Thus, here a complete determination of the
amplitudes from the present data is not possible. 
The situation is far worse for the reactions $\ppb\to\SSb$ where only
differential cross sections are available, and only for one momentum.

%%%%%%%%%%%%%%%%%%%%%%%%%%%%%%%%%%%%
\section{Results for $\NNb$}
\label{sec:NNb}

\begin{figure}[t]
    \centering
    \includegraphics[width=0.99\linewidth]{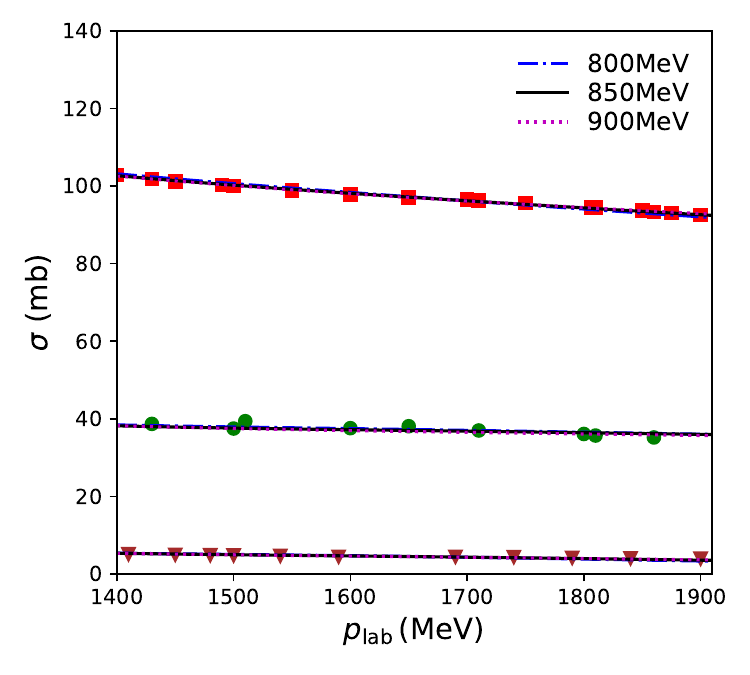}
    \caption{Total, integrated elastic, and charge-exchange cross sections for $\ppb$
    scattering, from top to bottom.
The black solid, purple dashed, and green dash-dotted lines are our results for the cutoff values  
    $\Lambda=800,850,900$ MeV, respectively. 
    Data (squares, circles, and inverted triangles) are from 
Refs.~\cite{Coupland:1977zr,Abrams:1969jm,Parker:1971pea,Cutts:1974vi}. }
    \label{fig:ppsigma}
\end{figure}

\begin{figure*}[h!]
    \centering
    \includegraphics[width=0.89\linewidth]{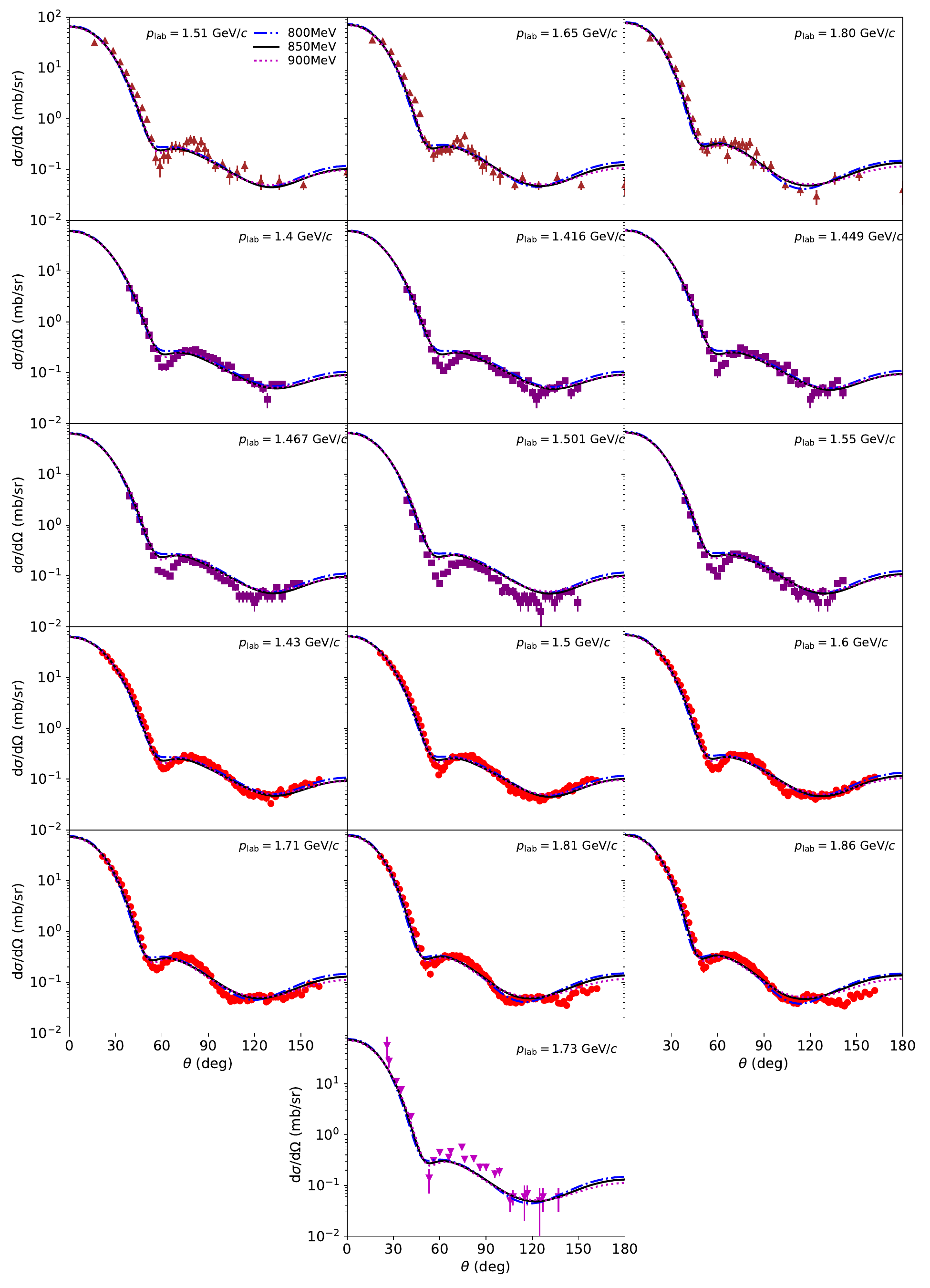}
    \caption{Differential cross section for $\ppb$ elastic scattering. Same
    description of curves as in Fig.~\ref{fig:ppsigma}.
    Data are from \cite{Parker:1971pea} (brown triangles), \cite{Kunne:1988kg}
    (purple squares), \cite{Eisenhandler:1976kf} (red circles), and 
    \cite{Daum:1968xea} (magenta inverted triangles).}
    \label{fig:ppbds}
\end{figure*}

\begin{figure*}[t]
    \centering
    \includegraphics[width=0.99\linewidth]{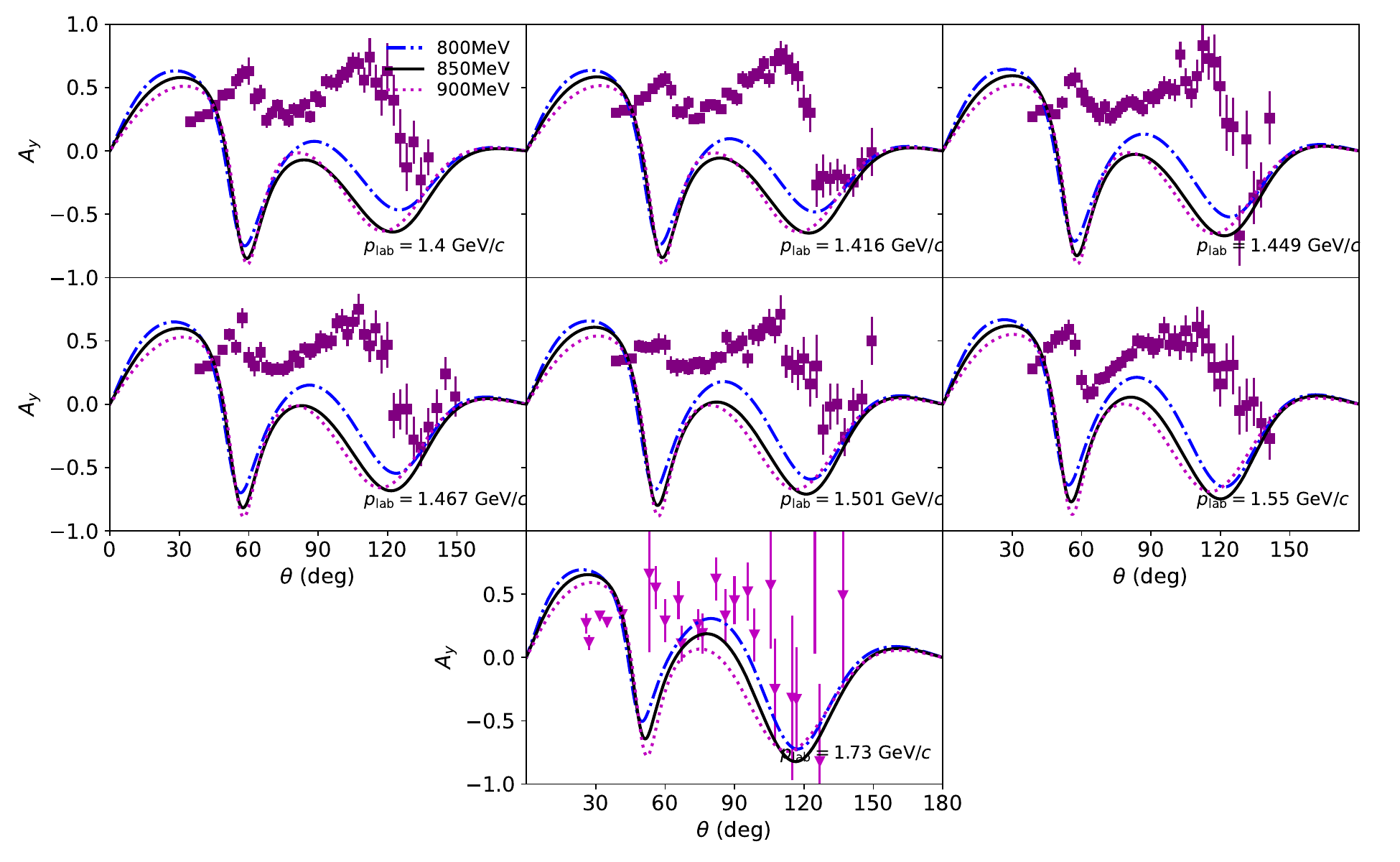}
    \caption{Analyzing power for $\ppb$ elastic scattering. 
    Same description of curves as in Fig.~\ref{fig:ppsigma}.
    Data are from \cite{Kunne:1988kg}
    (purple squares) and \cite{Daum:1968xea} (magenta inverted triangles).}
    \label{fig:ppbay}
\end{figure*}
%%%%%%%%%%%%%%%%%%%%%%%%%%%%%%%%%%%%%%%%%%%%%

To begin with we discuss the results for the $\NNb$ system. The $\NNb$ potential
has been fitted independently from the $\ppb\to\LLb$ data. Since the transition
potential must be fairly weak (the $\ppb\to\LLb$ cross section is roughly
three orders of magnitude smaller than that for $\ppb$ scattering), the 
influence of the coupling to $\LLb$ is basically negligible for the $\NNb$
interaction. Actually, this conjecture has been confirmed by simply evaluating
the $\NNb$ observables again after the coupling to the $\LLb$ channel has 
been included and the $\LLb$ and $\NNb\to\LLb$ potentials have been established
by a fit to the $\ppb\to\LLb$ data. 

The fit was performed to the $\ppb$ total, elastic, and charge exchange cross
sections, as well as to the differential cross sections and analyzing powers of 
$\ppb$ scattering, within 
the relevant momentum range from $p_{\rm lab}=1.4$~GeV$/c$ to 1.9~GeV$/c$.
In the fitting procedure initially all three terms (central, LS, tensor) 
in the optical potential Eq.~\eqref{eq:Vopt} were considered. However, it turned 
out that including the tensor term did not lead to any noticeable improvements,
specially also not for the analyzing powers, and, thus, it was omitted. 
Also, the parameters in the optical potential for isospin $I=0$ and $1$ 
could be set to be equal. 
The parameters of the $\NNb$ potential, the LECs of the
elastic part and the strengths and range of the annihilation part, are
summarized in Table~\ref{tab:paraNNb} in \ref{app:Parameters}. 
In the actual calculation partial waves
up to a total angular momentum of $J=15$ are taken into account. This is
necessary to obtain converged results for the differential cross sections
in backward direction and also for other spin-dependent observables. 

A comparison of the results with $\ppb$ data for integrated cross sections,  differential cross sections,  % Fig.~\ref{fig:ppbds} 
and analyzing powers
is presented in Figs.~\ref{fig:ppsigma} to \ref{fig:ppbay}
for the momentum region of $1.4$ to $1.9$~GeV/c. 
Fig.~\ref{fig:ppsigma} shows also a comparison with the total and the
integrated charge-exchange ($\ppb\to\bar nn$) cross sections. Unfortunately,
for the latter reaction, detailed experimental information, i.e. differential 
observables, are not available in the relevant region. 

As one can see, the integrated cross sections and also the differential
cross sections for $\ppb$ elastic scattering are very well reproduced by
the $\NNb$ potential. In particular, the diffraction peak in the 
differential cross sections in forward direction is perfectly described,
and even the location of the two minima are well in line with the 
experiment, see Fig.~\ref{fig:ppbds}. Given that the values of the 
differential cross sections extend over more than two orders of magnitude
this is quite remarkable. In this context, notice also the excellent
description of the experimental cross section for 
$\ppb\to\bar nn$, despite the fact that it is just a small
fraction of the total cross section. The charge-exchange observables reflect the difference in the
isospin $I=1$ and $0$ components of the $\NNb$ amplitude. A proper description of those
observables requires that $V_{\NNb} (I=1)\approx V_{\NNb} (I=0)$, and the $\NNb$ potentials used in
the present study are clearly in line with this. 
Finally, it is worth noting that there is almost
no cutoff dependence of our results for those observables. As already
mentioned, we vary the cutoff between $800$ and $900$~MeV. 

The situation is much less satisfying for the asymmetry $A_y$. In this
case the results are only (and only roughly) in line with the older data 
at $1.73$~GeV/s from Ref.~\cite{Daum:1968xea}, but not with the 
newer measurements by the PS172 Collaboration at LEAR \cite{Kunne:1988kg}. 
Indeed, also in past studies of the reaction $\ppb\to\LLb$ it had turned
out to be difficult to reproduce $A_y$ when describing annihilation effects
with a simple optical potential like that in Eq.~(\ref{eq:Vopt}), 
see the attempts by the J\"ulich \cite{Haidenbauer:1991kt} and by the 
Salamanca \cite{GarciaOrtega:2012fmb} groups. Better results have been 
reported in by the Paris group~\cite{LaFrance:1988if} and by the Nijmegen group
\cite{Timmermans:1992fu}. In the former study the Paris $\NNb$ potential
\cite{Cote:1982gr} is adopted while in the latter the so-called 
P-matrix formalism \cite{Jaffe:1978bu} is used. In that case the 
effect of the potential at short distances, below a boundary $b$
(here in the order of $1.1$ to $1.3$ fm), is parameterized in terms of
the wave function at $b$ that follows from a square well potential. 
Then with one parameter for the imaginary part of the potential and 
10 real potential parameters not only the differential $\ppb$ cross
section but also $A_y$ can be described \cite{Timmermans:1992fu}. 
 
The above study indicates that there is clearly a delicate interplay 
between the various partial waves which cannot be generated by a simple 
optical potential. Obviously,
an improvement of the results for $A_y$ is possible, but only at the
expense of introducing many more parameters. 
On the other hand, the results of the J\"ulich group indicate that the
results for $\ppb\to\LLb$ are not too sensitive to the details of the employed 
$\NNb$ interaction. This observation is supported by the aforementioned
study of the Nijmegen group where likewise not much sensitivity of the
$\ppb\to\LLb$ results to the details of the $\NNb$ interaction have been
found, and a simple description of $\NNb$ annihilation had been adopted
for the actual calculations. 
In view of that, we refrain here from a further fine-tuning of the 
$\NNb$ potential. However, we did generate alternative variants of the
$\NNb$ interaction and examined their influence on the results for 
$\NNb\to\LLb$. 

$\chi$EFT is an expansion of the potential in powers
of a small scale over a large scale, $Q/\Lambda$,
where $Q$ stands for the relative
momentum of the baryons and/or the pion mass and 
$\Lambda$ is usually identified with
the $\rho$ or nucleon mass. A more quantitative value for the latter can 
be established by considering the so-called chiral breakdown scale, see
\cite{Epelbaum:2014efa}, 
which was found to be in the order of $600$~MeV or even smaller.
In typical applications to the $NN$ or $YN$ interactions the upper limit 
is set by the pion-production threshold, so that 
$Q\equiv p\approx 300$~MeV/c. In the study of the
reactions $\NNb\to\YYb$ the relevant momenta in the
$\NNb$ system exceed this limit significantly, 
see Table~\ref{tab:thr}. Therefore, it is clear that the $\NNb$
potential established in the present work should be considered as being 
of phenomenological nature. Still, we want to emphasize that using $\chi$EFT
allows us to construct the important long-range part of the interaction 
in an unambiguous way, in line with constraints from chiral symmetry and
beyond the one-pion exchange tail.

%%%%%%%%%%%%%%%%%%%%%%%%%%%%%%%%%%%%%%%%%%%%%
\section{Results for $\NNb\to\LLb$ and $\LLb$}
\label{sec:LLb}

\subsection{Fitting procedure}
\label{sec:fit}

In the fitting procedure we consider two different energy ranges. In the
first case $\ppb\to\LLb$ data up to $p_{\rm lab}=1546$~MeV$/c$ are taken into
account. Thus, we include the lowest momentum where, besides 
integrated and 
differential cross sections and polarizations, experimental information on
the spin correlation parameters $C_{ij}$ and on the observables $D_{nn}$ 
and $K_{nn}$ are available. The corresponding range of excess energies in the 
$\LLb$ system is $\varepsilon\le 40$ MeV and the respective upper limit of
the kinetic energy in the laboratory system is $T_{\rm lab}\approx 80$~MeV$/c$. 
This choice is well motivated from a physics point of view and  
it is also in line with past applications of $\chi$EFT up to NLO 
to other $BB$ interactions. 
For example, Dai et al.~\cite{Dai:2017ont} (studying the $\NNb$
interaction) set the maximum energy in the fitting at NLO (N$^2$LO) and 
N$^3$LO to $T_{\rm lab} = 125$ and $175$ MeV, respectively. 
Reinert et al.~\cite{Reinert:2017usi} (in the $NN$ case) set it to
$100$ ($125$) and $200$~MeV, respectively. 
Therefore, we expect that
we should be able to describe the data rather well over this energy region.

In the second scenario $\ppb\to\LLb$ data up to $p_{\rm lab}=1642$~MeV$/c$ are
included, so that the first and so far only experimental information on 
triple-spin observables can be taken into account in the fit. In this case 
the access energies $\varepsilon\le 75$~MeV are covered, i.e. the region 
up to $T_{lab}\approx 150$~MeV.
Judging from the aforementioned applications to $NN$ and $\NNb$ this is 
already at the limit where NLO interactions could work. Thus, it 
will be interesting to see whether a quantitative description of the data
can be achieved. Anyway, based on the results of Ref.~\cite{Carlsson:2015vda}
where an NLO $NN$ potential had been fitted do data over different energy
ranges, we do expect a certain increase in the achievable $\chi^2$ minimum. 

\begin{figure*}[t]
    \centering
    \includegraphics[width=0.90\linewidth]{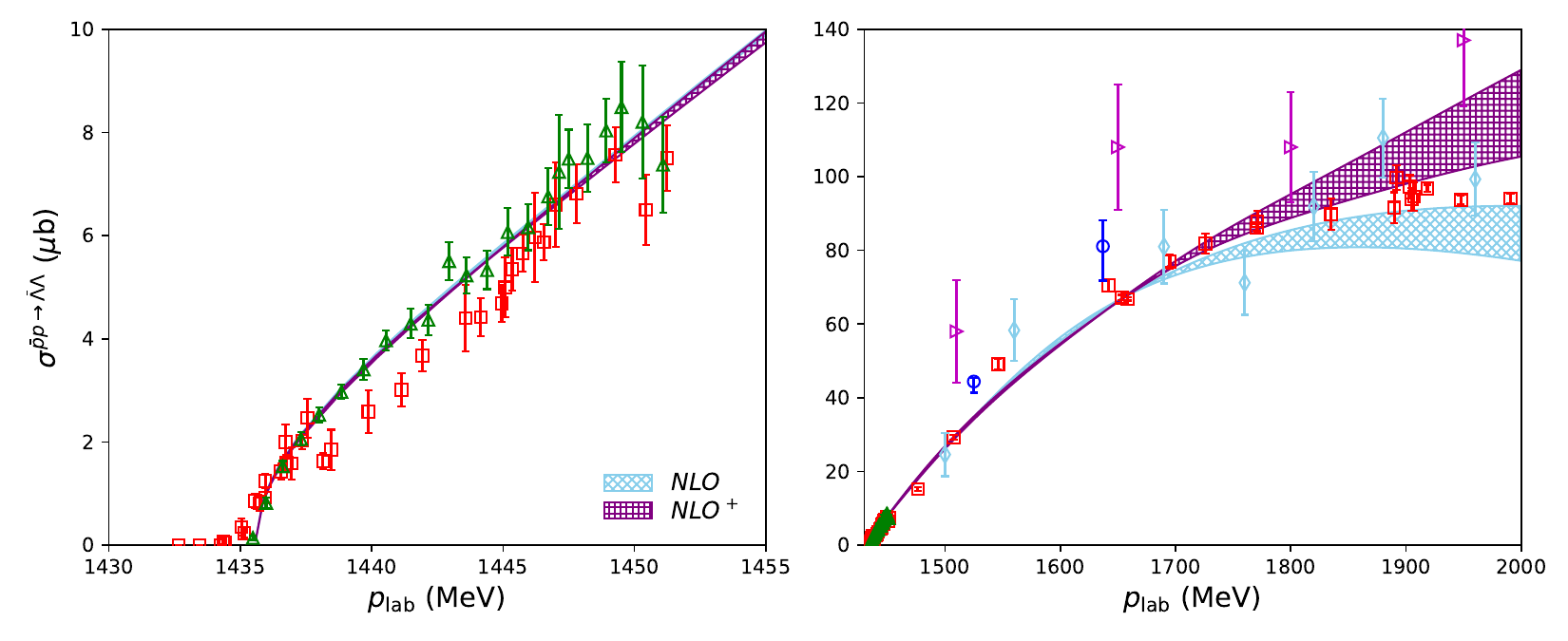}
    \caption{Cross section of the reaction $\ppb\to\LLb$ near threshold (left) and over a
    larger energy range (right). 
    The blue (grey) and purple (black) bands are the results of our NLO (low energy) and NLO$^+$ fits. 
    LEAR data are taken from 
Refs.~\cite{Barnes:1989je,Barnes:1994bh,Barnes:2000be,Barnes:1987aw,Barnes:1990bs,Barnes:1996si,Barnes:1996cf,PS185:2006yyx,Todenhagen:1995,Kraft:1994,Sefzick:1992,Broders:1995}.
    Pre-LEAR data (sideward triangles, diamonds) are from Refs.~\cite{Oh:1973ny,Jayet:1978yq}.
    }
    \label{fig:ppbLLbsigma}
\end{figure*}

Alternatively, to explore the situation in more detail we consider
a scenario that we call NLO$^+$, following the notation introduced in Ref.~\cite{Reinert:2017usi}. 
NLO$^+$ in our case means that also contact terms in $D$-waves and in the 
$^3P_2$-$^3F_2$ transition are considered, and eventually also those in $F$ 
waves are included, in order to check in how far the agreement with the
data can be improved. The additional terms (LECs) are introduced in the
$\NNb\to\LLb$ transition potential and in the $\LLb$ potential itself. 
With such terms it might be possible to achieve
reasonable results for, say,  
$\varepsilon\le 100$~MeV 
($T_{lab}\approx 210$~MeV), i.e. for momenta
$p_{lab}\le 1726$~MeV/c.
At this stage no attempt is made to establish a full N$^2$LO interaction
since the sub-leading meson-baryon vertices involving 
hyperons that enter at this order are only poorly constrained or not known.

As discussed in Sect.~\ref{sec:LLbdata} most of the 
PS185 data have to be re-normalized in view of revisions after their publication. 
This concerns the cross sections (to correct for an incomplete
Monte Carlo used in the analysis of earlier experiments) as well as the 
spin-dependent observables (due to a new determination of the $\Lambda$ weak 
decay constant). The fits in the present work have been performed to the
appropriately readjusted data. 
In the actual fitting procedure an additional re-normalization of each data 
set is allowed, in order to take into account an overall normalization
uncertainty, follow the strategy of the
Nijmegen group \cite{Timmermans:1992fu,Zhou:2012ui}. 
In practice, the data on integrated cross sections are
directly used for determining the overall magnitude 
and the energy dependence of the reaction amplitude.  
At the same time, all differential observables are supplemented with
normalization factors which are then adjusted, for each set, together 
with the parameters of the $\NNb\to\LLb$ and $\LLb$ potentials. 
For the evaluation of the $\chi^2$ value at the minimum only
the differential observables are taken into account,
in accordance with the arguments in Ref.~\cite{Timmermans:1992fu}. Only published
data are considered. 
Since the integrated cross section at $1637$~MeV/c
is noticeably larger than the ones reported for 
near-by momenta we included cross-section data up to
$1726$~MeV/c in the fitting procedure in order to
stabilize the energy dependence. 

    \begin{table}[htbp!]
    \centering
        \caption{Fitting range and achieved $\chi^2$ for the NLO and NLO$^+$ interactions, for
        different cutoffs. 
    }
        \renewcommand{\arraystretch}{1.5}
    \begin{tabular}{|c| c c c|}
         \hline
         \multicolumn{4}{|c|}{NLO (297 data; up to $1546$ MeV/c)} \\
         \hline
         Cutoff (MeV)      & 800 & 850 & 900 \\ \hline
         $\chi^2$   &   440.13  &  403.42    &    393.84 \\
         $\chi^2$/data & $1.48$ & $1.36$ & $1.33$ \\
         \hline\hline
        \multicolumn{4}{|c|}{NLO$^+$ (589 data; up to $1642$ MeV/c)} \\
         \hline
         $\chi^2$  &  957.75   &   1006.54  &   1211.91  \\
        $\chi^2$/data & $1.63$ & $1.71$ & $2.06$ \\
        \hline
    \end{tabular}
        \renewcommand{\arraystretch}{1.0}
    \label{tab:chi2}
\end{table}

Information on the achieved minimal $\chi^2$ in the various fits is 
summarized in Table~\ref{tab:chi2}.
The $\chi^2$/data in the regular NLO fit, including the
$297$ data points up to $1546$~MeV/c,
is $1.48$, $1.36$, and $1.33$ for the cutoffs
$800$, $850$, and $900$~MeV, respectively.
With the NLO$^+$ potential, where $589$ data up to $1642$ MeV/c are
included in the fit, the corresponding
values are $1.63$, $1.71$, and $2.06$.
Evaluating the $\chi^2$/data for the latter interaction 
considering only data up to $1546$ MeV/c yields $1.46$, $1.42$, and $1.44$. 
The scenario where the NLO potential is fitted to all data up to
$1642$ MeV/c did not produce satisfying results. In this case the
$\chi^2$/data was found to be in the order of $4$ to $5$. Therefore,
in the following we do not discuss this scenario in detail. In the few cases where results of it
are mentioned, it will be denoted by NLO(he). 

The actual parameters of the interactions, i.e. the
LECs and the strengths and range parameters of the
optical potential are summarized in tables
in~\ref{app:Parameters}. 

\begin{figure*}[t]
    \centering
    \includegraphics[width=0.99\linewidth]{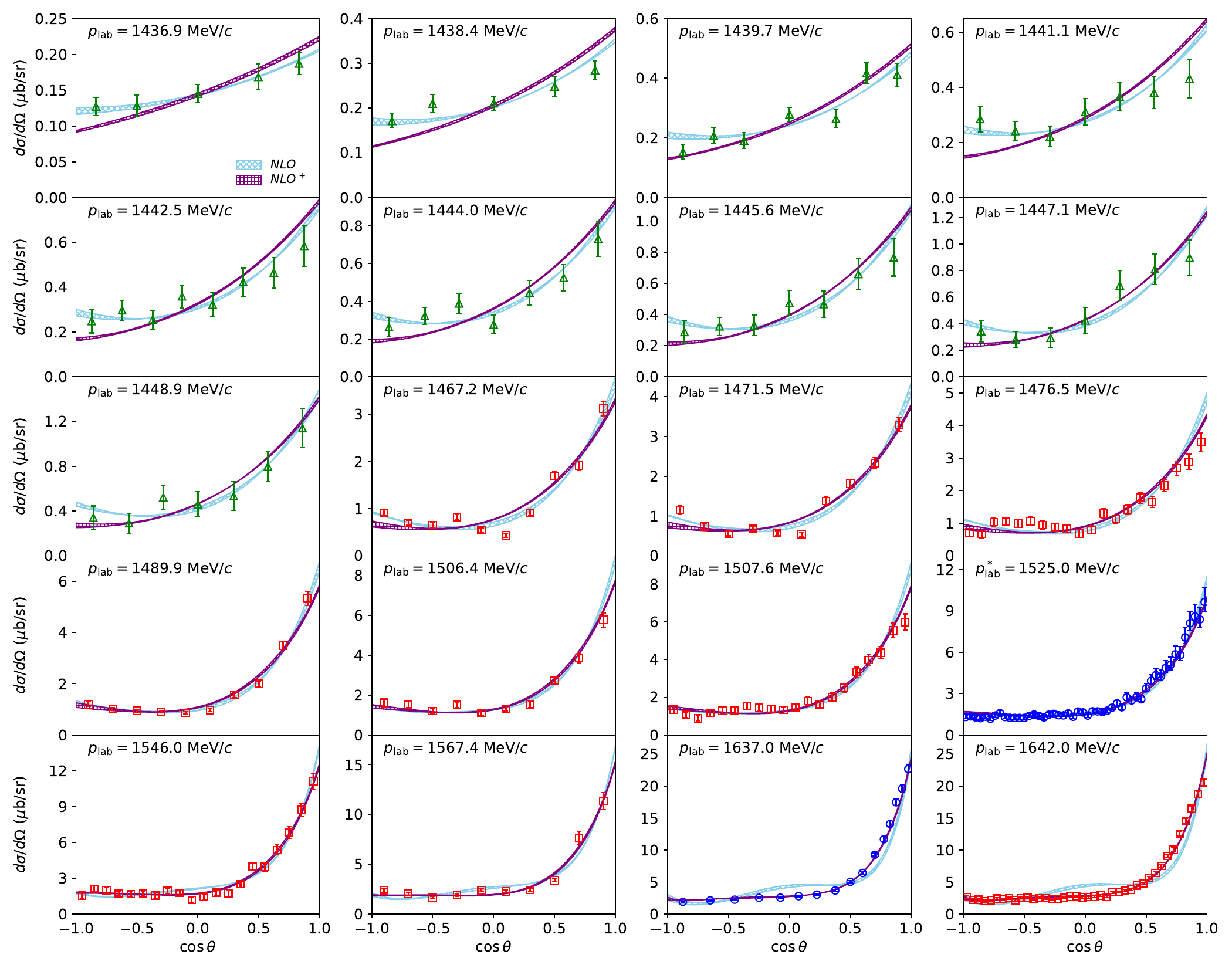}
    \caption{Differential cross sections of the reaction $\ppb\to\LLb$. The blue (grey) and purple (black) bands are the results of our NLO (low energy) and NLO$^+$ fits.
Data are taken from 
Refs.~\cite{Barnes:1994bh,Barnes:2000be,Barnes:1987aw,Barnes:1990bs,Barnes:1996si,PS185:2006yyx}. In case of unpublished data taken from theses 
\cite{Kingsberry:2002} the momenta are marked with an asterisk. These data sets were not included in the fitting procedure and in the evaluation of the $\chi^2$.
    }
    \label{fig:ppbLLbdsigma}
\end{figure*}

In the past, the Nijmegen group \cite{Timmermans:1992fu} and 
David Bugg \cite{Bugg:2004rj} have performed a $\chi^2$ analysis and 
we can compare with their results. In the work of the Nijmegen group
data up to $1546.2$~MeV/c were considered. The reported $\chi^2$
minimum is $180.6$ for a set of 157 data, which
amounts to $\chi^2/{\rm data} =1.15$. This is slightly better than 
our result. However, one should keep in mind that the data base used by
them differs from ours. As discussed in Sect.~\ref{sec:LLbdata} the cross
sections had to be re-scaled, the data at very low momenta are superseded
by new measurements, 
and observables like $D_{nn}$ and $K_{nn}$ at $1525$~MeV/c
were not available at the time of their calculation. 
Bugg reported for his fit a $\chi^2$ of $1377$ for $1201$ data points, 
i.e. $\chi^2/{\rm data} =1.15$. In his analysis data up to $1990.6$~MeV/c 
were considered. Thus, the number of data included by him is much larger
than in our case. On the other hand, observables like 
 $D_{nn}$ and $K_{nn}$ at $1525$~MeV/c are missing. 

%%%%%%%%%%%%%%%%%%%%%%%%%%%%%%%%%%%%%%%%%

\begin{figure*}[t]
    \centering
    \includegraphics[width=0.85\linewidth]{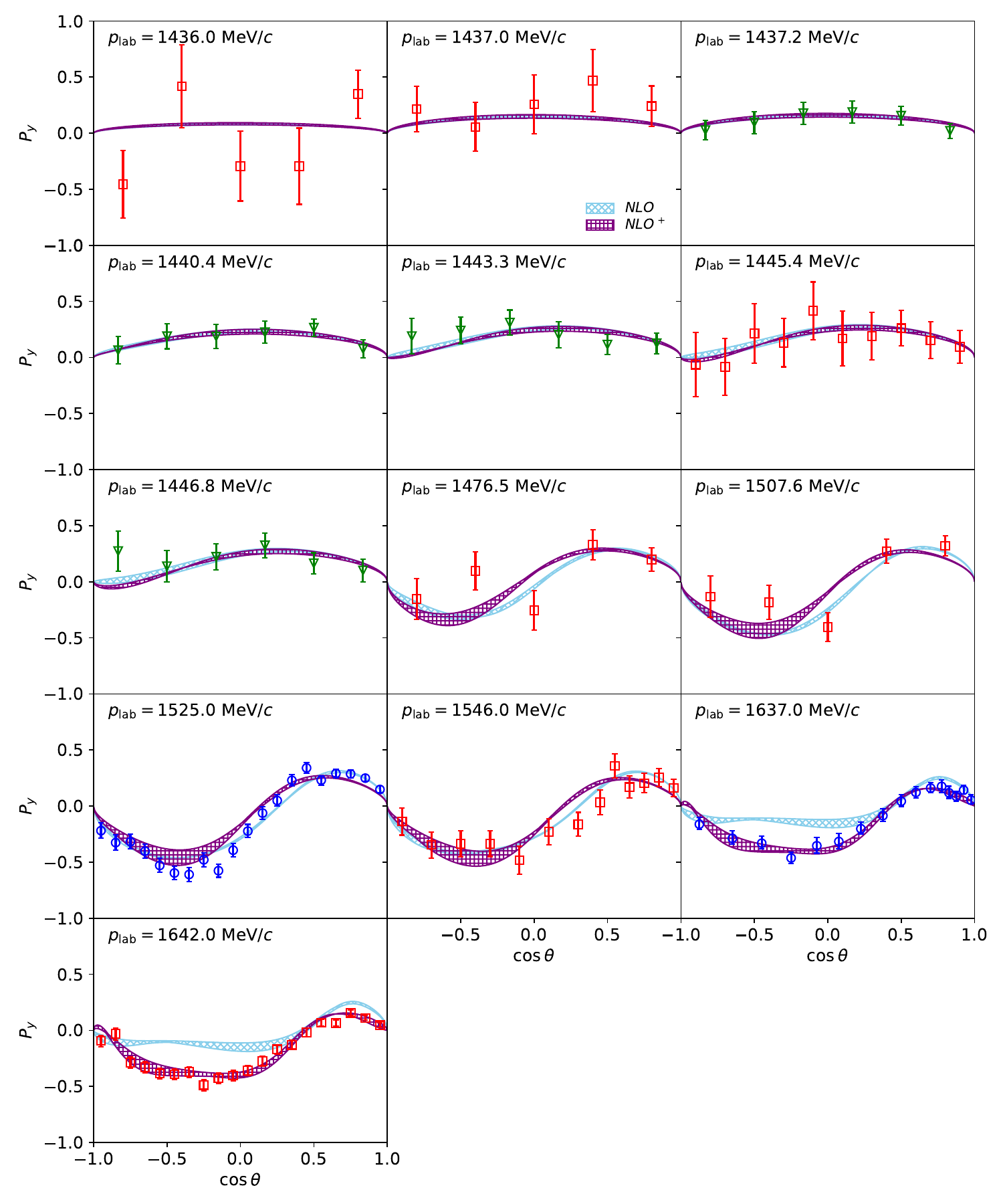}
    \caption{Polarizations for the reaction $\ppb\to\LLb$. Same description of the
    bands as in Fig.~\ref{fig:ppbLLbdsigma}.
Data are taken from 
Refs.~\cite{Barnes:1989je,Barnes:2000be,Barnes:1987aw,Barnes:1990bs,Barnes:1996si,Bassalleck:2004dz,PS185:2006yyx}.
    }
    \label{fig:ppbLLbPy}
\end{figure*}

\subsection{Results for $\NNb\to\LLb$}
Results for the 
integrated cross section are presented in Fig.~\ref{fig:ppbLLbsigma}.
The left side shows the situation close to the threshold which is 
at $p_{lab}= 1435.3$~MeV/c, i.e. for excess energies below $6$~MeV. 
The PS185 data from 2000
\cite{Barnes:2000be} are indicated by green triangles, while their
earlier data \cite{Barnes:1989je,Barnes:1994bh} correspond to the red
squares. The latter have been re-scaled by the factor $1.1$, as discussed
in Sect.~\ref{sec:LLbdata}. 
The results for the NLO and NLO$^+$ potentials are shown by
blue (grey) and purple (black) bands, respectively, where the bands reflect 
the variation of the results with the cutoff for $\Lambda = 800-900$~MeV.  
As one can see from the figures, both potentials describe the data in the
threshold region very well and are almost indistinguishable. Differences
are only visible at higher momenta where NLO$^+$ yields 
results well in line with the PS185 data up to $\approx 1770$~MeV/c. 

\begin{figure*}[t]
    \centering
    \includegraphics[width=0.99\linewidth]{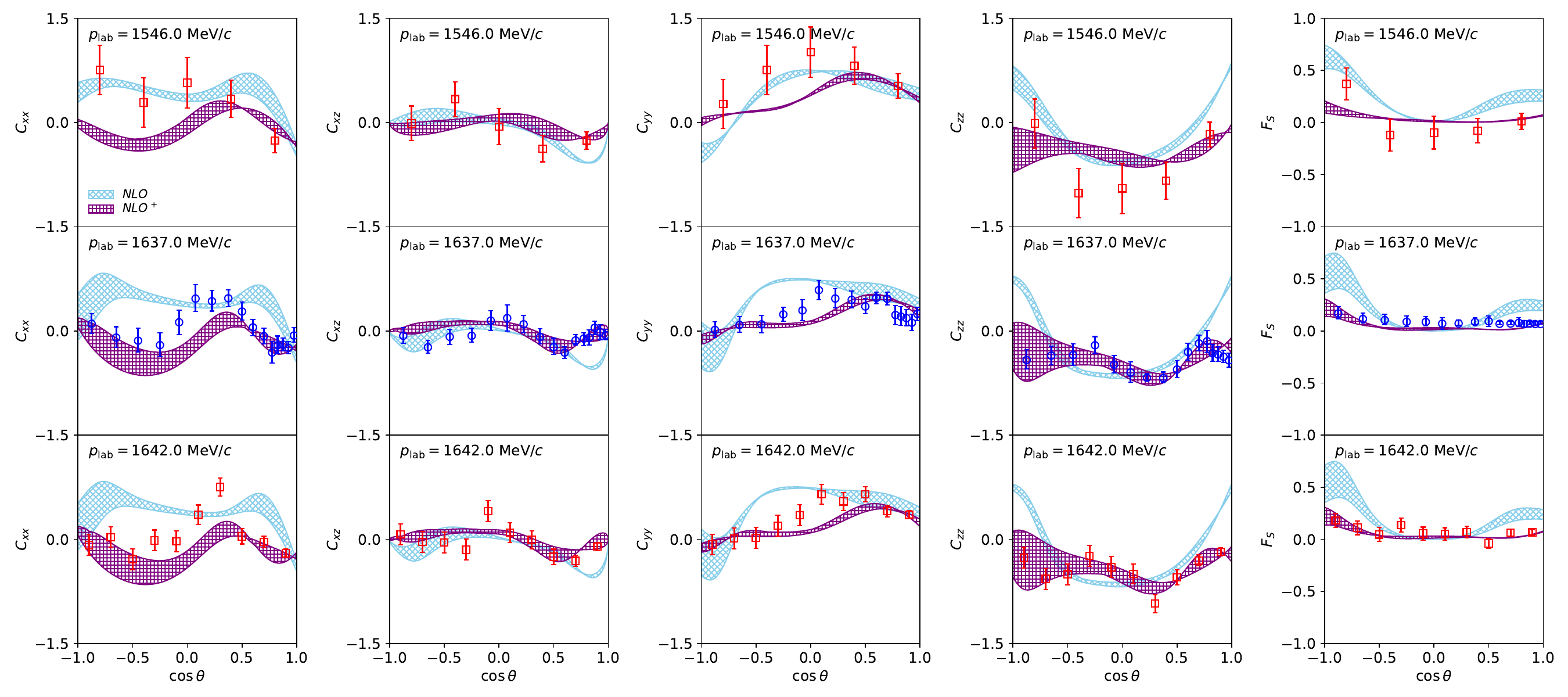}
    \caption{Spin-correlation parameters $C_{ij}$ for the reaction $\ppb\to\LLb$. Same description of
    bands as in Fig.~\ref{fig:ppbLLbdsigma}.
    Data are taken from 
Refs.~\cite{Barnes:1990bs,Barnes:1996si,PS185:2006yyx}.}
    \label{fig:ppbLLbSC}
\end{figure*}

\begin{figure*}[t]
    \centering
    \includegraphics[width=0.99\linewidth]{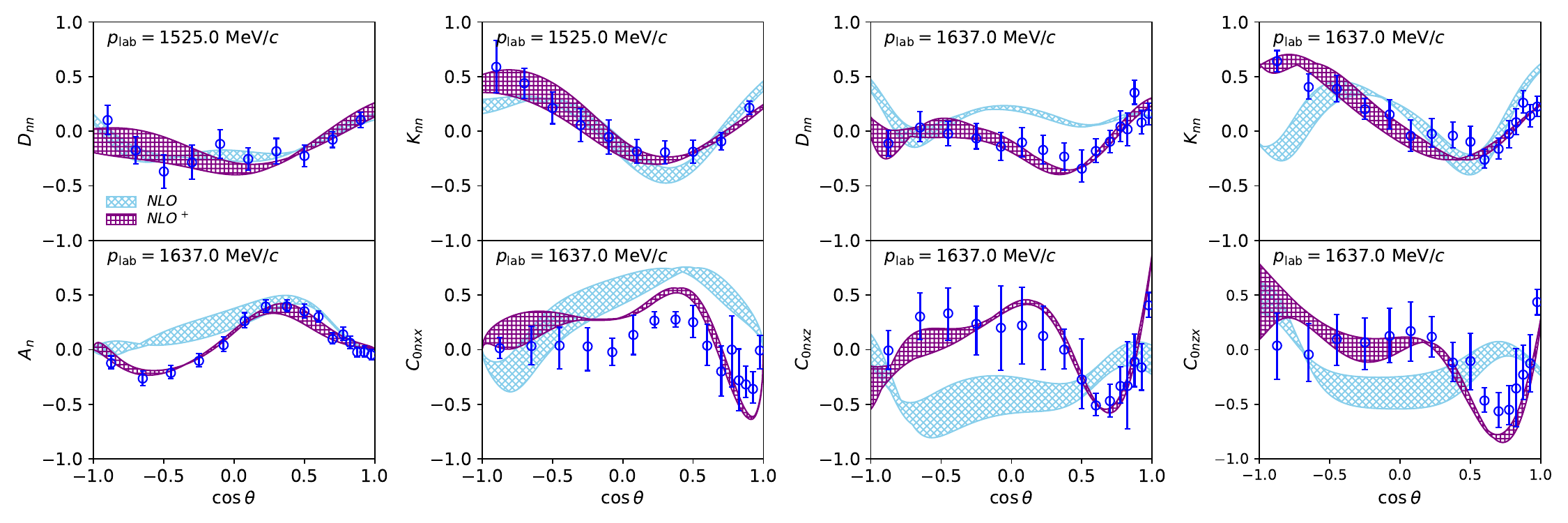}
    \caption{Spin-transfer and triple-spin 
    observables for the reaction $\ppb\to\LLb$.  Same description of
    bands as in Fig.~\ref{fig:ppbLLbdsigma}.
    Data are taken from 
Refs.~\cite{Bassalleck:2004dz,PS185:2006yyx}.
    }
    \label{fig:ppbLLbSC1}
\end{figure*}

Results for differential cross sections can be found in Fig.~\ref{fig:ppbLLbdsigma}. Obviously already at very low momenta the
data exhibit a noticeable angular dependence, a clear signal that higher
partial waves play a role right from the reaction threshold.
Note that $1436.9$~MeV/c corresponds to an excess energy of only $0.48$~MeV. 
At higher energies the cross section is characterized by a pronounced peak
in forward direction and a practically flat behavior in the backward region.
The NLO as well as the NLO$^+$ potential reproduce the angular dependence
at energies close to the threshold within the experimental uncertainty. 
The NLO$^+$ potential yields also an excellent description of the 
differential cross section at higher momenta, while NLO does show some
deficiencies. The latter is not surprising, given that the NLO interaction
has been fitted only to the data up to $1546$~MeV/c. 
It is worth noting that there is practically no cutoff
dependence of the achieved results.

Results for polarizations are summarized in Fig.~\ref{fig:ppbLLbPy}. 
One can see that the experimental results deviate from zero already at very 
low energies. This again signals that higher partial waves play a role 
right from the reaction threshold. This feature was already indicated
by the older data at $1436$ and $1437$~MeV, see Fig.~\ref{fig:ppbLLbPy} 
upper left side, despite of a large experimental uncertainty, and confirmed
by the new measurement from 2000 \cite{Barnes:2000be} at $1437.2$~MeV/c 
and upwards. Like in case of the angular distributions,
the results for NLO agree well with the
data up to $1546$~MeV/c, while the NLO$^+$ potential
reproduces the experiment also at higher energies.
There is some dependence on the cutoff, especially
at higher momenta, however, the variations are
small and comparable to the experimental uncertainty. 

\begin{figure*}[t]
    \centering
    \includegraphics[width=0.99\linewidth]{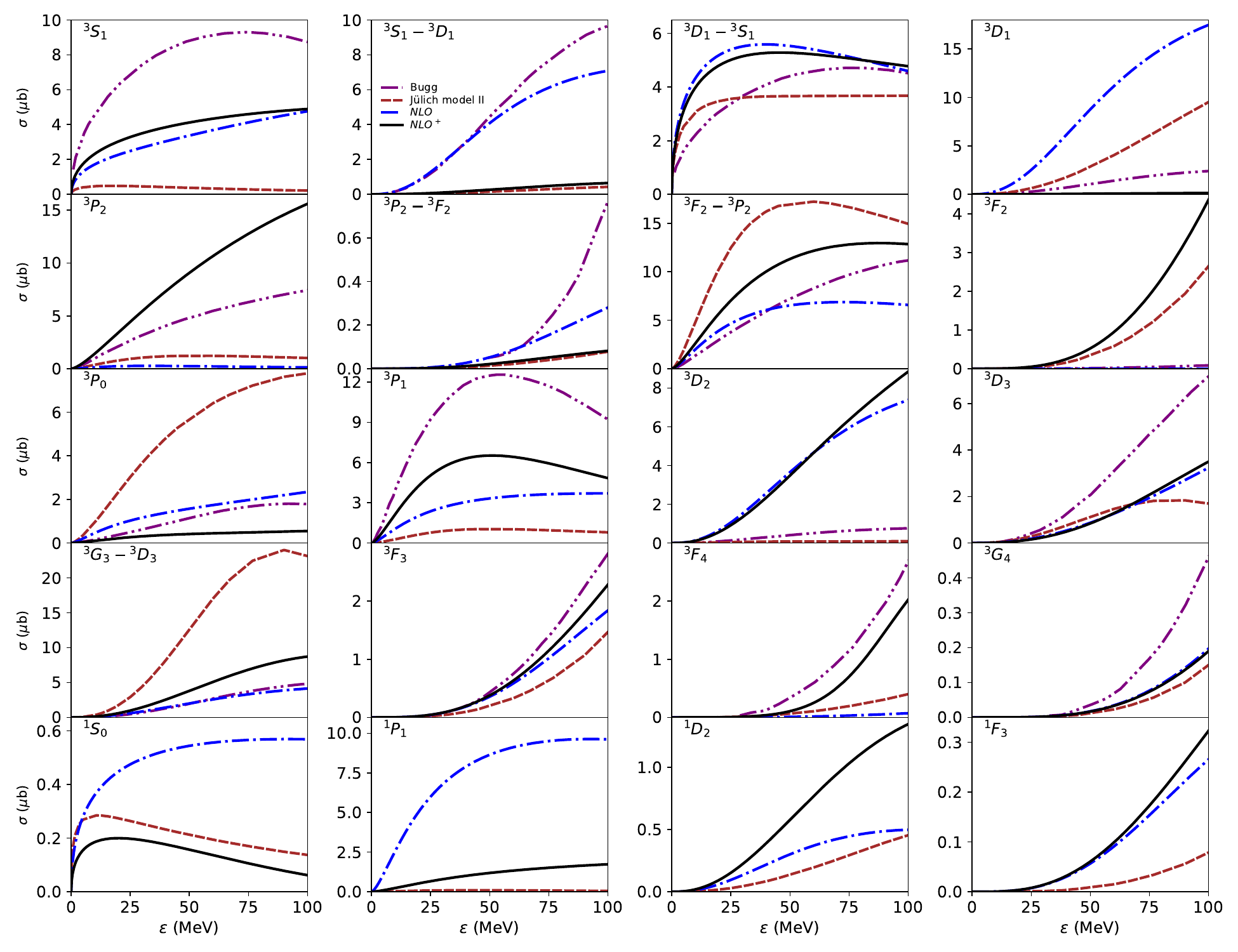}
    \caption{Partial-wave contributions to the 
    $\ppb\to\LLb$ cross section. Results are shown for the NLO (dashed lines) and 
    NLO$^+$ (solid lines) potentials (for the 
    cutoff $850$~MeV), for Bugg's phase-shift analysis \cite{Bugg:2004rj} 
    (dash-dot-dotted lines), and
    for the J\"ulich $\LLb$ model II \cite{Haidenbauer:1991kt} 
    (dashed lines).
    }
    \label{fig:LLb}
\end{figure*}

Spin-correlation parameters and the singlet 
fraction are presented in Fig.~\ref{fig:ppbLLbSC}.
Here one can observe a drastic difference between
the result from the NLO and the NLO$^+$ potentials
already at the lowest momentum where data are
available. We believe that this is in part a consequence of 
inconsistencies in the data. As already noted in Ref.~\cite{Timmermans:1992fu} 
the data at $1546$~MeV/c do not satisfy the requirement $C_{xx}=-C_{yy}$ 
at backward angles, which follows from rotational invariance and 
is, of course, fulfilled by the calculation. Since the NLO potential is 
fitted up to $1546$~MeV/c, only the $C_{ij}$ data at that momentum enter
and the solution we obtain is seemingly biased by the possibly wrong 
value of $C_{xx}$ in backward direction. 
The NLO$^+$ fit, on the other 
hand, involves also the data at $1637$ and $1642$~MeV/c. Since those 
data have smaller uncertainty they dominate the $\chi^2$ minimization. As a 
consequence, the $C_{ij}$ at those momenta are fairly well reproduced as one 
can see in Fig.~\ref{fig:ppbLLbSC}. Note that the data at
those two momenta fulfill the requirements at backward angles within the
experimental uncertainty.
Regarding the singlet fraction, the predictions from the NLO potential
exhibit a sizable deviation from $F_S=0$, in forward as well as in backward
direction. Again we attribute this to the conflicting situation for 
$C_{xx}$ and $C_{yy}$. The NLO$^+$ results show a small but noticeable deviation
from the value zero, at small and large angles, and in line with the
experimental information. 

Finally, spin-transfer and triple-spin observables are shown in 
Fig.~\ref{fig:ppbLLbSC1}.
For those observables the results for NLO agree well with the
data at $1525$~MeV/c, while the NLO$^+$ potential
reproduces the experiment also at $1637$ and $1642$~MeV/c.
The dependence on the cutoff is somewhat larger
than for the angular distributions and polarizations. 
However, again the variations are of an order
comparable to the experimental uncertainty.

    \begin{table*}[htbp!]
\renewcommand{\arraystretch}{1.5}
\centering
    \caption{Partial-wave cross sections at $p_{\rm lab}=1446$ MeV$/c$ (top) and $p_{\rm lab}=1546$ MeV$/c$ (bottom). The total empirical cross sections are $6.16\pm0.45$~$\mu$b and $49.06\pm1.50$~$\mu$b, respectively. Empty entries mean that the partial wave was either not considered or is very small.}
    \begin{tabular}{|c |c c c |c c c|c |c|c|}
    \hline\hline
    & \multicolumn{3}{c|}{NLO} & \multicolumn{3}{c|}{NLO$^+$} & J\"ulich II& Nijmegen & Bugg \\
    & 800 & 850& 900 & 800 & 850& 900 & \cite{Haidenbauer:1991kt} & \cite{Timmermans:1992fu}& \cite{Bugg:2004rj}\\
        \hline
        \multicolumn{10}{c}{$p_{\rm lab}=1446$ MeV$/c$} \\ \hline
        $^3D_1\to {}^3S_1$   & 3.42  & 3.02  & 2.58  &  2.93 &  2.76 & 2.54 & 2.29 &  2.9 & 1.34 \\
        $^3F_2\to {}^3P_2$   & 0.53  & 0.57  & 0.61  &  0.64 &  0.70 & 0.61 & 1.23 &  0.7 & 0.33 \\
        $^3G_3\to {}^3D_3$   & 0.01  & 0.01  & 0.01  &  0.01 &  0.01 & 0.01 & 0.03 &      & 0.00 \\
        $^1S_0$              & 0.20  & 0.25  & 0.23  &  0.01 &  0.14 & 0.14 & 0.26 &      &      \\
        $^1P_1$              & 0.63  & 0.72  & 0.80  &  0.07 &  0.06 & 0.05 & 0.01 &      &      \\
        $^3S_1$              & 0.84  & 1.14  & 1.52  &  1.39 &  1.53 & 1.82 & 0.37 &  0.3 & 2.80 \\
        $^3P_0$              & 0.13  & 0.14  & 0.13  &  0.02 &  0.03 & 0.04 & 0.23 &  0.1 & 0.04 \\
        $^3P_1$              & 0.39  & 0.31  & 0.29  &  0.47 &  0.60 & 0.64 & 0.07 &  0.5 & 1.05 \\
        $^3P_2$              & 0.05  & 0.04  & 0.05  &  0.65 &  0.35 & 0.31 & 0.10 &  0.4 & 0.26 \\
        $^3D_1$              & 0.03  & 0.03  & 0.02  &  0.00 &  0.00 & 0.00 & 0.01 &      & 0.00 \\
        % $^1D_2$              & 0.00  & 0.00  & 0.00  &  0.00 &  0.00 & 0.00 & 0.00 &      &       \\
        $^3D_2$              & 0.01  & 0.01  & 0.01  &  0.01 &  0.01 & 0.01 & 0.00 &      & 0.00 \\
        % $^3D_3$              & 0.00  & 0.00  & 0.00  &  0.00 &  0.00 & 0.00 & 0.00 &      & 0.00 \\
        % $^3F_2$              & 0.00  & 0.00  & 0.00  &  0.00 &  0.00 & 0.00 & 0.00 &      & 0.00 \\
        % $^3F_3$              & 0.00  & 0.00  & 0.00  &  0.00 &  0.00 & 0.00 & 0.00 &      & 0.00 \\
        $^3S_1\to {}^3D_1$   & 0.02  & 0.02  & 0.01  &  0.00 &  0.00 & 0.00 & 0.00 &      & 0.00 \\
        % $^3P_2\to {}^3F_2$   & 0.00  & 0.00  & 0.00  &  0.00 &  0.00 & 0.00 & 0.00 &      & 0.00 \\
        % $^3F_4$              & 0.00  & 0.00  & 0.00  &  0.00 &  0.00 & 0.00 & 0.00 &      & 0.00 \\
        % $^3G_4$              & 0.00  & 0.00  & 0.00  &  0.00 &  0.00 & 0.00 & 0.00 &      & 0.00 \\
        \hline
        Total                & 6.27  & 6.26  & 6.25  &  6.21 &  6.19 & 6.17 & 4.58 &  4.86 & 5.83 \\
        \hline
        \multicolumn{10}{c}{$p_{\rm lab}=1546$ MeV$/c$} \\ \hline
        $^3D_1\to {}^3S_1$   &  6.44  &  5.59  &  4.66  &  6.35 &  5.27 &  4.39 &  3.65 &  4.0 &  4.04 \\
        $^3F_2\to {}^3P_2$   &  5.37  &  6.00  &  6.76  &  9.39 &  9.88 &  8.45 & 16.08 &  8.9 &  5.78 \\
        $^3G_3\to {}^3D_3$   &  1.15  &  1.36  &  1.64  &  2.21 &  2.42 &  2.57 &  7.76 &  9.6 &  1.24 \\
        $^1S_0$              &  0.37  &  0.52  &  0.50  &  0.00 &  0.18 &  0.19 &  0.23 &      &       \\
        $^1P_1$              &  6.47  &  7.80  &  9.54  &  1.09 &  0.99 &  0.78 &  0.10 &  0.1 &       \\
        $^3S_1$              &  2.36  &  3.00  &  3.95  &  3.25 &  3.82 &  4.66 &  0.42 &  0.7 &  8.18 \\
        $^3P_0$              &  1.39  &  1.37  &  1.25  &  0.23 &  0.37 &  0.68 &  4.70 &  0.7 &  0.86 \\
        $^3P_1$              &  3.19  &  3.03  &  3.02  &  4.31 &  6.27 &  7.93 &  1.01 &  5.3 & 11.76 \\
        $^3P_2$              &  0.29  &  0.26  &  0.39  & 10.05 &  7.14 &  6.48 &  1.17 &  5.1 &  3.97 \\
        $^3D_1$              &  7.09  &  5.91  &  4.17  &  0.04 &  0.04 &  0.05 &  1.69 &  0.5 &  0.68 \\
        $^1D_2$              &  0.19  &  0.21  &  0.23  &  0.34 &  0.37 &  0.45 &  0.08 &      &       \\
        $^3D_2$              &  2.61  &  2.46  &  2.47  &  1.78 &  2.24 &  2.66 &  0.07 &  1.4 &  0.29 \\
        $^3D_3$              &  0.59  &  0.51  &  0.47  &  0.91 &  0.46 &  0.22 &  0.72 &  3.2 &  1.16 \\
        $^3F_2$              &  0.00  &  0.00  &  0.00  &  0.33 &  0.23 &  0.10 &  0.15 &      &  0.01 \\
        $^3F_3$              &  0.15  &  0.17  &  0.19  &  0.16 &  0.18 &  0.20 &  0.09 &  0.2 &  0.18 \\
        $^3S_1\to {}^3D_1$   &  3.72  &  2.85  &  1.53  &  0.19 &  0.17 &  0.17 &  0.09 &  0.8 &  2.89 \\
        $^3P_2\to {}^3F_2$   &  0.01  &  0.02  &  0.04  &  0.01 &  0.01 &  0.02 &  0.01 &  0.1 &  0.00 \\
        $^3F_4$              &  0.00  &  0.00  &  0.00  &  0.11 &  0.03 &  0.01 &  0.03 &      &  0.18 \\
        $^3G_4$              &  0.01  &  0.01  &  0.01  &  0.01 &  0.01 &  0.01 &  0.00 &      &  0.00 \\
        \hline
        Total                & 41.42  & 41.10  & 40.82  & 40.76 & 40.07 & 40.03 & 38.06 & 40.5 & 41.20 \\ \hline\hline
    \end{tabular}
    \label{tab:PW1}
    \renewcommand{\arraystretch}{1.0}
\end{table*}

\begin{table*}[htbp!]
\renewcommand{\arraystretch}{1.5}
\centering
    \caption{Partial-wave cross sections at $p_{\rm lab}=1642$ MeV$/c$. The total empirical cross section is 
    $70.51\pm1.65$~$\mu$b. Empty entries mean that the partial wave was either not considered or is very small.}
    \begin{tabular}{|c |c c c |c c c|c|c|}
    \hline\hline
    & \multicolumn{3}{c|}{NLO} & \multicolumn{3}{c|}{NLO$^+$} & J\"ulich II &  Bugg \\
    & 800 & 850& 900 & 800 & 850& 900 & \cite{Haidenbauer:1991kt} &  \cite{Bugg:2004rj}\\
        \hline
        $^3D_1\to {}^3S_1$   &  5.86  &  5.17  &  4.25  &  7.19 &  5.09 &  3.65 &  3.67 &  4.71 \\
        $^3F_2\to {}^3P_2$   &  6.01  &  6.87  &  7.87  & 13.14 & 12.77 &  9.92 & 16.80 &  9.54 \\
        $^3G_3\to {}^3D_3$   &  2.63  &  3.23  &  4.08  &  4.97 &  6.62 &  7.90 & 21.53 &  3.58 \\
        $^1S_0$              &  0.37  &  0.57  &  0.58  &  0.02 &  0.11 &  0.13 &  0.17 &       \\
        $^1P_1$              &  7.49  &  9.47  & 12.18  &  1.56 &  1.49 &  1.16 &  0.08 &       \\
        $^3S_1$              &  3.34  &  4.07  &  5.18  &  3.64 &  4.57 &  5.70 &  0.29 &  9.29 \\
        $^3P_0$              &  2.04  &  1.94  &  1.69  &  0.22 &  0.48 &  0.97 &  7.08 &  1.64 \\
        $^3P_1$              &  3.30  &  3.63  &  3.89  &  3.95 &  6.02 &  8.76 &  0.96 & 11.86 \\
        $^3P_2$              &  0.20  &  0.17  &  0.31  & 13.92 & 12.49 & 11.25 &  1.17 &  6.18 \\
        $^3D_1$              & 16.18  & 13.80  & 10.52  &  0.10 &  0.10 &  0.12 &  5.78 &  1.82 \\
        $^1D_2$              &  0.38  &  0.44  &  0.51  &  0.88 &  1.00 &  1.30 &  0.28 &       \\
        $^3D_2$              &  5.89  &  5.79  &  6.11  &  4.47 &  6.13 &  7.92 &  0.08 &  0.62 \\
        $^3D_3$              &  2.02  &  1.84  &  1.79  &  3.07 &  1.96 &  1.09 &  1.77 &  4.46 \\
        $^3F_2$              &  0.00  &  0.00  &  0.00  &  2.31 &  1.74 &  0.81 &  1.05 &  0.04 \\
        $^3F_3$              &  0.81  &  0.95  &  1.08  &  0.93 &  1.07 &  1.21 &  0.58 &  1.30 \\
        $^3S_1\to {}^3D_1$   &  7.79  &  5.99  &  3.39  &  0.53 &  0.47 &  0.44 &  0.28 &  7.44 \\
        $^3P_2\to {}^3F_2$   &  0.08  &  0.14  &  0.22  &  0.06 &  0.05 &  0.11 &  0.04 &  0.19 \\
        $^3F_4$              &  0.03  &  0.03  &  0.03  &  1.71 &  0.62 &  0.20 &  0.18 &  1.05 \\
        $^3G_4$              &  0.06  &  0.07  &  0.08  &  0.05 &  0.07 &  0.08 &  0.05 &  0.15 \\
        \hline
        Total                & 64.47  & 64.18  & 63.76  & 62.72 & 62.85 & 62.73 & 61.84 & 63.89 \\
        \hline\hline
    \end{tabular}
    \label{tab:PW3}
    \renewcommand{\arraystretch}{1.0}
\end{table*}

Additional results of the NLO$^+$ potential, for energies up to $1835$~MeV/c, 
can be found in \ref{app:highe}. One can see that there is still good agreement
with experiment up to the data at $1726$~MeV/c. But then contributions from
higher partial become increasingly important, beyond the ones that can be
controlled within the present NLO$^+$ scenario. 

Let us now take a look at the partial-wave contributions to the 
$\ppb\to\LLb$ cross section. These are shown in Fig.~\ref{fig:LLb} as a function
of the excess energy, and summarized for selected energies in 
Tables~\ref{tab:PW1} ($1446$~MeV/c, $1546$~MeV/c),
and \ref{tab:PW3} ($1642$~MeV/c).
We compare our results with
others from the literature, notably with the 
partial-wave analyses of Bugg~\cite{Bugg:2004rj}
and Timmermans et al.~\cite{Timmermans:1992fu}.
We include also the results for one of the J\"ulich meson-exchange 
potentials \cite{Haidenbauer:1991kt} (model II)
for illustration. 
The Salamanca Group provided also partial-wave
cross sections for their calculation within a constituent quark model~\cite{Ortega:2011zza} and compared it to the results of Bugg. 

Obviously, there is a sizable variation between 
the partial-wave cross sections of the considered calculations. Of course, 
the contributions from singlet transitions are very small, as reflected in 
the measured singlet fraction, cf. Fig.~\ref{fig:ppbLLbSC}, 
and therefore the corresponding amplitudes cannot be fixed well
\cite{Timmermans:1992fu,Bugg:2004rj}, which explains the large spread of 
the results. 
But also in the other partial waves there is at most qualitative agreement 
in some cases and drastic differences in others. Even when comparing
the NLO and NLO$^+$ results quantitative agreement is only observed in a few partial waves. 
The discrepancies could be due to the different fitting scenarios and/or 
a consequence of a more limited set of data. For example, the complete 
data set at $1637$~MeV/c was not available in the 1990ties when
the Nijmegen and J\"ulich groups established 
their potentials. They are also not included in
our NLO fit. The data for $D_{nn}$ and $K_{nn}$ at $1525$~MeV/c 
were likewise published after 2000. They were already available 
at the time when Bugg performed his analysis though, for reasons 
unknown to us, those data were not included by him. 
Of course, it is also possible that the presently
available data simply do not allow for an unambiguous
determination of the partial-wave amplitudes.

\begin{table*}[htbp!]
    \centering
    \caption{Breakdown of the $\chi^2$ contributions for observables
    of the reaction $\ppb\to\LLb$ at selected momenta.}
    \begin{tabular}{|c | c | c | c | c c | c c|}
    \hline\hline
       $p_{\rm lab}$(MeV$/c$)  & Ref.  & No. type & normalization  & \multicolumn{2}{c}{NLO} & \multicolumn{2}{c}{NLO$^+$} \\
       &       &          &             &  $\chi^2_{\rm min}$ & $\chi^2/$data &  $\chi^2_{\rm min}$ & $\chi^2/$data     \\ \hline
       1446.0          & \cite{Barnes:2000be}     &  7$d\sigma$  &  1.0198  &   5.75 &  0.82 &   4.76 &  0.68 \\
       (1447.1,1446.8) & \cite{Barnes:2000be}     &  6$P_y$      &  0.7198  &   3.62 &  0.60 &   4.96 &  0.83 \\
       1546.0          & \cite{Barnes:1990bs}     & 20$d\sigma$  &  0.8286  &  66.15 &  3.30 &  22.39 &  1.12 \\
       (1525.0)        & \cite{Barnes:1990bs}     & 13$P_y$      &  0.8741  &  10.02 &  0.77 &  18.63 &  1.43 \\
                       & \cite{Bassalleck:2004dz} & 20$P_y$      &  0.9084  &  30.34 &  1.52 &  31.36 &  1.57 \\
                       & \cite{Barnes:1990bs}     &  5$C_{xx}$   &  1.0000  &   8.76 &  1.75 &  11.55 &  2.31 \\
                       & \cite{Barnes:1990bs}     &  5$C_{xz}$   &  1.0000  &   8.20 &  1.64 &   3.21 &  0.64 \\
                       & \cite{Barnes:1990bs}     &  5$C_{yy}$   &  1.0000  &   3.17 &  0.63 &   5.80 &  1.16 \\
                       & \cite{Barnes:1990bs}     &  5$C_{zz}$   &  1.0000  &  12.86 &  2.57 &   5.42 &  1.08 \\
                       & \cite{Bassalleck:2004dz} & 10$D_{nn}$   &  1.0000  &   3.33 &  0.33 &  10.36 &  1.04 \\
                       & \cite{Bassalleck:2004dz} & 10$K_{nn}$   &  1.0000  &  12.45 &  1.25 &   5.27 &  0.53 \\
        1642.0         & \cite{Barnes:1996si}     & 40$d\sigma$  &  0.9122  &        &       & 124.92 &  3.12 \\
        (1637.0)       & \cite{PS185:2006yyx}     & 16$d\sigma$  &  0.7772  &        &       &  59.49 &  3.72 \\
                       & \cite{Barnes:1996si}     & 20$P_y$      &  0.9310  &        &       &  31.61 &  1.58 \\
                       & \cite{PS185:2006yyx}     & 16$P_y$      &  0.8631  &        &       &  14.98 &  0.94 \\
                       & \cite{Barnes:1996si}     & 10$C_{xx}$   &  1.0000  &        &       &  48.71 &  4.87 \\
                       & \cite{Barnes:1996si}     & 10$C_{xz}$   &  1.0000  &        &       &  15.25 &  1.53 \\
                       & \cite{Barnes:1996si}     & 10$C_{yy}$   &  1.0000  &        &       &  22.58 &  2.26 \\
                       & \cite{Barnes:1996si}     & 10$C_{zz}$   &  1.0000  &        &       &  11.02 &  1.10 \\
                       & \cite{PS185:2006yyx}     & 16$C_{xx}$   &  1.0000  &        &       &  35.51 &  2.22 \\
                       & \cite{PS185:2006yyx}     & 16$C_{xz}$   &  1.0000  &        &       &  27.65 &  1.73 \\
                       & \cite{PS185:2006yyx}     & 16$C_{yy}$   &  1.0000  &        &       &  25.14 &  1.57 \\
                       & \cite{PS185:2006yyx}     & 16$C_{zz}$   &  1.0000  &        &       &  24.21 &  1.51 \\
                       & \cite{PS185:2002tao}     & 16$D_{nn}$   &  1.0000  &        &       &  10.35 &  0.65 \\
                       & \cite{PS185:2002tao}     & 16$K_{nn}$   &  1.0000  &        &       &  16.57 &  1.04 \\
                       & \cite{PS185:2006yyx}     & 16$A_n$      &  1.0000  &        &       &  18.10 &  1.13 \\
                       & \cite{PS185:2006yyx}     & 16$C_{0nmm}$ &  1.0000  &        &       &  52.04 &  3.25 \\
                       & \cite{PS185:2006yyx}     & 16$C_{0nml}$ &  1.0000  &        &       &  17.35 &  1.08 \\
                       & \cite{PS185:2006yyx}     & 16$C_{0nlm}$ &  1.0000  &        &       &  28.05 &  1.75 \\
                       \hline\hline              
    \end{tabular}
    \label{tab:chisq}
\end{table*}

%%%%%%%%%%%%%%%%%%%%%%%%%%%%%%%%%%
\begin{figure*}[t]
    \centering
    \includegraphics[width=0.99\linewidth]{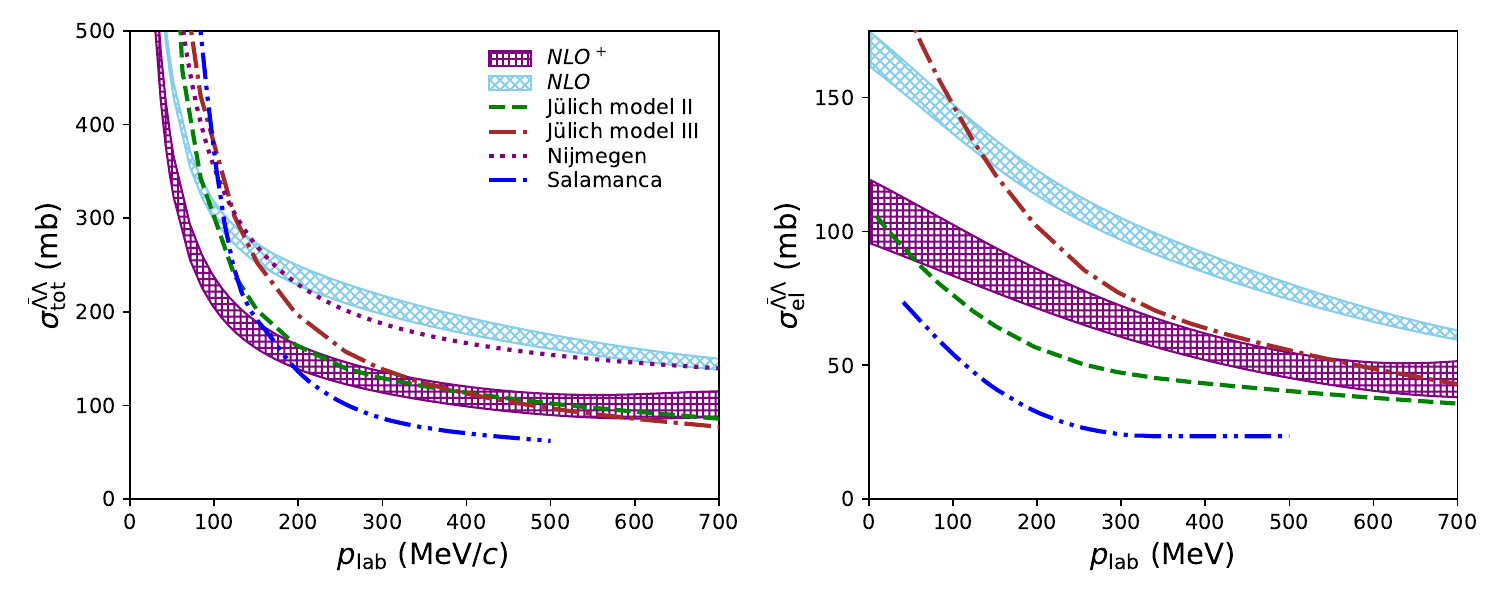}
\caption{Total and integrated elastic cross sections for the reaction 
$\LLb\to\LLb$. The blue (grey) and red (black) bands correspond to the
NLO and NLO$^+$ potentials, respectively. The dashed and dash-dotted line
are the results for the variants II and III of the J\"ulich meson-exchange 
$\LLb$ potential \cite{Haidenbauer:1991kt,Haidenbauer:1992wp}.
Predictions by the Nijmegen \cite{Timmermans:1992fu}
and Salamanca \cite{GarciaOrtega:2012fmb} groups are indicated by the
dotted and dash-double dotted lines, respectively.
    }
    \label{fig:LLbsigma}
\end{figure*}

From the results listed in Tables~\ref{tab:PW1} 
and \ref{tab:PW3} one can read off which 
partial waves dominate the cross section at specific momenta. 
Past calculations in the meson-exchange framework
have shown a dominance of transitions 
$L_{\LLb}=L_{\ppb}-2$, generated by the strong
tensor forces due to $K$~($K^*,...)$ exchanges. Indeed, 
at the lowest momentum considered, $1446$~MeV/c ($\varepsilon = 3.68$~MeV), 
the $^3D_1\to{}^3S_1$ transition dominates the cross section in all
calculations, except for the Bugg analysis where the largest contribution
comes from the $^3S_1$. 
The picture is more varied at $1546$~MeV/c ($39.1$~MeV). Here only the
J\"ulich model yields a single dominant partial wave, namely the
$^3F_2\to{}^3P_2$ transition. In all other cases there are two
(Nijmegen, Bugg) or even three (NLO, NLO$^+$) contributions that are 
comparably large. Still, in all results the $^3F_2\to{}^3P_2$ transition 
does play an important role.
The situation at $1642$~MeV/c ($73$~MeV) is
similar. In general, there are several contributions
of comparable magnitude where the largest one
comes from the $^3P_1$ (Bugg), $^3G_3\to{}^3D_3$
(J\"ulich II), $^3D_1$ (NLO), and $^3P_2$
(NLO$^+$).

An overview of the achieved total $\chi^2$ and the $\chi^2$/data has
been already given in Table~\ref{tab:chi2} in Sect.~\ref{sec:fit}. 
As additional information, a breakdown of the $\chi^2$ contributions
for the individual observables for a selection of momenta is provided 
in Table~\ref{tab:chisq}. Also listed is the normalization applied to
the individual data sets.
Timmermans et al. \cite{Timmermans:1992fu} assumed a $5$~\% normalization 
error for all data.
According to the PS185 paper from 2000 \cite{Barnes:2000be} the overall
normalization uncertainty in the earlier experiments could be
approximately $15$~\%. As one can see from Table~\ref{tab:chisq} in most cases 
the normalization factors we find in the fitting procedure lie below the 
latter value. Interestingly, no normalization is
required for the $C_{ij}$ and the other double- and triple-spin observables.

% Discuss some special values/observables!?

%%%%%%%%%%%%%%%%%%%%%%%%%%%%%%%%%%
\subsection{Predictions for $\LLb$ scattering}

Due to the absence of genuine $\LLb$ scattering data the $\LLb$ interaction as
well as the $\NNb\to\LLb$ transition potential have to be established from
fitting to the $\ppb\to\LLb$ data. Thus, the results for $\LLb$ scattering can 
be considered as actual predictions. However, to which extend 
is the $\LLb$ interaction really determined by the analysis of the 
production data? In the following we compare our present results for $\LLb$
with our past ones \cite{Haidenbauer:1991kt}, and with predictions from
other groups that can be found in the literature \cite{Timmermans:1992fu,GarciaOrtega:2012fmb}. 
Clearly, agreement can be expected on a qualitative level only because there
should be a certain degree of model dependence. And, in addition, the
considered interactions do not describe the $\ppb\to\LLb$ data on the same 
quantitative level. 

In Fig.~\ref{fig:LLbsigma} we show total and integrated elastic 
cross sections for $\LLb$ for the NLO and NLO$^+$ interactions  
and compare them with results from the 
J\"ulich meson-exchange potential, and from the 
Nijmegen \cite{Timmermans:1992fu} 
and Salamanca groups~\cite{GarciaOrtega:2012fmb}.
obviously, the predictions from the J\"ulich $\LLb$ potentials 
more or less coincide with the NLO$^+$ cross sections at momenta above, say,
$300$~MeV/c. There is a larger difference between the NLO and NLO$^+$ 
predictions in the total as well as in the elastic cross section. 
The Nijmegen group published only results for the total cross section, 
cf. the dotted line. Interestingly, those are close 
to the ones from the NLO potential. The results from the Salamanca group
(dashed-double dotted lines), based on the constituent-quark model, exhibit 
a stronger momentum dependence and, accordingly, the cross sections at higher momenta are fairly small. 

\begin{figure}
    \centering
    \includegraphics[width=0.99\linewidth]{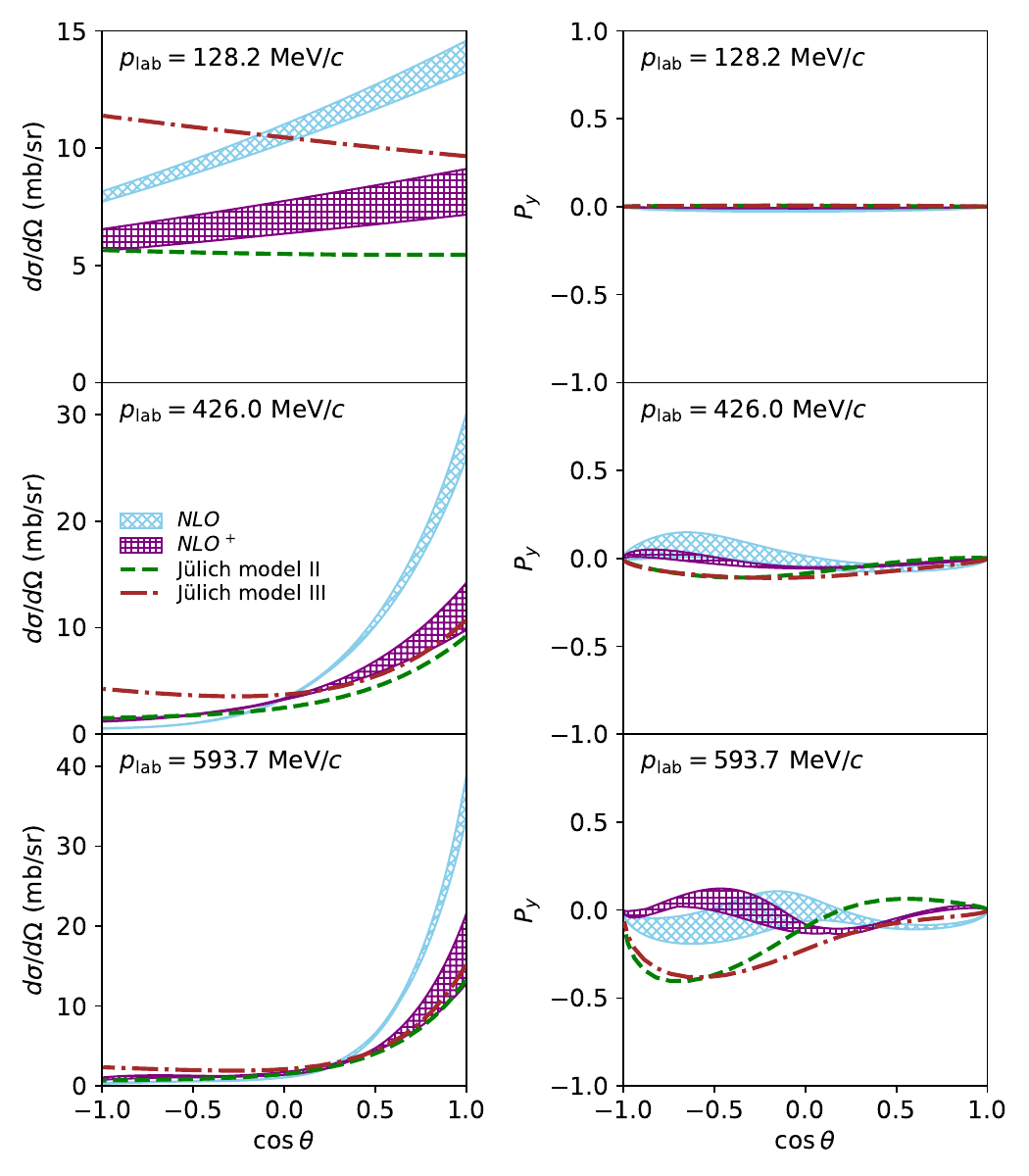}
    \caption{Differential cross sections and polarization for the reaction $\LLb\to\LLb$.
    Same description of the curves as in Fig.~\ref{fig:LLbsigma}.}
    \label{fig:LLbdsPy}
\end{figure}

Selected results for the differential cross section
and the polarization are shown in Fig.~\ref{fig:LLbdsPy}. 
Note that the lab momentum $128$~MeV/c in the $\LLb$ system corresponds to
$1446$~MeV/c in the $\NNb$ system,  $426$~MeV/c to 
$1546$~MeV/c, and $594$~MeV/c to $1642$~MeV/c.
One can see that the difference between the NLO and NLO$^+$ interactions is 
primarily due to a different behavior in forward direction,
where the prediction of the former potential is
strongly enhanced. Again, the results for NLO$^+$
and the phenomenological J\"ulich potentials
are quite similar at the higher momenta. There are differences in the
polarization, especially at the highest considered momentum. However, overall
$P_y$ is predicted to be small by all interactions considered. 

Results of the NLO and NLO$^+$ potentials for the $\LLb$ scattering lengths 
in the $^1S_0$ and $^3S_1$ partial waves are summarized in Table~\ref{tab:ere}.
Evidently, the real part is predicted to be in the order of $1$~fm while the
imaginary part is around $0.3$~fm.
Comparing those values with the ones found for the $\NNb$ interaction \cite{Dai:2017ont} (Table~3 in that paper) one can see that the real parts
are similar while the imaginary parts of the $\LLb$ scattering lengths
are roughly a factor two smaller, which suggests that there is less annihilation 
in the $\LLb$ system. The J\"ulich potentials predict 
somewhat larger values for the imaginary part in the $^3S_1$
and significantly larger values in the $^1S_0$ partial wave. The latter 
is not too surprising because, as noted above, the singlet amplitudes are
not that well constrained by the $\ppb\to\LLb$ data. 

\begin{table}[htbp]
    \centering
        \caption{$\LLb$ scattering lengths (in fm) in the $^1S_0$ and $^3S_1$ partial waves
 predicted by by the $\chi$EFT potentials presented in this
 work and different meson-exchange $\LLb$ potentials (I-IV) \cite{Haidenbauer:1991kt,Haidenbauer:1992wp}.  
        NLO and NLO(he) refer to the NLO
        potentials fitted to data up to $p_{\rm lab}=$1546 MeV$/c$ and 1642 MeV$/c$, respectively.
        }
    \begin{tabular}{|l | l | c | c|}
    \hline\hline
                           &  Cutoff (MeV) & $a (^1S_0)$  & $a (^3S_1)$   \\ \hline
    \multirow{3}{*}{NLO} &    800           & $0.99 -{\rm i}0.33$ & $1.19 -{\rm i}0.28$ \\
                           &    850           & $1.12 -{\rm i}0.28$ & $1.14 -{\rm i}0.28$ \\
                           &    900           & $1.15 -{\rm i}0.36$ & $1.08 -{\rm i}0.27$ \\ \hline
   \multirow{3}{*}{NLO(he)} &    800           & $1.12 -{\rm i}0.33$ & $1.11 -{\rm i}0.26$ \\
                           &    850           & $1.10 -{\rm i}0.28$ & $1.09 -{\rm i}0.25$ \\
                           &    900           & $1.02 -{\rm i}0.29$ & $1.03 -{\rm i}0.23$ \\ \hline
   \multirow{3}{*}{NLO$^+$} &    800           & $0.75 -{\rm i}0.38$ & $0.99  -{\rm i}0.24$ \\
                           &    850           & $0.78 -{\rm i}0.40$ & $0.96 -{\rm i}0.23$ \\
                           &    900           & $0.67 -{\rm i}0.42$ & $0.88 -{\rm i}0.20$ \\                        
                           \hline \hline
\multicolumn{2}{|l|}{J\"ulich I}  &  $0.32 -{\rm i}  0.52$  & $ 0.74 -{\rm i}  0.56$ \\
\multicolumn{2}{|l|}{J\"ulich II} &  $0.67 -{\rm i}  1.14$  & $ 0.66 -{\rm i}  0.37$ \\
\multicolumn{2}{|l|}{J\"ulich III} &  $1.42 -{\rm i}  1.15$  & $ 1.00 -{\rm i}  0.44$ \\
\multicolumn{2}{|l|}{J\"ulich IV} &  $1.56 -{\rm i}  1.40$  & $ 0.98 -{\rm i}  0.65$ \\                         \hline\hline
    \end{tabular}
    \label{tab:ere}
\end{table}

As already said, there are no direct measurements of $\LLb$ scattering
observables. However, two-particle $\LLb$ momentum correlation
functions, that have been measured by the ALICE Collaboraton in Pb–Pb 
collisions \cite{ALICE:2019igo} and in $pp$ collisions \cite{ALICE:2021cyj}, 
do allow a glimpse on the $\LLb$ interaction, at least on a qualitative
level. Indeed, the ALICE Collaboration has deduced spin-averaged values 
for the $\LLb$ scattering length from an analysis of their 
$\LLb$ correlation functions. 
The values reported are $(0.90\pm 0.16)-{\rm i}(0.40\pm 0.18)$~fm
from data obtained in Pb-Pb collisions \cite{ALICE:2019igo}
and $(1.54\pm 0.03) - {\rm i} (0.43 \pm 0.08)$~fm \cite{Sarti:2025sdo}
for data from $pp$ collisions \cite{ALICE:2021cyj}. Interestingly, those
values are roughly in line with the predictions from the NLO and NLO$^+$
potentials. 
Nonetheless, it should be mentioned that there are same caveats associated 
with the underlying approach, femtoscopy,
see the discussion in the Appendix of Ref.~\cite{Haidenbauer:2023zcu} and
also the more fundamental critique in Ref.~\cite{Epelbaum:2025aan}.

\section{Summary and outlook}
\label{sec:Summary}
This paper had two goals. First, the antibaryon-baryon interaction has
been reviewed in the context of the hyperon-pro\-duction reactions
$\NNb\to\YYb$ ($Y=\Lambda,\, \Sigma,\,\Xi$)
with focus on the production of the $\LLb$ system. 
An overview of the experimental situation for $\NNb\to\YYb$ from threshold up to antiproton laboratory momenta of $2$~GeV/c has been
given and past theoretical approaches and results have been recounted.

In addition, and as principal part, a new calculation of the reaction 
$\ppb\to\LLb$ has been presented which is performed along the 
lines of SU(3) chiral EFT. 
The description of the $\BBb$ interaction in terms of potentials derived 
within SU(3) chiral EFT is outlined in detail and 
constraints from SU(3) flavor symmetry on the 
$\BBb$ interaction in the various reaction channels are indicated. 
The concrete application to the reaction $\ppb\to\LLb$ is performed
utilizing potentials for the $\NNb$ and $\LLb$ channels, and for the 
$\NNb\to\LLb$ transition, up to next-to-leading order in the chiral 
expansion, supplemented by a phenomenological part to account for 
annihilation processes. An excellent description of data from the PS185 
experiment at CERN up to lab momenta of $1.55$~GeV/c is achieved. 
In particular, for the 297 data points in that range a $\chi^2$/data of
around $1.4$ is obtained. 
When including higher-order contact terms in the analysis 
the range of agreement can be extended up to momenta around $1.65$~GeV/c, 
where a complete set of scattering observables is available. Here the achieved
$\chi^2$/data is around $1.8$ for 589 data points. 

We also presented predictions for $\LLb$ scattering observables and we 
compared them to results from the literature, i.e. to calculations 
with other $\LLb$ potentials that were likewise established by fitting the PS185
data on the reaction $\ppb\to\LLb$. May be not too surprising, there is no
quantitative agreement, not even between the predictions of the two
NLO potentials fitted to different momentum ranges. However, qualitatively 
the results are fairly similar with differences usually below a factor of two in 
the total and the integrated elastic cross sections, depending, of course, on 
the lab momentum.  
Selected results for the differential cross section indicate a more and 
more pronounced peak in forward direction with increasing $\Lambda$ momentum, 
and a polarization that remains overall small. 

As next step it will be interesting to apply the $\LLb$ interaction established
here to study effects of the final-state interaction in other reactions
where a $\LLb$ pair is produced. Several such reactions have been already 
investigated by us~\cite{Haidenbauer:2023zcu,Haidenbauer:2020wyp}, 
using the phenomenological
J\"ulich meson-exchange models from the 1990ties \cite{Haidenbauer:1991kt},
and by others \cite{Milstein:2021hri,Salnikov:2023azb,Milstein:2026ffc}. 
Applying the present interaction based on chiral EFT permits
a test and validation of those results. 
Of specific interest is, of course, the reaction $\ee\to\LLb$ which can be
treated rigorously and where the involved $\LLb$ partial waves are 
restricted to the $^3S_1$ and $^3D_1$ essentially. 
Measurements by the BESIII Collaboration have not only established the
integrated cross section but also the differential cross section and polarization
at some energies \cite{BESIII:2019nep,BESIII:2025kyg}, so that the
electromagnetic form factors $G_E$ and $G_M$ could be completely determined,
that means their ratio and also their relative phase. It will be instructive
to confront the results based on the new $\LLb$ interactions with those data. 

Another interesting project for the future is the application of the framework introduced in the
present work to other hyperon-production reactions. Specifically, the
reaction $\ppb\to\LSbz$+c.c. is promising because in this case there is
also an extended set of data from the PS185 experiment, which includes
spin-correlation parameters at four laboratory momenta. 
For the reaction $pp\to\SSb$ the experimental information is, however, 
very limited, at least at the moment. Future experiments, say at the 
planned FAIR facility, could possibly change this.

\iffalse
\section{Refs. to be added/discussed somewhere} 

aspects of the spin formalism 
\cite{Alberg:2009ccs} 
\\
Electromagnetic form factors of hyperons in the time-like region 
\cite{Dai:2024lau,Haidenbauer:2020wyp} and the possibly unusual threshold
behavior in the $\LLb$ case indicated by the BESIII measurement. \\
Recent data on $J/\psi\to \Lambda\bar\Sigma^0\eta+\mathrm{c.c}$ 
by BESIII \cite{BESIII:2026vrm}
indicate negligible FSI enhancement, contrary to what has been seen in
the reaction $\ee\to\LSb$+c.c. \\
$\ee\to\OOb$ with FSI \cite{Rahmani:2026uwu}. 

\vskip 0.2cm \noindent 
Treatment is also guided by the work on 
$\ee\to\LcLcb$ \cite{Dai:2017fwx,Guo:2024pti}.
(The former uses $n=6$ for the cutoff, the latter $n=2$!)
\fi

\section*{Acknowledgments}
We would like to thank B.~Bassalleck and T.~Sefzick
for helping us to get access to several PhD theses completed
in the course of the PS185 experiment. 
The work of UGM was supported in part by the CAS President's International
Fellowship Initiative (PIFI) (Grant No.~2025PD0022),  by ERC AdG EXOTIC (grant No. 101018170) and
by Deutsche For\-schungsgemeinschaft (DFG) under Grant No. 525056915
    under Germany's Excellence Strategy -- EXC 3107 -- Project-ID~533766364.
The work of QHY was supported in part by the  ERC AdG EXOTIC (grant No. 101018170).

\newpage

%\begin{widetext}
\appendix
\section{Parameters}
\label{app:Parameters}

Parameters for the $\NNb$ potential (Table~\ref{tab:paraNNb}), 
and for the $\LLb$ potentials and the $\NNb\to\LLb$ transition
potentials obtained in the different fitting scenarios described in 
Sect.~\ref{sec:fit}
(Tables~\ref{tab:paraNLcouple}, \ref{tab:my_label1}, 
and \ref{tab:my_label2}). 
The explicit form of the potentials can be found in Sect.~\ref{sec:BBbar}.
In particular, the expressions for the contact terms that involve the
low-energy constants are given by Eqs.~(\ref{VC}) and the
optical potential is specified in Eq.~(\ref{eq:Vopt}). 

\begin{table}[htbp!]
\renewcommand{\arraystretch}{1.3}
    \centering
        \caption{Parameters for the $\NNb$ potential. 
            $\tilde{C}_{{}^{2I+1,2S+1}L_J}$ and $C_{{}^{2I+1,2S+1}L_J}$ denote the low-energy constants of the contact terms
            in the different isospin- and spin channels.
    The values of the $\tilde C$'s are in $10^4$ ${\rm GeV}^{-2}$ 
the ones of the $C$'s in $10^4$ ${\rm GeV}^{-4}$.
    $U_C$, $W_C$ and $U_{LS}$, $W_{LS}$ denote the strength parameters of the (isospin-independent) 
    optical potential. Their values are given in MeV.
    The range parameter $r_0$ is in fm. 
 %   There are 975 data included in the fitting.
    }
    \begin{tabular}{c c c c}
         \hline\hline
         Cutoff (MeV)      & 800&850&900 \\ \hline
        $\tilde{C}_{{}^{11}S_0}$  & -7.2835 & -4.4557 & -3.1643  \\
        $C_{{}^{11}S_0}$          &  5.2720 &  3.8067 &  2.7367  \\
        $C_{{}^{13}P_0}$          &  3.3101 &  3.3101 &  3.3101 \\
        $C_{{}^{11}P_1}$          & -0.0741 & -0.0678 & -0.0682 \\
        $C_{{}^{13}P_1}$          &  0.2190 &  0.2190 &  0.2190 \\
        $\tilde{C}_{{}^{13}S_1}$  &  0.5629 &  0.6850 &  1.3388 \\
        $C_{{}^{13}S_1}$          & -2.0136 & -1.9535 & -1.7409 \\
        $C_{{}^{1}\epsilon_1}$    &  0.5737 &  0.5102 &  0.3077 \\
        $C_{{}^{13}P_2}$          &  1.5050 &  1.5270 &  1.5478 \\
        \hline
        $\tilde{C}_{{}^{31}S_0}$  &  0.2093 &  0.1689 &  0.1213 \\
        $C_{{}^{31}S_0}$          &  1.1749 &  1.2728 &  1.3641 \\
        $C_{{}^{33}P_0}$          &  0.7844 &  0.7848 &  0.7950 \\
        $C_{{}^{31}P_1}$          & -0.9094 & -0.9614 & -0.9925 \\
        $C_{{}^{33}P_1}$          & -0.4884 & -0.5310 & -0.5828 \\
        $\tilde{C}_{{}^{33}S_1}$  &-10.0000 &-10.0000 &-10.0000 \\
        $C_{{}^{33}S_1}$          & -0.6075 & -0.1184 & -0.1577 \\
        $C_{{}^{3}\epsilon_1}$    &  2.6144 &  2.7386 &  2.6517 \\
        $C_{{}^{33}P_2}$          &  0.0045 & -0.0110 & -0.0155 \\
        \hline
        $U_C$                            &   -41.7 &  -43.7  &  -51.3  \\
        $W_C$                            &  -115.3 & -151.4  & -197.4  \\
        $U_{LS}$                         &   -46.7 &  -55.2  &  -62.7  \\
        $W_{LS}$                         &    22.8 &   28.1  &   34.4  \\
        $r_0$                     &    0.85 &   0.79  &   0.75  
        % \hline
        % $\chi^2$                         & 22014.2 & 16693.2 & 14985.1         
        \\ \hline\hline
    \end{tabular}
    \label{tab:paraNNb}
    \renewcommand{\arraystretch}{1.0}
\end{table}
\vfill

\begin{table}[htbp!]
\renewcommand{\arraystretch}{1.3}
    \centering
        \caption{Parameters for the $\NNb$-$\LLb$ transition
        potential (top) and for the $\LLb$ potential (bottom)
        up to NLO, which was
    fitted to the data up to $p_{\rm lab}=$ 1546 MeV$/c$. 
 Same description of parameters and same units as in
 Table~\ref{tab:paraNNb}.
}
    \begin{tabular}{c c c c}
          \hline\hline
        Cut-off(MeV) & 800 & 850 & 900 \\ \hline
        \multicolumn{4}{c}{$\NNb$-$\LLb$} \\ \hline
        $\tilde{C}_{{}^{1}S_0}$ & -2.0688 & -2.0688 & -2.0688 \\
        $C_{{}^{1}S_0}$         &  1.5304 &  0.9107 &  0.6948 \\
        $C_{{}^{3}P_0}$         &  2.2517 &  1.9133 &  1.7271 \\
        $C_{{}^{1}P_1}$         &  0.3715 &  0.2990 &  0.2496 \\
        $C_{{}^{3}P_1}$         & -0.6545 & -0.6054 & -0.4535 \\
        $\tilde{C}_{{}^{3}S_1}$ & -1.6848 & -1.6541 & -1.5688 \\
        $C_{{}^{3}S_1}$         &  0.1701 &  0.1528 &  0.3962 \\
        $C_{\epsilon_1}$        & -0.1718 & -0.2094 & -0.3965 \\
        $C_{{}^{3}P_2}$         & -0.4039 & -0.3489 & -0.1103 \\
        \hline
        \multicolumn{4}{c}{$\LLb$} \\ \hline
        $\tilde{C}_{{}^{1}S_0}$     & 10.0000 &  9.9925 &  2.1220 \\
        $C_{{}^{1}S_0}$             &-10.8632 & -3.8808 & -0.4647 \\
        $C_{{}^{3}P_0}$             & 14.9999 &  7.8048 &  4.8972 \\
        $C_{{}^{1}P_1}$             & -4.8148 & -3.1893 & -2.1159 \\
        $C_{{}^{3}P_1}$             & -4.6821 & -2.7612 & -1.2785 \\
        $\tilde{C}_{{}^{3}S_1}$     & 10.0000 & 10.0000 & 10.0000 \\
        $C_{{}^{3}S_1}$             & -5.4696 & -4.5396 & -3.6103 \\
        $C_{\epsilon_1}$            &  1.0689 &  0.9925 &  0.7962 \\
        $C_{{}^{3}P_2}$             & 15.0000 & 15.0000 & 15.0000 \\
        \hline
        $U_C$                & -1074.5 & -1074.5 & -1074.5 \\
        $W_C$                & -2714.7 & -2714.7 & -2714.7 \\
        $U_{LS}$             &  1000.0 &  1000.0 &  1000.0 \\
        $W_{LS}$             &   404.2 &   404.2 &   404.2 \\
        $r_0$                &    0.53 &    0.52 &    0.50 \\
%        $(\chi^2/N)_{1642}$         &   11.56 &   10.21 &   9.75 \\
        \hline\hline
    \end{tabular}
    \label{tab:paraNLcouple}
    \renewcommand{\arraystretch}{1.0}
\end{table}
\vspace{3cm}
$\phantom{x}$

\begin{table}[htbp]
\renewcommand{\arraystretch}{1.3}
    \centering
        \caption{Parameters for the $\NNb$-$\LLb$ transition
        potential (top) and for the $\LLb$ potential (bottom)
        that involves additional contact terms in 
        higher partial waves (NLO$^+$). The potential 
        is fitted to the data up to $p_{\rm lab}=$ 1642 MeV$/c$. 
 Same description of parameters and same units as in
 Table~\ref{tab:paraNNb}.
}
    \begin{tabular}{c c c c} 
         \hline\hline
        Cut-off(MeV) & 800 & 850 & 900 \\ \hline
        \multicolumn{4}{c}{$\NNb$-$\LLb$} \\ \hline
        $\tilde{C}_{{}^{1}S_0}$ &  2.9249 &  2.4438 &  1.7538 \\
        $C_{{}^{1}S_0}$         & -0.8262 & -0.2989 &  0.0445 \\
        $C_{{}^{3}P_0}$         &  1.1859 &  1.5301 &  1.5060 \\
        $C_{{}^{1}P_1}$         &  0.0176 &  0.0220 &  0.0247 \\
        $C_{{}^{3}P_1}$         & -1.3600 & -1.5657 & -1.6821 \\
        $\tilde{C}_{{}^{3}S_1}$ & -0.0591 &  0.0106 &  0.1747 \\
        $C_{{}^{3}S_1}$         & -0.4018 & -0.4711 & -0.4930 \\
        $C_{\epsilon_1}$        &  0.2846 &  0.2498 &  0.2295 \\
        $C_{{}^{3}P_2}$         & -0.0846 &  0.0847 & -0.5637 \\
        \hline
        $D_{{}^{3}D_2}$         & -0.1215 & -0.1265 & -0.1288 \\
        $D_{\epsilon_2}$        & -0.3877 &  0.4419 &  1.1238 \\
        $D_{{}^{3}D_3}$         & -0.1347 & -0.1252 & -0.0987 \\
        $E_{{}^{3}F_2}$         & -2.9769 & -2.2200 & -2.7276 \\
        $E_{\epsilon_3}$        & -0.0789 & -0.0669 & -0.0301 \\
        $E_{{}^{3}F_4}$         & -0.0257 & -0.0170 & -0.0114 \\
        \hline
        \multicolumn{4}{c}{$\LLb$} \\ \hline
        $\tilde{C}_{{}^{1}S_0}$     & -1.6490 & -0.9996 & -0.3334 \\
        $C_{{}^{1}S_0}$             &  0.5258 &  0.2259 &  0.1722 \\
        $C_{{}^{3}P_0}$             &  9.8503 &  5.4727 &  3.0389 \\
        $C_{{}^{1}P_1}$             & -0.4555 & -0.0883 &  0.2679 \\
        $C_{{}^{3}P_1}$             &-12.9550 &-12.9550 &-12.9550 \\
        $\tilde{C}_{{}^{3}S_1}$     &  1.0299 &  0.9910 &  0.9379 \\
        $C_{{}^{3}S_1}$             & -1.7309 & -1.4861 & -1.3082 \\
        $C_{\epsilon_1}$            & -0.0321 &  0.0114 &  0.0204 \\
        $C_{{}^{3}P_2}$             &  1.1750 &  1.2308 &  0.7790 \\
        \hline
        $D_{\epsilon_2}$            & -1.1279 & -2.8424 & -2.7817 \\
        $E_{{}^{3}F_2}$             & 13.5000 & 13.5000 & 13.5000 \\
        $E_{\epsilon_3}$            &  1.1941 &  0.7178 &  0.3384 \\
        $E_{{}^{3}F_4}$             & -1.2732 & -0.8731 & -0.5610 \\
        \hline
        $U_C$                &    58.4 &   193.9 &    87.0 \\
        $W_C$                & -1810.4 & -1810.4 & -1810.4 \\
        $U_{LS}$             &   381.8 &   381.8 &   381.8 \\
        $W_{LS}$             &   233.7 &   276.6 &   323.2 \\
        $r_0$                &    0.46 &    0.46 &    0.42 \\
        \hline\hline
    \end{tabular}
    \label{tab:my_label1}
    \renewcommand{\arraystretch}{1.0}
\end{table}

\begin{table}[htbp]
\renewcommand{\arraystretch}{1.3}
    \centering
        \caption{Parameters for the $\NNb$-$\LLb$ transition
        potential (top) and
        for the $\LLb$ potential (bottom) up to NLO, which
    was fitted to the data up to $p_{\rm lab}=$ 1642 MeV$/c$
    and is denoted by NLO(he). 
 Same description of parameters and same units as in
 Table~\ref{tab:paraNNb}.
 }
    \begin{tabular}{c c c c} 
         \hline\hline
        Cut-off(MeV) & 800 & 850 & 900 \\ \hline
        \multicolumn{4}{c}{$\NNb$-$\LLb$} \\ \hline
        $\tilde{C}_{{}^{1}S_0}$ &  4.5329 &  6.4313 &  3.5011 \\
        $C_{{}^{1}S_0}$         & -1.2237 & -1.3002 & -0.4192 \\
        $C_{{}^{3}P_0}$         &  2.3915 &  2.1238 &  1.9334 \\
        $C_{{}^{1}P_1}$         &  0.7995 &  0.6641 &  0.9928 \\
        $C_{{}^{3}P_1}$         & -0.1788 & -0.5811 & -0.5782 \\
        $\tilde{C}_{{}^{3}S_1}$ & -0.4777 & -0.7178 & -0.9643 \\
        $C_{{}^{3}S_1}$         & -0.3749 & -0.3395 & -0.3699 \\
        $C_{\epsilon_1}$        &  0.3614 &  0.3828 &  0.4175 \\
        $C_{{}^{3}P_2}$         & -0.8780 & -0.3394 & -0.3504 \\
        \hline
        \multicolumn{4}{c}{$\LLb$} \\ \hline
        $\tilde{C}_{{}^{1}S_0}$     &  0.7814 &  0.7814 &  0.7814 \\
        $C_{{}^{1}S_0}$             & -1.3614 & -1.3647 & -0.5816 \\
        $C_{{}^{3}P_0}$             & 15.0000 &  9.3718 &  5.6829 \\
        $C_{{}^{1}P_1}$             & -9.3538 & -6.8440 & -6.5493 \\
        $C_{{}^{3}P_1}$             & -0.8676 & -2.2050 & -1.9614 \\
        $\tilde{C}_{{}^{3}S_1}$     &  1.8125 &  2.2079 &  2.8605 \\
        $C_{{}^{3}S_1}$             & -2.4510 & -2.0585 & -1.8248 \\
        $C_{\epsilon_1}$            & -0.8034 & -0.5222 & -0.5510 \\
        $C_{{}^{3}P_2}$             & -0.6262 &  0.1851 & -1.9459 \\
        \hline
        $U_C$                &   587.5 &   587.5 &   587.5 \\
        $W_C$                & -5000.0 & -5000.0 & -5000.0 \\
        $U_{LS}$             &  -369.5 &  -369.5 &  -369.5 \\
        $W_{LS}$             &  -500.0 &  -500.0 &  -500.0 \\
        $r_0$                &    0.46 &    0.46 &    0.44 \\
        \hline\hline
    \end{tabular}
    \label{tab:my_label2}
    \renewcommand{\arraystretch}{1.0}
\end{table}

%\doublepagebreak

\clearpage

 \section{SU(3) relations for the $\bar BB$ interactions}
\label{app:SU3}
In this appendix we provide the SU(3) relations for the interactions in 
all $\bar BB$ systems from strangeness $S=2$ to $S=-2$.
Those relations can be readily deduced from the isoscalar factors for the ${\rm SU(3)}$ structure relevant for scattering of two octet baryons,
namely the direct product $8$ $\otimes$ $8$ and its decomposition
$8$ $\otimes$ $8$ = $1$ $\oplus$ $8_a$ $\oplus$ $8_s$ $\oplus$ $10^*$ $\oplus$ $10$ $\oplus$ $27$
\cite{deSwart:1963pdg}. 
The corresponding factors can be found in Table II of Ref.~\cite{deSwart:1963pdg}.   
With those one can express all the 
$C_{^1S_0,{\nu}}$, $C_{^3S_1,{\nu }},\ldots$, in Eqs.~(\ref{VC})
($\nu$= $\NNb\to \NNb$, $\LLb\to \LLb$,
..., $\bar\Xi N\to \bar \Xi N$)
by coefficients corresponding to the {$\mathrm{SU}(3)$} irreducible representations (irreps): 
$C^1$, $C^{8_a}$, $C^{8_s}$, $C^{10^*}$, $C^{10}$, $C^{27}$. 
However, contrary to the $YN$ case considered in 
Refs.~\cite{Haidenbauer:2013oca,Haidenbauer:2019boi,Haidenbauer:2023qhf,Dover:1990id},
antibaryons and baryons do not belong to the same multiplet and, thus, there is no general Pauli principle that provides additional
constraints. Therefore, in general, all SU(3) irreps
will contribute to a specific partial wave. Moreover, mixing
between the $8_s$ and $8_a$, denoted here by $8_{as}$, is possible. 
Indeed, the situation is similar to that in meson-baryon scattering where
the particles are likewise from different multiplets.  
The particular combinations of LECs in the various $\bar BB\to \bar BB$ channels, valid for any partial wave, are summarized in Tab.~\ref{Tab:SU3}.
For example, for the potential in an $S$-wave 
($^1S_0$ or $^3S_1$) of the $\LLb \to \LLb$ channel we get
\begin{eqnarray}
V_{\LLb \to \LLb} &=& \frac1{40}\left[27\tilde C^{27}_\alpha + 
8\tilde C^{8_s}_\alpha + 5\tilde C^{1}_\alpha \right. 
\nonumber \\
&&\left.+ (27C^{27}_\alpha + 8C^{8_s}_\alpha + 5C^{1}_\alpha)(p^2+p'^2)\right] \ , \nonumber \\
&&
\end{eqnarray}
while for a $P$-wave ($^3P_0$, $^1P_1$, ...) it is 
\begin{eqnarray}
V_{\LLb \to \LLb} &=& \frac1{40}\left[(27C^{27}_\alpha + 8C^{8_s}_\alpha + 5C^{1}_\alpha)\,pp'\right] \ . \nonumber \\
&&
\end{eqnarray}

The fact that SU(3) symmetry is much less restrictive for 
$\bar BB$ than for $BB$ has consequences.  
While there are only three independent
LECs for a specific partial wave in the $BB$ case so that constraints
from the symmetry have already an impact when there are more than 3 channels,
there are 6 or 7 independent LECs for $\bar BB$ so that evidently SU(3)
restrictions arise only if one considers more than 6 or 7 channels.
In any case particle-antiparticle symmetry requires that 
$C^{10} \equiv C^{10^*}$ as one can see from Table~\ref{Tab:SU3}
by comparing, e.g. $\bar N\Lambda$ and $\bar\Lambda N$. 

We checked the consistency of our results with SU(3) symmetry in a 
two-step process. First we calculated the interaction from single-meson
exchange (with an averaged meson mass) in selected $\BBb$ channels 
based on the coupling constants in Eq.~(\ref{eq:SU3})
and the isospin coefficients in Table~\ref{Tab:IFOBE}, and then used 
them to determine the respective LECs in the corresponding   
representation in terms of SU(3) irreps based on Table~\ref{Tab:SU3}. 
Afterwards we evaluated the interaction in other
$\BBb$ channels, from single-meson exchange and from
the SU(3) representation with the LECs fixed in step
one, and compared them. The same procedure was also carried out for 
the individual contributions from two-meson exchange. 

\newpage
\onecolumn
	\renewcommand{\arraystretch}{1.3}
\begin{longtable}{c|c|c|ccccccc}
		\caption{The $SU(3)$ structure of the $\bar BB$ interaction potentials} \label{Tab:SU3}\\ 
		\hline\hline
		                            &                          Channel                           &    Isospin    &          $27$           &          $10$          &         $10^*$         &         $8_s$          &     $8_a$      &          $1$          &        $8_{as}$         \\ \hline
		 \endfirsthead
		 \multicolumn{10}{r}{$\cdots$ continued from previous page}\\
		 \hline\hline
		                            &                          Channel                           &    Isospin    &          $27$           &          $10$          &         $10^*$         &         $8_s$          &     $8_a$      &          $1$          &        $8_{as}$         \\ \hline
		 \endhead
		 \hline\hline
		 \multicolumn{10}{r}{$\cdots$ continued on next page} \\
		 \endfoot
		 
		 \hline\hline
		 \endlastfoot
		 
		 \multirow{2}{*}{$S=2$}  &         \multirow{2}{*}{$\bar{\Xi}N\to\bar{\Xi}N$}         &       0       &            0            &           0            &           1            &           0            &       0        &           0           &            0            \\
		                         &                                                            &       1       &            1            &           0            &           0            &           0            &       0        &           0           &            0            \\ \hline
		 \multirow{13}{*}{$S=1$} &             $\bar{\Lambda}N\to\bar{\Lambda}N$              & $\frac{1}{2}$ &     $\frac{9}{20}$      &           0            &     $\frac{1}{4}$      &     $\frac{1}{20}$     & $\frac{1}{4}$  &           0           &  $\frac{\sqrt{5}}{10}$  \\ \cline{2-10}
		                         &              $\bar{\Lambda}N\to\bar{\Sigma}N$              & $\frac{1}{2}$ &     $-\frac{3}{20}$     &           0            &     $\frac{1}{4}$      &     $\frac{3}{20}$     & $-\frac{1}{4}$ &           0           &  $\frac{\sqrt{5}}{10}$  \\ \cline{2-10}
		                         &      \multirow{2}{*}{$\bar{\Sigma}N\to\bar{\Sigma}N$}      & $\frac{1}{2}$ &     $\frac{1}{20}$      &           0            &     $\frac{1}{4}$      &     $\frac{9}{20}$     & $\frac{1}{4}$  &           0           & $-\frac{3\sqrt{5}}{10}$ \\
		                         &                                                            & $\frac{3}{2}$ &      $\frac{1}{2}$      &     $\frac{1}{2}$      &           0            &           0            &       0        &           0           &            0            \\ \cline{2-10}
		                         &            $\bar{\Lambda}N\to\bar{\Xi}\Lambda$             & $\frac{1}{2}$ &     $\frac{9}{20}$      &           0            &     $-\frac{1}{4}$     &     $\frac{1}{20}$     & $-\frac{1}{4}$ &           0           &            0            \\
\hline
                                                          
	\end{longtable}
 \twocolumn
	\renewcommand{\arraystretch}{1.0}

 \onecolumn
\section{Isospin coefficients}
\label{app:Iso}

\normalfont{
In this appendix we provide tables with the isospin coefficients in the one- and 
two-meson exchange diagrams. 

%\bf{We need to explain} that $\pi K$ means $\pi K$ + $K\pi$, etc. 

	\renewcommand{\arraystretch}{1.0}
	\begin{table*}[htbp!]
		\centering
  		\caption{Isospin factors for one pseudoscalar meson exchange for 
     the strangeness $S=0$ sector.}
		\begin{tabular}{c|ccc|cc|ccc}
			\hline\hline
			                  Channels                    & \multicolumn{3}{c}{$\pi$ exchange} & \multicolumn{2}{c}{$K$ exchange} & \multicolumn{3}{c}{$\eta$ exchange} \\ \hline
			                                              &    $I=0$    &    $I=1$    & $I=2$  &    $I=0$    &       $I=1$        & $I=0$ & $I=1$ &        $I=2$        \\ \hline
			           $\bar{N}N\to \bar{N}N$             &      3      &    $-1$     &   0    &      0      &         0          &   1   &   1   &          0          \\
			$\bar{\Lambda}\Lambda\to\bar{\Lambda}\Lambda$ &      0      &      0      &   0    &      0      &         0          &   1   &   0   &          0          \\
			     $\bar{N}N\to \bar{\Lambda}\Lambda$       &      0      &      0      &   0    & $\sqrt{2}$  &         0          &   0   &   0   &          0          \\
			$\bar{\Lambda}\Lambda\to \bar{\Sigma}\Sigma$  & $-\sqrt{3}$ &      0      &   0    &      0      &         0          &   0   &   0   &          0          \\
			 $\bar{\Sigma}\Sigma\to \bar{\Sigma}\Sigma$   &      2      &      1      &  $-1$  &      0      &         0          &   1   &   1   &          1          \\
			      $\bar{N}N\to \bar{\Sigma}\Sigma$        &      0      &      0      &   0    & $-\sqrt{6}$ &        $-2$        &   0   &   0   &          0          \\
\hline\hline
		\end{tabular}
     \label{Tab:IFOBE}
	\end{table*}
	\renewcommand{\arraystretch}{1.0}

\renewcommand{\arraystretch}{1.0}
	\begin{table*}[htbp!]
		\centering
  		\caption{Isospin factors for contributions from the football diagram for $S=0$}
		\begin{tabular}{c|c|cccccc}
			\hline\hline
			                         Channels                          & Isospin & $\pi\pi$ & $\pi\eta$ & $\eta\eta$ & $KK$  &    $K\pi$     &    $K\eta$    \\ \hline
			          \multirow{2}{*}{$\bar{N}N\to\bar{N}N$}           &    0    &  $-24$   &     0     &     0      & $-48$ &       0       &       0       \\
			                                                           &    1    &    8     &     0     &     0      & $-32$ &       0       &       0       \\
			      $\bar{\Lambda}\Lambda\to\bar{\Lambda}\Lambda$        &    0    &    0     &     0     &     0      &   0   &       0       &       0       \\
			            $\bar{N}N\to\bar{\Lambda}\Lambda$              &    0    &    0     &     0     &     0      &   0   & $-18\sqrt{2}$ & $-18\sqrt{2}$ \\
			       $\bar{\Lambda}\Lambda\to\bar{\Sigma}\Sigma$         &    0    &    0     &     0     &     0      &   0   &       0       &       0       \\
			\multirow{3}{*}{$\bar{\Sigma}\Sigma\to\bar{\Sigma}\Sigma$} &    0    &  $-64$   &     0     &     0      & $-32$ &       0       &       0       \\
			                                                           &    1    &  $-32$   &     0     &     0      & $-16$ &       0       &       0       \\
\hline\hline
		\end{tabular}
\label{Tab:IFFootball}
	\end{table*}
	\renewcommand{\arraystretch}{1.0}

   \renewcommand{\arraystretch}{1.0}
   {\small
   \begin{longtable}{ccccccccccccc c}
   			\caption{Isospin factors for contributions from the left triangle diagram for~$S=0$}\\ 
   			\hline\hline
   			${\rm Channel}$ & $I$ & $B_{ir}$ & $\pi\pi$ & $\pi\eta$ & $\eta\pi$ & $\eta\eta$ & $\pi K$ & $\eta K$ & $B_{ir}$ & $KK$ & $\bar{K}\bar{K}$ & $\bar{K}\pi$ & $\bar{K}\eta$ \\ \hline
   			\endfirsthead
   			\multicolumn{9}{r}{$\cdots$ continued from previous page}\\
   			\hline\hline
   			Channel & I & $B_{ir}$ & $\pi\pi$ & $\pi\eta$ & $\eta\pi$ & $\eta\eta$ & $\pi K$ & $\eta K$ & $B_{ir}$ & $KK$ & $\bar{K}\bar{K}$ & $\bar{K}\pi$ & $\bar{K}\eta$ \\ \hline
   			\endhead
   			\hline\hline
   			\multicolumn{9}{r}{$\cdots$ continued on next page} \\
   			\endfoot
   			
   			\hline\hline
   			\endlastfoot
                $\NNb\to\NNb$ & $0$ & $N$       & $12$         & $0$ & $0$ & $0$ & $0$          & $0$          & $\Lambda$ & $6$         & $0$         & $0$          & $0$          \\
        		              &     &           &              &     &     &     &              &              & $\Sigma$  & $6$         & $0$         & $0$          & $0$          \\
        		              & $1$ & $N$       & $-4$         & $0$ & $0$ & $0$ & $0$          & $0$          & $\Lambda$ & $2$         & $0$         & $0$          & $0$          \\
        		              &     &           &              &     &     &     &              &              & $\Sigma$  & $10$        & $0$         & $0$          & $0$          \\
        		$\LLb\to\LLb$ & $0$ &           &              &     &     &     &              &              &           &             &             &              &              \\
        		$\NNb\to\LLb$ & $0$ & $\Lambda$ & $0$          & $0$ & $0$ & $0$ & $0$          & $3\sqrt{2}$  & $N$       & $0$         & $0$         & $-3\sqrt{6}$ & $-3\sqrt{2}$ \\
        		              &     & $\Sigma$  & $0$          & $0$ & $0$ & $0$ & $3\sqrt{6}$  & $0$          &           &             &             &              &              \\
        		$\LLb\to\SSb$ & $0$ &           &              &     &     &     &              &              &           &             &             &              &              \\
\end{longtable}
}
\twocolumn
   \renewcommand{\arraystretch}{1.0}
   
   \renewcommand{\arraystretch}{1.2}
   \onecolumn
   {\small
   \begin{longtable}{cccccccccccccc}
   	\caption{Isospin factors for contributions from the right triangle diagram for $S=0$}\\ 
   	\hline\hline
   	${\rm Channel}$ & $I$ & $\bar{B}_{il}$ & $\pi\pi$ & $\pi\eta$ & $\eta\pi$ & $\eta\eta$ & $\pi\bar{K}$ & $\eta\bar{K}$ & $\bar{B}_{il}$ & $KK$ & $\bar{K}\bar{K}$ & $K\pi$ & $K\eta$ \\ \hline
   	\endfirsthead
   	\multicolumn{9}{r}{$\cdots$ continued from previous page}\\
   	\hline\hline
   	Channel & I & $\bar{B}_{ir}$ & $\pi\pi$ & $\pi\eta$ & $\eta\pi$ & $\eta\eta$ & $\pi\bar{K}$ & $\eta\bar{K}$ & $\bar{B}_{ir}$ & $KK$ & $\bar{K}\bar{K}$ & $K\pi$ & $K\eta$ \\ \hline
   	\endhead
   	\hline\hline
   	\multicolumn{9}{r}{$\cdots$ continued on next page} \\
   	\endfoot
   	           
   	\hline\hline
   	\endlastfoot
                $\NNb\to\NNb$ & $0$ & $\bar N$      & $12$         & $0$ & $0$ & $0$ & $0$          & $0$          & $\bar\Lambda$ & $0$         & $6$         & $0$          & $0$          \\
        		              &     &               &              &     &     &     &              &              & $\bar\Sigma$  & $0$         & $6$         & $0$          & $0$          \\
        		              & $1$ & $\bar N$      & $-4$         & $0$ & $0$ & $0$ & $0$          & $0$          & $\bar\Lambda$ & $0$         & $2$         & $0$          & $0$          \\
        		              &     &               &              &     &     &     &              &              & $\bar\Sigma$  & $0$         & $10$        & $0$          & $0$          \\
        		$\LLb\to\LLb$ & $0$ &               &              &     &     &     &              &              &               &             &             &              &              \\
        		$\NNb\to\LLb$ & $0$ & $\bar\Lambda$ & $0$          & $0$ & $0$ & $0$ & $0$          & $3\sqrt{2}$  & $\bar N$      & $0$         & $0$         & $-3\sqrt{6}$ & $-3\sqrt{2}$ \\
        		              &     & $\bar\Sigma$  & $0$          & $0$ & $0$ & $0$ & $3\sqrt{6}$  & $0$          &               &             &             &              &              \\
        		$\LLb\to\SSb$ & $0$ &               &              &     &     &     &              &              &               &             &             &              &              \\
\end{longtable}
}
\twocolumn
   \renewcommand{\arraystretch}{1.0}
   
    \renewcommand{\arraystretch}{1.2}
    \onecolumn
    {\tiny
   \begin{longtable}{ccccccccccccccc}
   	\caption{Isospin factors for contributions from the planar box diagram for $S=0$}\\ 
   	\hline\hline
   	${\rm Channel}$ & $I$ & $\bar{B}_{il}B_{ir}$ & $\pi\pi$ & $\pi\eta$ & $\eta\pi$ & $\eta\eta$ & $\pi\bar{K}$ & $\eta\bar{K}$ & $\bar{B}_{il}B_{ir}$ & $KK$ & $\bar{K}\bar{K}$ & $K\bar{K}$ & $K\pi$ & $K\eta$ \\ \hline
   	\endfirsthead
   	\multicolumn{9}{r}{$\cdots$ continued from previous page}\\
   	\hline\hline
   	Channel & I & $\bar{B}_{il}B_{ir}$ & $\pi\pi$ & $\pi\eta$ & $\eta\pi$ & $\eta\eta$ & $\pi\bar{K}$ & $\eta\bar{K}$ & $\bar{B}_{il}B_{ir}$ & $KK$ & $\bar{K}\bar{K}$ & $K\bar{K}$ & $K\pi$ & $K\eta$ \\ \hline
   	\endhead
   	\hline\hline
   	\multicolumn{9}{r}{$\cdots$ continued on next page} \\
   	\endfoot
   	 
   	\hline\hline 
   	\endlastfoot
                $\NNb\to\NNb$ & $0$ & $\NNb$        & $9$          & $3$         & $3$         & $1$ & $0$          & $0$         & $\LLb$        & $0$          & $2$          & $0$  & $0$          & $0$         \\
        		              &     &               &              &             &             &     &              &             & $\SSb$        & $0$          & $6$          & $0$  & $0$          & $0$         \\
        		              & $1$ & $\NNb$        & $1$          & $-1$        & $-1$        & $1$ & $0$          & $0$         & $\SSb$        & $0$          & $4$          & $0$  & $0$          & $0$         \\
        		              &     &               &              &             &             &     &              &             & $\LSb,\,\SLb$ & $0$          & $2$          & $0$  & $0$          & $0$         \\
        		$\LLb\to\LLb$ & $0$ & $\LLb$        & $0$          & $0$         & $0$         & $1$ & $0$          & $0$         & $\NNb$        & $2$          & $0$          & $0$  & $0$          & $0$         \\
        		              &     & $\SSb$        & $3$          & $0$         & $0$         & $0$ & $0$          & $0$         & $\XXb$        & $0$          & $2$          & $0$  & $0$          & $0$         \\
        		$\NNb\to\LLb$ & $0$ & $\LLb$        & $0$          & $0$         & $0$         & $0$ & $0$          & $\sqrt{2}$  & $\NNb$        & $0$          & $0$          & $0$  & $3\sqrt{2}$  & $\sqrt{2}$  \\
        		              &     & $\SSb$        & $0$          & $0$         & $0$         & $0$ & $3\sqrt{2}$  &             &               &              &              & $0$  &              &             \\
\end{longtable}}
   \twocolumn
   \renewcommand{\arraystretch}{1.0} 
   
   \renewcommand{\arraystretch}{1.2}
   \onecolumn
   {\tiny
   \begin{longtable}{ccccccccccccccc}
   	\caption{Isospin factors for contributions of the cross box diagram for $S=0$}\\ 
   	\hline\hline
   	${\rm Channel}$ & $I$ & $\bar{B}_{il}B_{ir}$ & $\pi\pi$ & $\pi\eta$ & $\eta\pi$ & $\eta\eta$ & $\pi\bar{K}$& $\eta\bar{K}$ & $\bar{B}_{il}B_{ir}$ & $KK$ & $\bar{K}\bar{K}$ & $K\bar{K}$ & $K\pi$ & $K\eta$ \\ \hline
   	\endfirsthead
   	\multicolumn{9}{r}{$\cdots$ continued from previous page}\\
   	\hline\hline
   	Channel & I & $\bar{B}_{il}B_{ir}$ & $\pi\pi$ & $\pi\eta$ & $\eta\pi$ & $\eta\eta$ & $\pi\bar{K}$& $\eta\bar{K}$ & $\bar{B}_{il}B_{ir}$ & $KK$ & $\bar{K}\bar{K}$ & $K\bar{K}$ & $K\pi$ & $K\eta$ \\ \hline
   	\endhead
   	\hline\hline
   	\multicolumn{9}{r}{$\cdots$ continued on next page} \\
   	\endfoot
    $\NNb\to\NNb$ & $0$ & $\NNb$        & $-3$        & $3$         & $3$         & $1$ & $0$          & $0$         &               &              &              & $0$  &              &             \\
        		              & $1$ & $\NNb$        & $5$         & $-1$        & $-1$        & $1$ & $0$          & $0$         &               &              &              & $0$  &              &             \\
        		$\LLb\to\LLb$ & $0$ & $\LLb$        & $0$         & $0$         & $0$         & $1$ & $0$          & $0$         & $\NXb$        & $2$          & $0$          & $0$  & $0$          & $0$         \\
        		              &     & $\SSb$        & $3$         & $0$         & $0$         & $0$ & $0$          & $0$         & $\XNb$        & $0$          & $2$          & $0$  & $0$          & $0$         \\
        		$\NNb\to\LLb$ & $0$ & $\LNb$        & $0$         & $0$         & $0$         & $0$ & $0$          & $\sqrt{2}$  & $\NLb$        & $0$          & $0$          & $0$  & $0$          & $\sqrt{2}$  \\
        		              &     & $\SNb$        & $0$         & $0$         & $0$         & $0$ & $3\sqrt{2}$  & $0$         & $\NSb$        & $0$          & $0$          & $0$  & $3\sqrt{2}$  & $0$         \\
        		$\LLb\to\SSb$ & $0$ & $\LSb$        & $0$         & $-\sqrt{3}$ & $0$         & $0$ & $0$          & $0$         & $\NXb$        & $2\sqrt{3}$  & $0$          & $0$  & $0$          & $0$         \\
        		              &     & $\SLb$        & $0$         & $0$         & $-\sqrt{3}$ & $0$ & $0$          & $0$         & $\XNb$        & $0$          & $2\sqrt{3}$  & $0$  & $0$          & $0$         \\
\end{longtable}}
   \twocolumn
   \renewcommand{\arraystretch}{1.0}
}

\onecolumn
\section{Predictions of the NLO$^+$ interaction for higher energies}
\label{app:highe}

\medskip

\normalfont{
 In this appendix, we show some predictions for differential cross sections, polarization observables and spin correlation parameters
 for the NLO$^+$ interaction for lab momenta larger than 1650~MeV/c, that is beyond the range of fitting. 
}
\begin{figure}[htb!]
    \centering
    \includegraphics[width=0.99\linewidth]{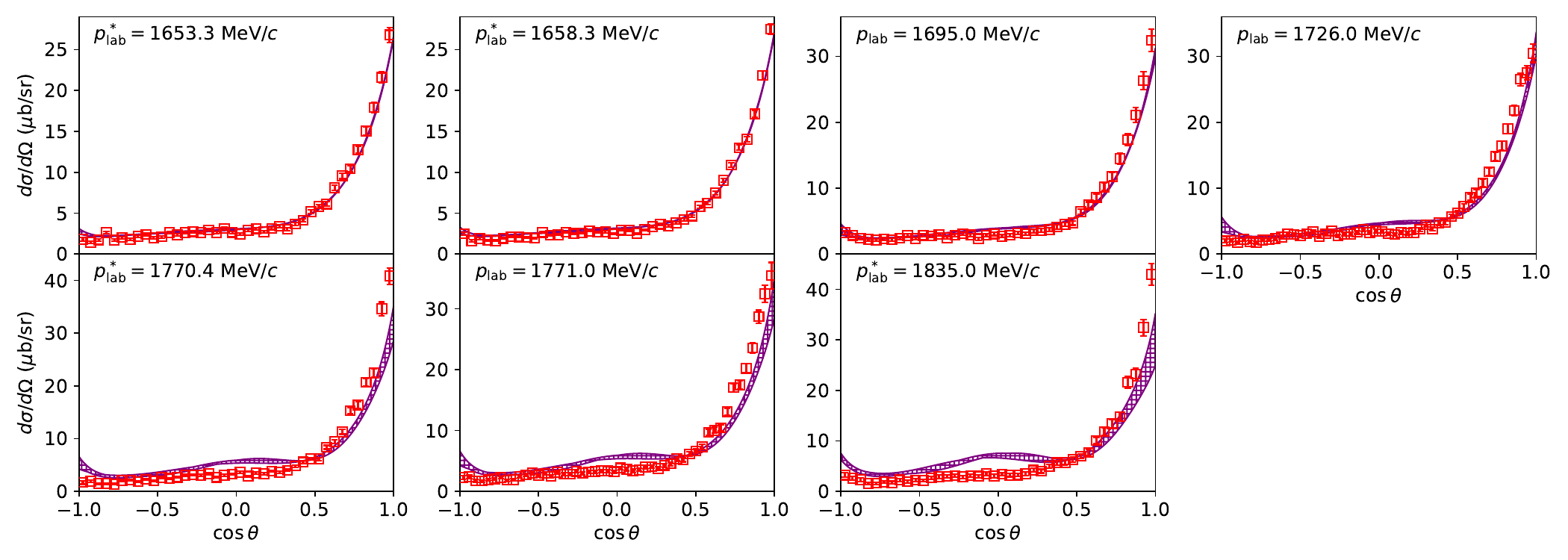}
    \caption{Prediction of the NLO$^+$ potentials for differential cross sections of the reaction $\ppb\to\LLb$ at higher energies. Data are taken from Refs.~\cite{Barnes:1990bs,Barnes:1996cf}.
    and from the theses \cite{Todenhagen:1995,Kraft:1994}. In the
    latter case the pertinent momenta are marked with an asterisk.
    }
    \label{fig:ppbLLbdsigma2}
\end{figure}

\begin{figure}[htb!]
    \centering
    \includegraphics[width=0.99\linewidth]{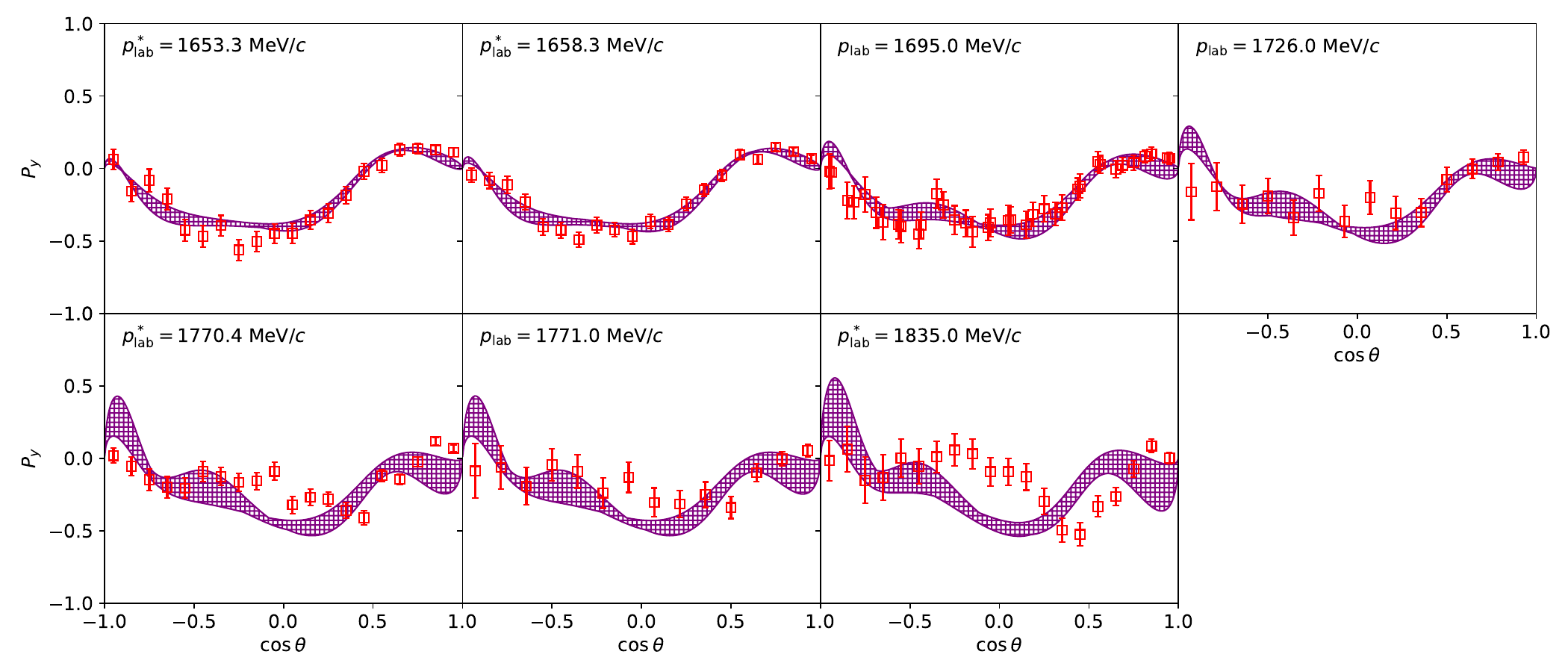}
    \caption{Prediction for polarization of the reaction $\ppb\to\LLb$ at higher energies.
    Same description of curves and data as in
    Fig.~\ref{fig:ppbLLbdsigma2}.}
    \label{fig:ppbLLbPy2}
\end{figure}

\begin{figure}
    \centering
    \includegraphics[width=0.99\linewidth]{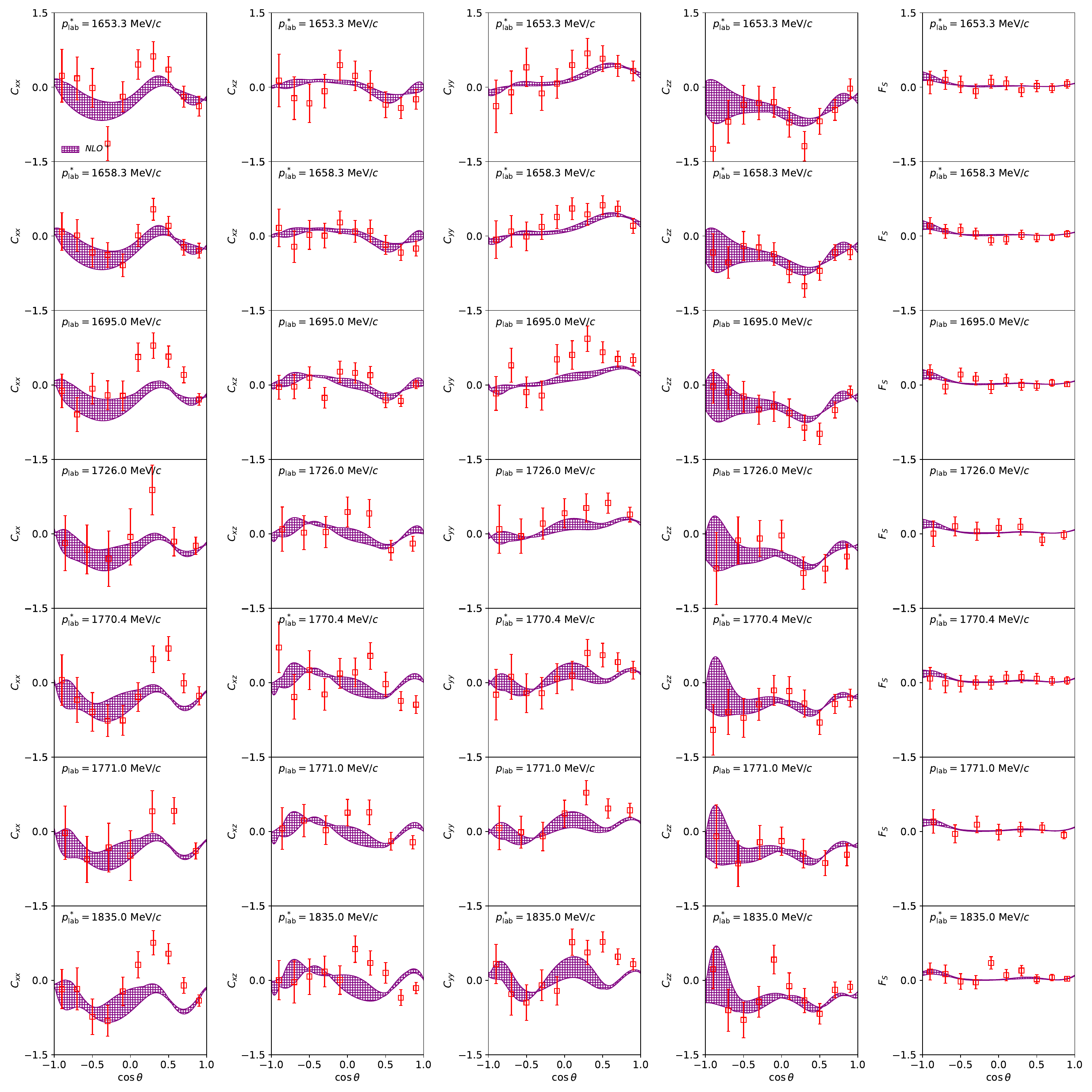}
    \caption{Prediction for spin correlation parameters $C_{ij}$ of the reaction $\ppb\to\LLb$ at higher energies.
    Same description of curves and data as in
    Fig.~\ref{fig:ppbLLbdsigma2}.}
    \label{fig:ppbLLbSC3}
\end{figure}

\twocolumn

 \bibliographystyle{unsrturl}
\bibliography{ref,refbb,refnn}

\end{document}